\documentclass[9pt,journal,twocolumn]{IEEEtran}

\usepackage[utf8]{inputenc}
\usepackage{graphicx}
\usepackage{epstopdf}
\usepackage{amsfonts}
\usepackage[cmex10]{amsmath}
\usepackage{mathtools}
\usepackage{framed,xcolor}
\usepackage{cite}
\definecolor{shadecolor}{gray}{0.90}

\DeclareMathAlphabet{\mathantt}{OT1}{antt}{li}{it}
\DeclareMathAlphabet{\mathantt}{OT1}{pzc}{m}{it}

\begin{document}
\title{Orbiter-to-orbiter tomography: a potential approach for small planetary objects}
\author{S.~Pursiainen,  M.~Kaasalainen
\thanks{Manuscript received February 22, 2016. Accepted for publication May 10,
2016. \footnotesize \textcopyright 2016 IEEE. Personal use of this material is permitted.
  Permission from IEEE must be obtained for all other uses, in any current or future 
  media, including reprinting/republishing this material for advertising or promotional 
  purposes, creating new collective works, for resale or redistribution to servers or 
  lists, or reuse of any copyrighted component of this work in other works.} \thanks{S.\ Pursiainen (corresponding author) and M.\ Kaasalainen are  with the Department of Mathematics, Tampere University of Technology, Finland.}}
\maketitle
\begin{abstract}
The goal of this paper is to advance mathematical and computational methodology for orbiter-to-orbiter radio
tomography of small planetary objects. In this study, an advanced full waveform forward model is coupled with a
total variation based inversion technique. We use a satellite formation model in which a single unit receives a signal
that is transmitted by one or more transponder satellites. Numerical results for a 2D domain are presented.
\end{abstract}

\IEEEpeerreviewmaketitle


\newcommand{\plotwidth}
{
  0.90\columnwidth
}

\newcommand{\figureMagRdm}[3]{%
\begin{figure}[h]
 \begin{center}
  \includegraphics[width=\plotwidth]{#1_rdmError} \\
RDM
  \includegraphics[width=\plotwidth]{#1_magError} \\
MAG
 \end{center}
 \caption{#3}
 \label{fig:#2}
\end{figure}}

\section{Introduction}

The goal of this paper is to advance mathematical and computational methodology for radio tomography of small planetary objects (SPOs) \cite{mann2008, michel2015, kaasalainen2013}. Tomographic imaging of SPO interiors can be seen as a natural future direction following the development of surface reconstruction methods that are nowadays extensively used in planetary research \cite{kaasalainen1992, kaasalainen2006,kaasalainen2011}.  The aim is to reconstruct the target SPO's internal permittivity distribution.  This is an ill-posed and non-linear inverse problem \cite{kaipio2004} in which even a slight amount of noise in the data or in the forward simulation can lead to very large errors in the final estimate. Moreover, the permittivity distribution to be recovered can be complicated including, e.g., anisotropicity \cite{nabighian1988}, surface layers \cite{benna2002}, voids \cite[p.\ 143]{belton2004} and dependence on temperature, humidity or electric field frequency \cite{herique2002}. 

The first attempt to recover the  interior of an SPO based on radio-frequency measurements has been made  in the CONSERT (COmet Nucleus Sounding Experiment by Radiowave  Transmission) experiment with Comet P67/Churyumov-Gerasimenko as the target  \cite{kofman2015,kofman2007,kofman2004,benna2004,kofman1998,herique2011,herique2011b,herique2010,herique1999,landmann2010,nielsen2001,barriot1999, pursiainen2013, pursiainen2014, pursiainen2014b}.  In CONSERT, the radio signal is transferred between the mothership {\em Rosetta} and its lander {\em Philae}. The full potential of radio tomography is yet to be discovered in the coming missions, which might utilize imaging techniques similar to the georadar surveys of today \cite{daniels2004}.  In this paper, we discuss a tomography approach which, in principle, can be applied in the future planetary missions to invert  data recorded by a multi-satellite system. Such an approach has previously been used, for example, in the exploration of the Earth's magnetosphere and ionosphere \cite{agrawal2014}. The mathematical basis and 2D test domain of  this study inherit from our recent, more concise study  of anomaly localization \cite{pursiainen2014}. 

Our focus is on the essential mathematical and computational aspects of waveform tomography, which include, among other things, simulating data for an extensive set of signal transmitter and receiver  positions, producing robust inverse estimates, and speeding up of computations for 3D tomographic imaging. Central in this study is a full waveform forward model coupling the finite-difference time-domain (FDTD) method \cite{schneider2012} with a deconvolution process which enables    computation of the actual and differentiated signal for a comprehensive set of data. The computational domain is discretized spatially utilizing the finite element method (FEM)  \cite{braess2007} which allows accurate adaptation of the forward model to geometrically complex interior and boundary structures.  The internal permittivity of the 2D test object is reconstructed using a total variation regularized inversion technique \cite{scherzer2008, stefan2008, kaipio2004} that is applicable for an arbitrary permittivity distribution defined on the finite element (FE) mesh.  

The numerical experiments included in this study cover both dense and sparse signal configurations and two different permittivity distributions. 
The results suggest that the present full waveform approach allows simultaneous reconstruction of both voids and a surface layer, e.g.\ dust or ice cover, if the data can be recorded for a sufficiently dense network of orbital transmitter and receiver positions. Additionally, to enable effective future 3D implementations of the present tomography strategy,  multiresolution and incomplete Cholesky speedups \cite{braess2007, golub1989}, are presented and discussed.

\section{Materials and methods}

\subsection{Forward model}
\label{section:forward_model} 

The forward model predicts the electric potential $u$ in the set $[0, T] \times \Omega$ in which $[0,T]$ is a time interval and $\Omega$ is the spatial part  containing the target SPO $\Omega_0$ together with the orbiter paths. Given a real-valued relative electric permittivity $\varepsilon_r$, real conductivity distribution $\sigma$, and the initial conditions $u|_{t = 0} = u_0$ and $(\partial u/ \partial t) |_{t  = 0} = u_1$,  the electric potential satisfies the following hyperbolic wave equation
\begin{equation} 
\label{pde1}
\varepsilon_r \frac{\partial^2 u}{\partial t^2} + \sigma \frac{\partial u}{\partial t}  - \Delta_{\vec x} u = \frac{\partial f}{\partial t} \quad \hbox{for all} \quad (t,{\vec x}) \in  [0, T] \times \Omega. 
\end{equation}
Here,  the right hand side is a signal of the form $\partial f(t, \vec{x} ) / \partial t= \delta_{\vec{p}}(\vec{x}) \partial \tilde{f}(t) / \partial t $ transmitted at point $\vec{p}$ with $\tilde{f}(t)$ denoting the time-dependent part of $f$  and $\delta_{\vec{p}}(\vec{x})$ a Dirac's delta function with respect to $\vec{p}$.  Definition ${\vec g} = \int_0^t \nabla u(\tau, \vec{x}) \, d \tau$  yields the first order system 
\begin{equation}\label{first-order_system}
\varepsilon_r \frac{\partial u}{\partial t} + \sigma {u} - \nabla \cdot {\vec g}  =   f \quad \hbox{and} \quad 
\frac{\partial {\vec g}}{\partial t}  -   \nabla u  =  0, \quad \hbox{in} \quad \Omega \times [0,T], 
\end{equation}
where ${\vec g} |_{t = 0}  = \nabla u_0$ and $u |_{t = 0} = u_1$. Multiplying the first and the second equation of (\ref{first-order_system}) by the test functions $v  : [0,T] \to H^1(\Omega)$ and ${\vec w} : [0,T] \to L_2(\Omega)$, respectively,  and integrating by parts leads to the weak formulation 
{\setlength\arraycolsep{1 pt} \begin{eqnarray} \label{weak_form_1}
\frac{\partial }{\partial t} \! \int_\Omega \! {\vec g} \cdot {\vec w} \, \hbox{d} \Omega \! - \! \int_\Omega \! {\vec w} \cdot \nabla u \, \hbox{d}  \Omega & = & 0, \\ 
\frac{\partial}{\partial t} \!  \int_\Omega \! \varepsilon_r  \, u \, v \, \hbox{d}  \Omega \! + \! \int_\Omega \!  \sigma  \, u \, v \, \hbox{d}  \Omega \! + \! \int_\Omega \!  {\vec g} \cdot \nabla v \, \hbox{d}  \Omega & = & \left\{ \begin{array}{ll} \tilde{f}(t), & \hbox{if } \vec{x} \! = \! \vec{p}, \\ 0, &  \hbox{else}.  \end{array} \right. \label{weak_form_2}
\end{eqnarray}} Under regular enough conditions the weak form has a unique solution $u : [0,T] \to H^1(\Omega)$ \cite{evans1998}. SI-unit values corresponding to $t,{\vec x}$, $\varepsilon_r$, $\sigma$ and ${\mathsf c}= \varepsilon_r^{-1/2}$ (signal velocity) can be obtained, respectively, via $(\mu_0 \varepsilon_0)^{1/2} s t  $, $s {\vec x}$, $\varepsilon_0 \varepsilon_r$, $(\varepsilon_0 / \mu_0)^{1/2} s^{-1} \sigma$, and  $(\varepsilon_0  \mu_0)^{-1/2} {\mathsf c}$ with a suitably chosen spatial scaling factor $s$ (meters), $\varepsilon_0 = 8.85 \cdot 10^{-12}$ F/m and $\mu_0 = 4 \pi \cdot 10^{-7}$ b/m.  

\subsection{Forward simulation}
\label{section:forward_simulation}

The spatial domain $\Omega$ can be discretized utilizing a set of finite elements (FEs) $\mathcal{T} = \{ \mathtt{T}_1, \mathtt{T}_2, \ldots, \mathtt{T}_m \}$ equipped with piecewise linear basis functions $\varphi_1,\varphi_2, \ldots, \varphi_n \in H^1(\Omega)$ \cite{braess2007}.  These are called nodal functions as their degrees of freedom coincide with the nodes $\vec{r}_1, \vec{r}_2, \ldots, \vec{r}_n$ of the FE mesh  $\mathcal{T}$. The associated finite element (FE) approximations of the potential and gradient fields are of the form $u = \sum_{j = 1}^n p_j  \, \varphi_j$ and ${\vec g } = \sum_{k = 1}^\mathantt{d} g^{(k)} {\vec e}_k$, where  $g^{(k)} = \sum_{i = 1}^m q^{(k)}_i \, \chi_i$ is a sum of element indicator functions $\chi_1, \chi_2, \ldots, \chi_m \in L_2(\Omega)$ and $\mathantt{d}$ is the number of spatial dimensions (2 or 3).   

Defining test functions $v  : [0,T] \to \mathcal{V} \subset b^1(\Omega)$ and ${\vec w} : [0,T] \to \mathcal{W} \subset L_2(\Omega)$ with $\mathcal{V} = \hbox{span}\{ \varphi_1,\varphi_2, \ldots, \varphi_n \}$ and $\mathcal{W}=\hbox{span} \{\chi_1, \chi_2, \ldots, \chi_m \}$ the weak form can be expressed in the discretized (Ritz-Galerkin) form \cite{braess2007}, that is, 
{\setlength\arraycolsep{2 pt} \begin{eqnarray}
\label{system1}
\frac{\partial}{\partial t}   {\bf A} {\bf q}^{(k)} - {\bf B}^{(k)}  {\bf  p} + {\bf T}^{(k)} {\bf q}^{(k)}  & = & 0,   \\
\frac{\partial}{\partial t}  {\bf C} {\bf p} + {\bf R} {\bf p}  + {\bf S} {\bf p} + \sum_{k = 1}^\mathantt{d} {{\bf B}^{(k)}}^T {\bf q}^{(k)} & = & {\bf f}, 
\label{system2}
\end{eqnarray}}
with ${\bf p} = (p_1,p_2,\ldots, p_n)$, ${\bf q}^{(k)} = (q^{(k)}_1, q^{(k)}_2, \ldots, q^{(k)}_m)$, ${\bf A} \in \mathbb{R}^{m \times m}$,  ${\bf B} \in \mathbb{R}^{m \times n}$, ${\bf C} \in \mathbb{R}^{n  \times n}$, ${\bf S} \in \mathbb{R}^{n  \times n}$, ${\bf T} \in \mathbb{R}^{m  \times m}$ and 
{\setlength\arraycolsep{2 pt} \begin{eqnarray} 
A_{i,i} &   = & \int_{\mathtt{T}_i}  \,  \hbox{d} \Omega, \qquad \quad \, \,  \, A_{i,j}  =  0, \quad \hbox{if} \quad i \neq j,  \\ 
f_{i} & = & \int_\Omega {f} \, \varphi_i  \, \hbox{d} \Omega, \quad \, \, \,  \, B_{i,j}^{(k)}  =  \int_{\mathtt{T}_i} \, \vec{e}_k \cdot \nabla \varphi_j \, \hbox{d} \Omega, \\  
C_{i,j} & = & \int_\Omega \varepsilon_r \, \varphi_i \varphi_j  \, \hbox{d} \Omega, \quad
R_{i,j} =  \int_\Omega \sigma \, \varphi_i \varphi_j  \, \hbox{d} \Omega, \\  
{T}_{i,i}^{(k)}  &  = & \zeta^{(k)}  \int_{\mathtt{T}_i}  \, \hbox{d} \Omega, \quad  T^{(k)}_{i,j}  = 0, \quad  \hbox{if} \quad i \neq j, \\ 
S_{i,j}  & =  & \int_\Omega \xi \, \varphi_i \varphi_j  \, \hbox{d} \Omega. 
\end{eqnarray}}
Here, ${\bf S}$ and ${\bf T}^{(k)}$ are additional matrices corresponding to a split-field perfectly matched layer (PML) defined for the outermost part $\{ {\vec x} \in \Omega \, | \, \varrho_1 \leq \max_k | x_k | \leq \varrho_2 \}$ of the computational domain to eliminate reflections from the boundary $\partial \Omega$ back to the inner part of $\Omega$ \cite{schneider2012}. The absorption parameters $\xi$ and $\zeta^{(k)}$ are of the form $\xi({\vec x}) = \varsigma$, if $\varrho_1 \leq \max_k | x_k | \leq \varrho_2$, and $\zeta^{(k)} ({\vec x})= \varsigma$, if $\varrho_1 \leq | x_k |  \leq \varrho_2$, and otherwise $\xi({\vec x}) = \zeta^{(k)}({\vec x}) = 0$. 

In this study, the standard finite difference approach within a $\Delta t$ spaced regular grid of $N$ time points is utilized to discretize the time interval $[0,T]$. A straightforward temporal discretization of (\ref{system1}) and (\ref{system2}) yields the following leap-frog time integration system 
{\setlength\arraycolsep{1 pt} \begin{eqnarray}
\label{leap-frog1}
{\bf q}^{(k)}_{\ell + \frac{1}{2}} & = & {\bf q}^{(k)}_{\ell - \frac{1}{2}} \! +  \! \Delta t  {\bf  A}^{-1} \Big(     {\bf B}^{(k)}  {\bf  p}_{\ell} \! - \! {\bf T}^{(k)} {\bf q}^{(k)}_{\ell - \frac{1}{2}} \Big),  \\ 
 {\bf p}_{\ell + 1}  & = & {\bf p}_{\ell}  \! + \!  \Delta t {\bf  C}^{-1} \Big( {\bf f}_\ell \! - \! {\bf R} {\bf p}_\ell \! - \!  {\bf S} {\bf p}_\ell \! - \! \sum_{k = 1}^\mathantt{d} {{\bf B}^{(k)}}^T {\bf q}^{(k)}_{\ell + \frac{1}{2}}  \Big), 
\label{leap-frog2}  
\end{eqnarray}}
$\ell = 1, 2, \ldots, N$, which can be used to simulate a signal propagating in $\Omega$  \cite{schneider2012,bossavit1999,yee1966}.

\subsubsection{Differentiated signal}
\label{section:linearized_forward_simulation}

The permittivity is here sought in the form $\varepsilon_r =  \varepsilon^{(bg)}_r + \varepsilon^{(p)}_r$, where  $\varepsilon^{(bg)}_r$ is a fixed background distribution and $\varepsilon^{(p)}_r = \sum_{j = 1}^M {c}_j {\chi}'_j$ a variable perturbation that is composed by  indicator functions of a coarse mesh ${\mathcal{T}'} = \{ \mathtt{T}'_1, \mathtt{T}'_2, \ldots, \mathtt{T}'_M  \}$. The density of  $\mathcal{T}$  is set by the geometrical constraints of the forward simulation, such as domain structure and applied wavelength, the resolution of $\mathcal{T}'$ is to be chosen based on the desired precision of the inversion results. Meshes  ${\mathcal{T}'}$ and ${\mathcal{T}}$ are assumed to be nested meaning that the nodes of ${\mathcal{T}'}$ belong to those of ${\mathcal{T}}$. 

For differentiating the signal with respect to the permittivity, we define  
{\setlength\arraycolsep{2 pt }\begin{eqnarray}
\label{aku_ankka}
{\bf b}_\ell &  = &  {\bf C}^{-1} \Big( \! {\bf R} {\bf p}_\ell \! + \!  {\bf S} {\bf p}_\ell \! + \! \sum_{k = 1}^\mathantt{d} {{\bf B}^{(k)}}^T {\bf q}^{(k)}_{\ell + \frac{1}{2}}  \Big)  \\    {\bf h}_\ell^{(i,j)}   & = &  \frac{\partial {\bf C}}{\partial c_j} {\bf b}_\ell^{(i)}  
\end{eqnarray}}
with  $(b^{(i)}_\ell)_j=  (b_\ell)_i$, if $j = i$ and $(b^{(i)}_\ell)_j = 0$, otherwise. The differentiated potential can be computed as the sum ${\partial {\bf p}_\ell}/{\partial c_j}  = \sum_{\vec{r}_i \in \mathtt{T}'_j}  {\bf d}_\ell^{(i,j)}$ (Appendix \ref{a_linearization}) where the terms can be obtained through the following auxiliary system 
{\setlength\arraycolsep{2 pt} \begin{eqnarray}
\label{linearized1}
{\bf r}^{(i, j, k)}_{\ell + \frac{1}{2}} & = & 
{\bf r}^{(i, j, k)}_{ \ell - \frac{1}{2}} +  \Delta t  {\bf  A}^{-1} \Big(  {\bf B}^{(k)}   {\bf d}^{(i, j)}_{\ell}  - {\bf T}^{(k)} {\bf r}^{(i, j, k)}_{\ell - \frac{1}{2}} \Big),  \\ 
{\bf d}^{(i, j)}_{\ell+1} & = & {\bf d}^{(i, j)}_{\ell}+  \Delta t {\bf  C}^{-1} \Big( {\bf h}_\ell^{(i,j)} - {\bf R}  {\bf d}^{(i, j)}_{\ell} - {\bf S} {{\bf d}^{(i,j)}_\ell} \nonumber\\ & & \qquad \qquad  \qquad \qquad \qquad - \sum_{k = 1}^\mathantt{d} {{\bf B}^{(k)}}^T {{\bf r}^{(i,j,k)}_{\ell + \frac{1}{2}}} \Big).
\label{linearized2} 
\end{eqnarray}}
This is otherwise identical to (\ref{leap-frog1})--(\ref{leap-frog2})  but instead of ${\bf f}$ has the node-specific vector ${\bf h}_\ell^{(i,j)} = ({\partial {\bf C}}/{\partial c_j}) {\bf b}_\ell^{(i)}$ working as the source. The solution of (\ref{linearized1})--(\ref{linearized2})  is an essential part of the following deconvolution strategy which yields the differentiated potential based on the reciprocity of the wave propagation \cite{altman1991}.


\subsubsection{Signal reciprocity and deconvolution in forward simulation}
\label{deconvolution}

\begin{figure*}
\begin{center}
\begin{minipage}{6cm} \begin{center} \includegraphics[width=5cm]{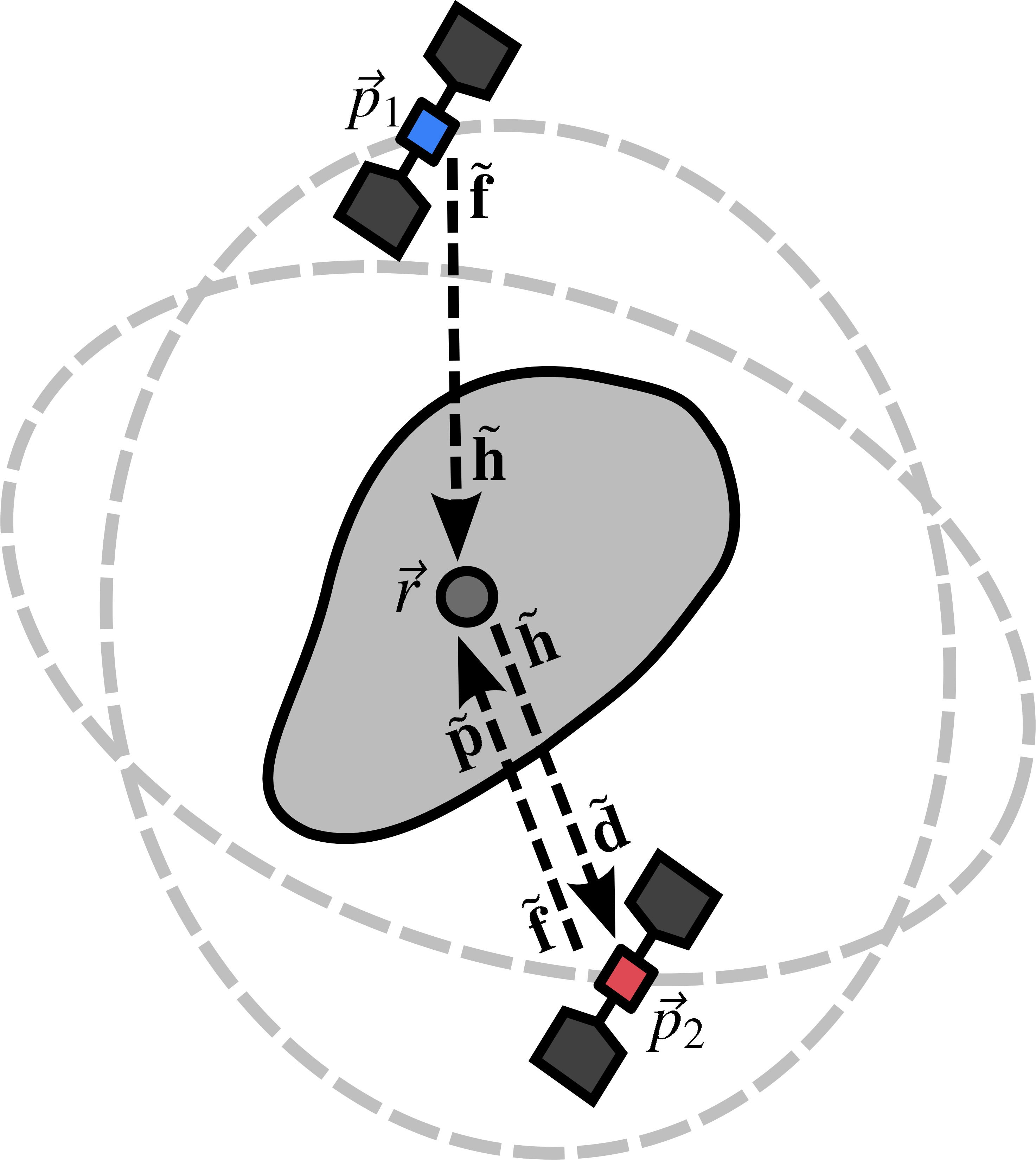} \end{center} \end{minipage}  
\begin{minipage}{6cm} \begin{center}
\includegraphics[width=5cm]{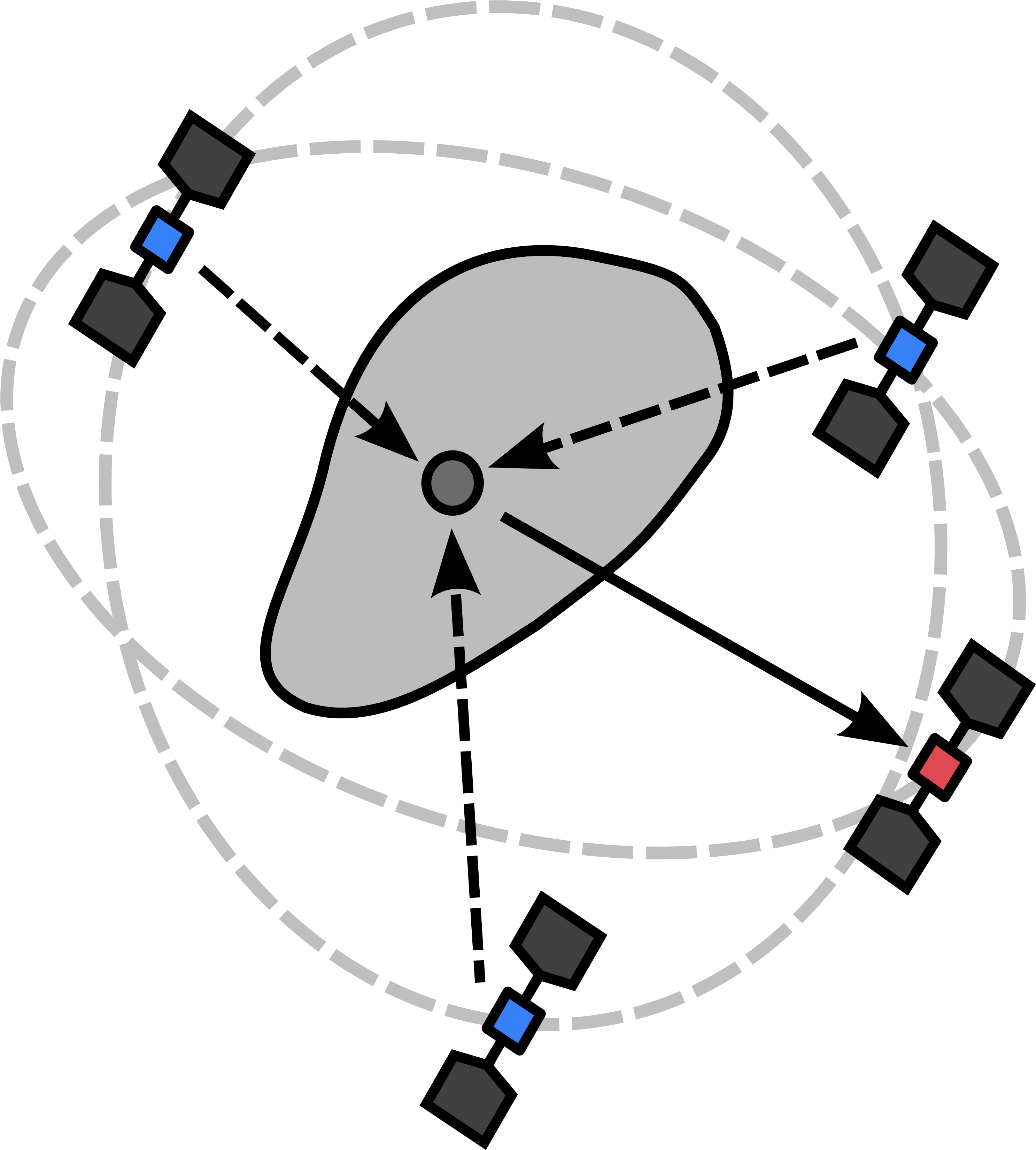} 
\end{center}
\end{minipage}
\end{center}
\caption{{\bf On left:}  A schematic picture visualizing computation of $\tilde{\bf d}$, i.e., time evolution of ${\bf d}^{(i,j)}$,  at $\vec{p}_2$ resulting from transmission $\tilde{\bf  f}$ made at $\vec{p_0}$. The following procedure can be used: (1) Compute $\tilde{\bf p}$  at $\vec{r}_i$ with the source $\tilde{\bf {f}}$ placed at $\vec{p}_2$ via, and approximate the corresponding Green's kernel $\tilde{\bf g}$  between $\vec{r}_i$ and  $\vec{p}_2$. (2) Calculate $\tilde{\bf h}$  at $\vec{r}_i$  with  $\tilde{\bf {f}}$  transmitted at $\vec{p}_1$, and calculate $\tilde{\bf d}$ as the convolution between $\tilde{\bf g}$ and $\tilde{\bf h}$. {\bf On right:} A schematic illustration of the applied synthetic satellite constellation model in which a single receiver (red)  records a signal that is transmitted  separately by one or more orbiters (blue).     \label{constellation_image}}
\end{figure*}

If an infinitely short monopolar pulse $\delta(t) \delta_{\vec{a}}(\vec{x})$  transmitted at point $\vec{a}$ leads to data $\mathcal{G}_{\vec{a}, \vec{b}} (t)$ received at $\vec{b}$, then an arbitrary data $\tilde{d}(t)$  resulting from transmission $\tilde{h}(t)$ obeys the following linear convolution relation
\begin{equation} \label{convolution} \tilde{d}(t) = \mathcal{G}_{\vec{a}, \vec{b}}  \ast_t \tilde{h} (t) = \int_{-\infty}^\infty \mathcal{G}_{\vec{a}, \vec{b}}  (t - \tau) \tilde{h}(\tau) \, \hbox{d} \tau, \end{equation} 
since the system (\ref{leap-frog1})--(\ref{leap-frog2}) is linear with respect to the source and $\tilde{h}(t) = \tilde{h} \ast_t \delta(t)$ at $\vec{a}$.  To approximate the Green's kernel $\mathcal{G}_{\vec{a}, \vec{b}} (t)$ based on (\ref{convolution}),  a regularized  deconvolution procedure can be applied. Moreover, the kernel is invariant under interchanging $\vec{a}$ and $\vec{b}$, i.e., it satisfies $\mathcal{G}_{\vec{a}, \vec{b}} = \mathcal{G}_{\vec{b}, \vec{a}}$, due to reciprocity of the wave propagation \cite{altman1991}. Consequently, the deconvolution process can rely on another transmitter-receiver  pair $\tilde{f}(t)$ and $\tilde{p}(t) = \mathcal{G}_{\vec{b}, \vec{a}}  \ast_t \tilde{f} (t)$ for a signal traveling from $\vec{b}$ to $\vec{a}$.

Assume  now that $\tilde{\bf d}^{(i,j)}$, the $N$-by-$1$ time evolution of ${\bf d}^{(i,j)}$ at $\vec{p}_2$, is to be computed corresponding to a monopolar transmission $\tilde{\bf f}$ made at $\vec{p}_1$. Denoting by $\tilde{\bf p}$ and  $\tilde{\bf h}$ the time evolution of ${\bf p}$ and  ${\bf h}^{(i,j)}$  at a given mesh node $\vec{r} \in \mathtt{T}'_j$   with the source placed at $\vec{p}_2$ and $\vec{p}_1$, respectively,  one can obtain  $\tilde{\bf d}^{(i,j)}$ by repeating the following three steps for each $\vec{r} \in \mathtt{T}'_j$  (Figure \ref{constellation_image}):
\begin{enumerate}
\item  Approximate the  Green's kernel  between $\vec{r}$ and  $\vec{p}_2$ in the following Tikhonov regularized form 
\begin{equation} 
\label{dc_eq}
   \tilde{\bf g}  = [{\bf K}_{\tilde{\bf f}}^T  {\bf K}_{\tilde{\bf f}}+ \nu {\bf I} ]^{-1} {\bf K}_{\tilde{\bf f}}^T  \left( \begin{array}{c} {\bf 0} \\  \tilde{\bf p}  \\ {\bf 0} \end{array}  \right)\end{equation} with \begin{equation} {\bf K}_{\tilde{\bf f}} =  
\left( \begin{array}{ccccc} 
\tilde{f}_1 & 0 &  0 & 0  & 0 \\ 
\tilde{f}_2 & \tilde{f}_1 &  0 & 0 & 0 \\
\vdots & \tilde{f}_2 &  \ddots & 0 & 0 \\ 
 \tilde{f}_N & \vdots &  \ddots  & \tilde{f}_1 & 0 \\ 
 0  & \tilde{f}_N &   & \tilde{f}_2  &  \tilde{f}_1 \\ 
\end{array} \right),
\end{equation} 
where $\nu$ is a regularization parameter,  ${\bf 0}$ denotes a $N$-by-$1$ zero-continuation of the signal and ${\bf K}_{\tilde{f}}$ is a $3N$-by-$3N$ convolution matrix. 
\item Calculate the convolution between $\tilde{\bf g}$ and $\tilde{\bf h}$ as given by $\tilde{\bf d}  =  {\bf P} \, {\bf K}_{\tilde{\bf h}}\tilde{\bf g}$, where ${\bf P}$ is a matrix picking the centermost $N$ entries from a $3N$-by-$1$ vector.
\item Update $\tilde{\bf d}^{(i,j)} \to \tilde{\bf d}^{(i,j)} + \tilde{\bf d}$.
\end{enumerate}
Central in this approach is that the differentiated signal ${\partial {\bf p}}/{\partial c_j}$ can be computed rapidly for each element of $\mathcal{T}'_j$, $j = 1, 2, \ldots, M$ by propagating two waves via (\ref{leap-frog1})--(\ref{leap-frog2}): 
\begin{enumerate}
\item ${\bf b}$  with ${\bf f}$ transmitted at $\vec{p}_1$, 
\item ${\bf p}$ with ${\bf f}$ transmitted at $\vec{p}_2$. 
\end{enumerate} 
This is an important aspect especially in a 3D geometry, where the number of discretization points can be, due to dimensionality,  very large compared to that of data gathering locations.

\subsubsection{Multiresolution approach and incomplete Cholesky decomposition}
\label{multiresolution}

In this study, the forward simulation process is speeded up using a multiresolution approach in which  the number of terms in the sum ${\partial {\bf p}}/{\partial c_j}  = \sum_{\vec{r}_i \in \mathtt{T}'_j}  {\bf d}^{(i,j)}$  is reduced by defining the system (\ref{linearized1})--(\ref{linearized2}) for the coarse (inversion) mesh $\mathcal{T}'$ instead of $\mathcal{T}$. Since the resolution of $\mathcal{T}'$ is here too low for propagating a wave, ${\bf b}$ and ${\bf p}$ needed for finding ${\bf d}^{(i,j)}$ (Section \ref{deconvolution}) are produced using $\mathcal{T}$. Due to the nested structure of these meshes the degrees of freedom (node values) for $\mathcal{T}'$ are included to those of  $\mathcal{T}$. As a result,  ${\bf b}$ and ${\bf p}$ can be transformed to the basis of $\mathcal{T}'$ via a direct restriction, similar to the FE multigrid approaches \cite{braess2007}. Notice that from the memory consumption viewpoint, it is advantageous that only subvectors of ${\bf b}$ and ${\bf p}$ need to be accessed during the computation.

In addition to the above multiresolution strategy, further speedup and memory saving in the leap-frog algorithm was obtained by decomposing the mass matrix ${\bf C}$ via the incomplete Cholesky strategy together with the symmetric approximate minimum degree permutation of the entries \cite{golub1989}.  Around  20 \% level of non-zero entries were included in the decomposition compared to a complete one. 

\subsection{Inversion procedure}
\label{inversion}

The present task to recover the permittivity can be formulated as 
\begin{equation} 
\label{lin_eq}
{\bf L}{\bf x}  + {\bf n} = {\bf y} - {\bf y}_{bg}, 
\end{equation} 
where ${\bf y}$ contains the actual data, ${\bf y}_{bg}$ simulated data for the background permittivity, ${\bf n}$ is a noise term containing modeling and measurement errors and ${\bf L}$ corresponds to the differentiated signal. An estimate of ${\bf x}$ can be produced  via  the following iterative regularization procedure
\begin{equation}
\label{tv_iteration}
{\bf x }_{\ell+1}  = ({\bf L}^T {\bf L} + \alpha {\bf D} {\bf \Gamma}_{\ell} {\bf D} )^{-1} {\bf L}^T {\bf y}, \quad {\bf \Gamma}_{\ell} = \hbox{diag} ( |{\bf D} {\bf x_{\ell}}|)^{-1}
\end{equation}
in which ${\bf \Gamma}_0 = {\bf I}$ and ${\bf D}$ is of the form
\begin{equation}
\label{d_mat}
D_{i,j} =  \beta \delta_{i,j}  + \frac{\ell^{(i,j)}  }{ \max_{i,j}   \ell^{(i,j)}  }(2\delta_{i,j} - 1) , \quad \! \! \! \delta_{i,j} =
\left\{ \begin{array}{ll}  1, & \hbox{if } j = i, \\ 0, & \hbox{otherwise}.  \end{array} \right. 
\end{equation} 
Here, the first term is a weighted identity operator limiting the total magnitude of ${\bf x}$, whereas the second one penalizes the jumps of ${\bf x}$ over the edges of $\mathcal{T}'$ multiplied with the edge length $\ell^{(i,j)} = \int_{\mathtt{T}'_i \cap \mathtt{T}'_j} \, \hbox{d} s$. The above inversion process (\ref{tv_iteration}) minimizes the function \begin{equation} F({\bf x}) =  \| {\bf L} {\bf x}  - {\bf y}_{bg} - {\bf y} \|^2_2 + 2 \sqrt{\alpha} \|{\bf D}{\bf x} \|_1, \label{f_eq} \end{equation} in which the second term equals to the total variation of ${\bf x}$, if $\beta = 0$ \cite{scherzer2008, stefan2008, kaipio2004}.  Characteristic to total variation regularization is that  a reconstruction has large connected areas close to constant, since the length of the boundary curve between the jumps is regularized. This is advantageous, for example, in the present context of  recovering  connected inclusions and boundary layers. If $\beta > 0$, then also the norm of $x$ will be regularized on each  iteration step. The validity of iteration (\ref{tv_iteration}) is shown in Appendix \ref{a_inversion}. 

\subsection{Numerical experiments}

\begin{table*} \caption{\label{signal_configurations_table} Specifications of the signal configurations (A)--(H). Spacing has been given in terms of the polar angle (radians), and sampling rate (SR) relative to the Nyquist rate (NR). Spatial SR refers to the density of receiver positions at the synthetic orbit $\mathcal{C}$.}
\centering \begin{tabular}{lrrrrrrrr} 
\hline
Configuration & (A) & (B) & (C) & (D) & (E) & (F) & (G) & (H) \\
\hline
Receiver positions & 32 & 32 & 128 & 128 & 32 & 32 & 128 & 128 \\
Transmitters & 1 & 1 & 1 & 1 & 3 & 3 & 3& 3\\
Receiver  spacing  & $\frac{\pi}{16}$ & $\frac{\pi}{16}$ & $\frac{\pi}{64}$ & $\frac{\pi}{64}$ & $\frac{\pi}{16}$ & $\frac{\pi}{16}$ & $\frac{\pi}{64}$ & $\frac{\pi}{64}$ \\
Transmitter  spacing  & $\frac{\pi}{8}$& $\frac{15\pi}{4}$& $\frac{\pi}{32}$ & $\frac{15\pi}{16}$ & $\frac{\pi}{8}$ & $\frac{15\pi}{4}$ & $\frac{\pi}{32}$& $\frac{15\pi}{16}$\\
Spatial SR w.r.t.\ NR & 0.53 & 0.53 & 2.1 & 2.1 & 0.53 & 0.53 & 2.1 & 2.1 \\
Temporal SR w.r.t.\ NR & 1.7 & 1.7 & 1.7 & 1.7 & 1.7 & 1.7 & 1.7 & 1.7 \\
\hline 
 \end{tabular}
\end{table*}

\begin{figure}
\begin{center}
\begin{minipage}{3.9cm}
\begin{center}
\includegraphics[width=3.7cm]{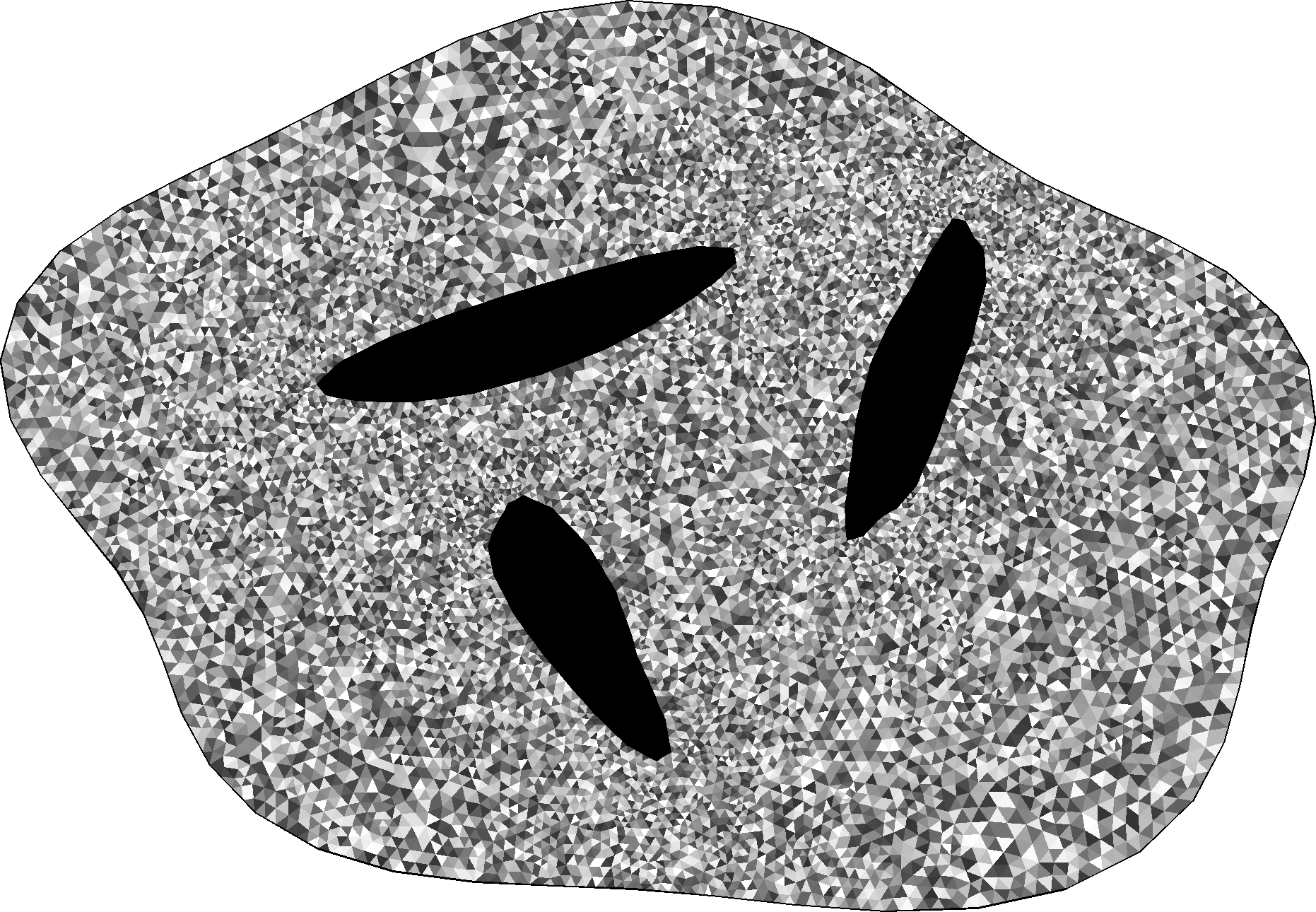} \\ (I)
\end{center}
\end{minipage} 
\begin{minipage}{3.9cm}
\begin{center} 
\includegraphics[width=3.7cm]{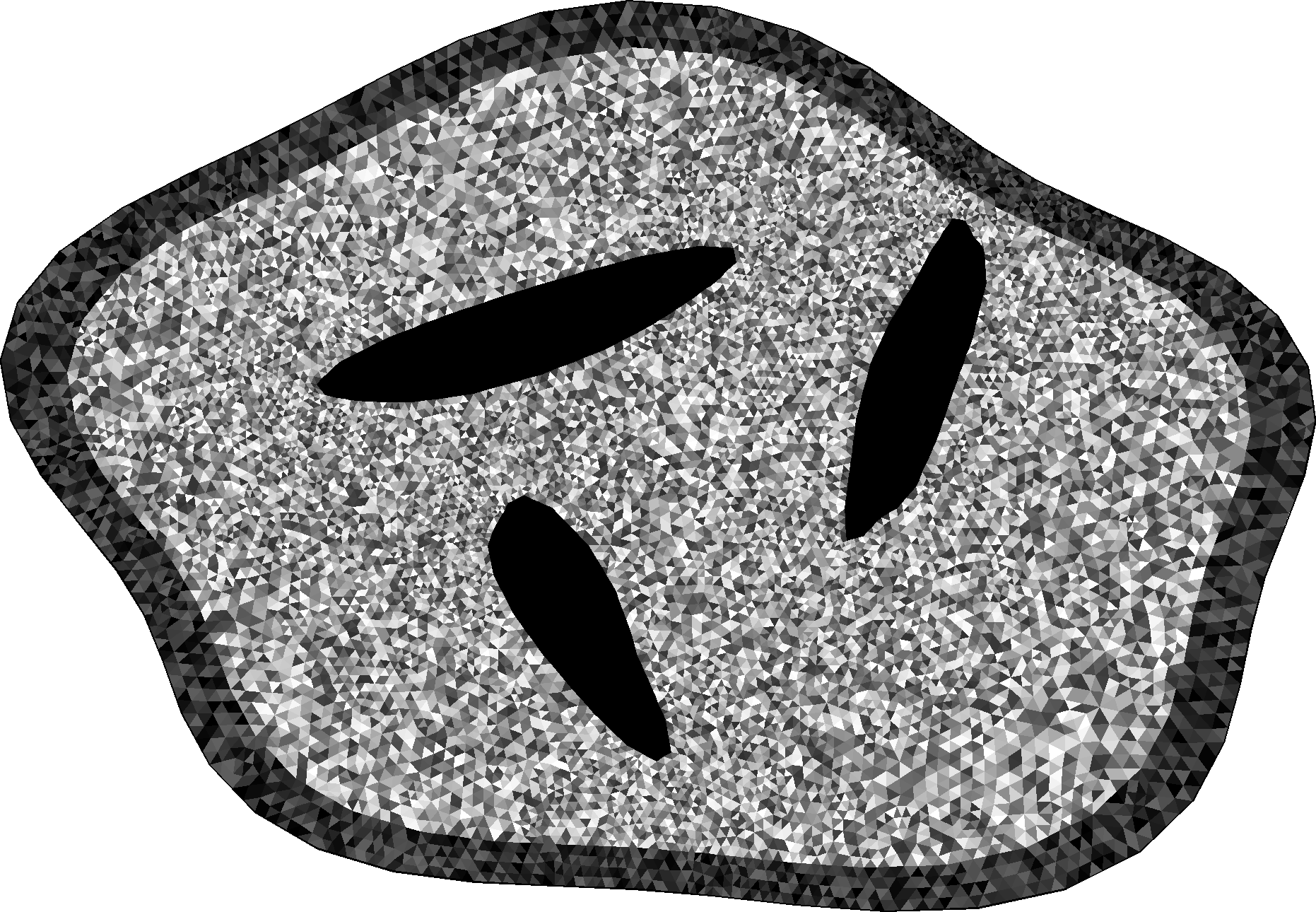} \\ (II)
\end{center} 
\end{minipage}  \\ \vskip0.2cm
\begin{minipage}{3.9cm}
\begin{center}
\includegraphics[width=3.7cm]{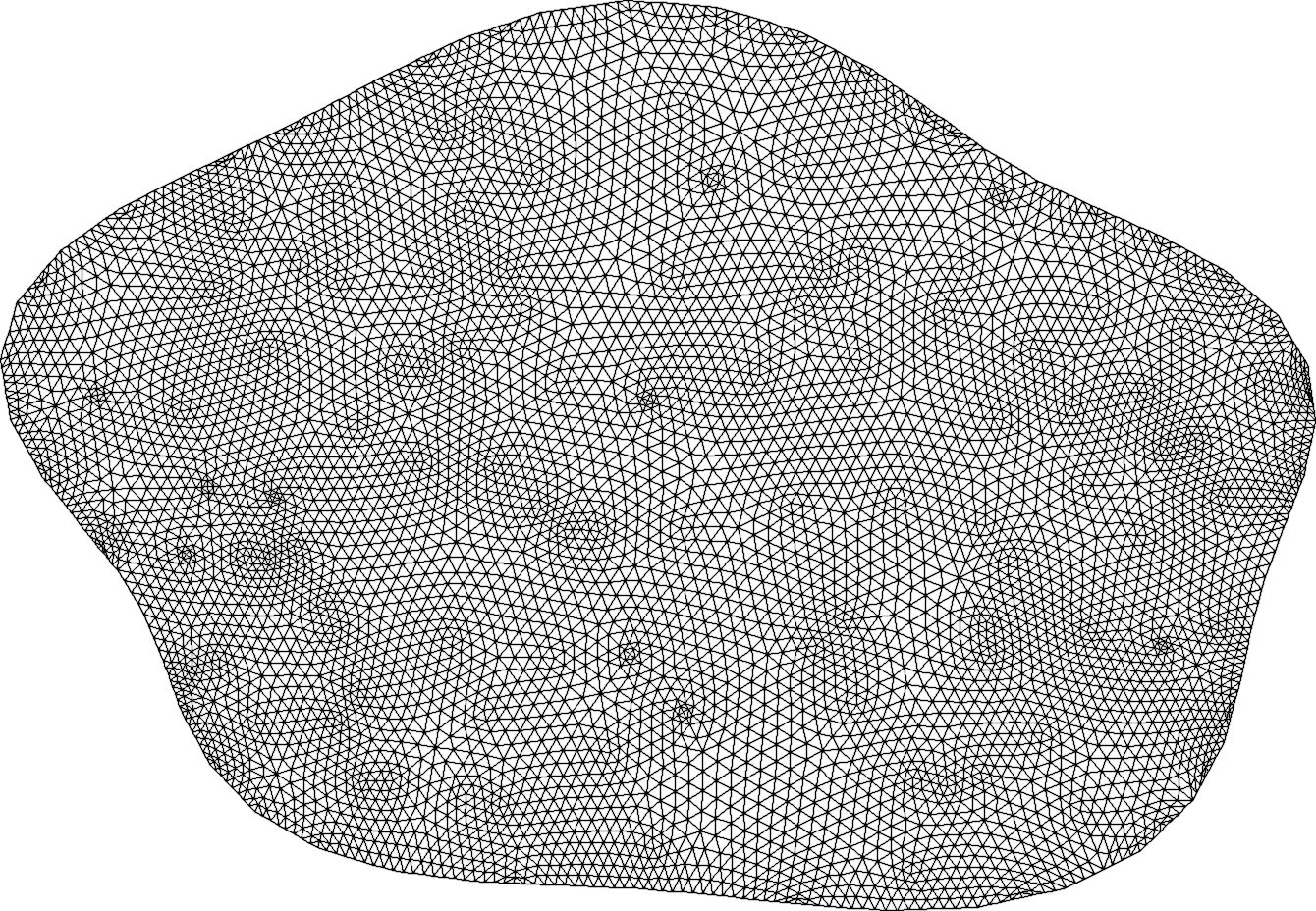} \\ $\mathcal{T}$
\end{center}
\end{minipage} 
\begin{minipage}{3.9cm}
\begin{center} 
\includegraphics[width=3.7cm]{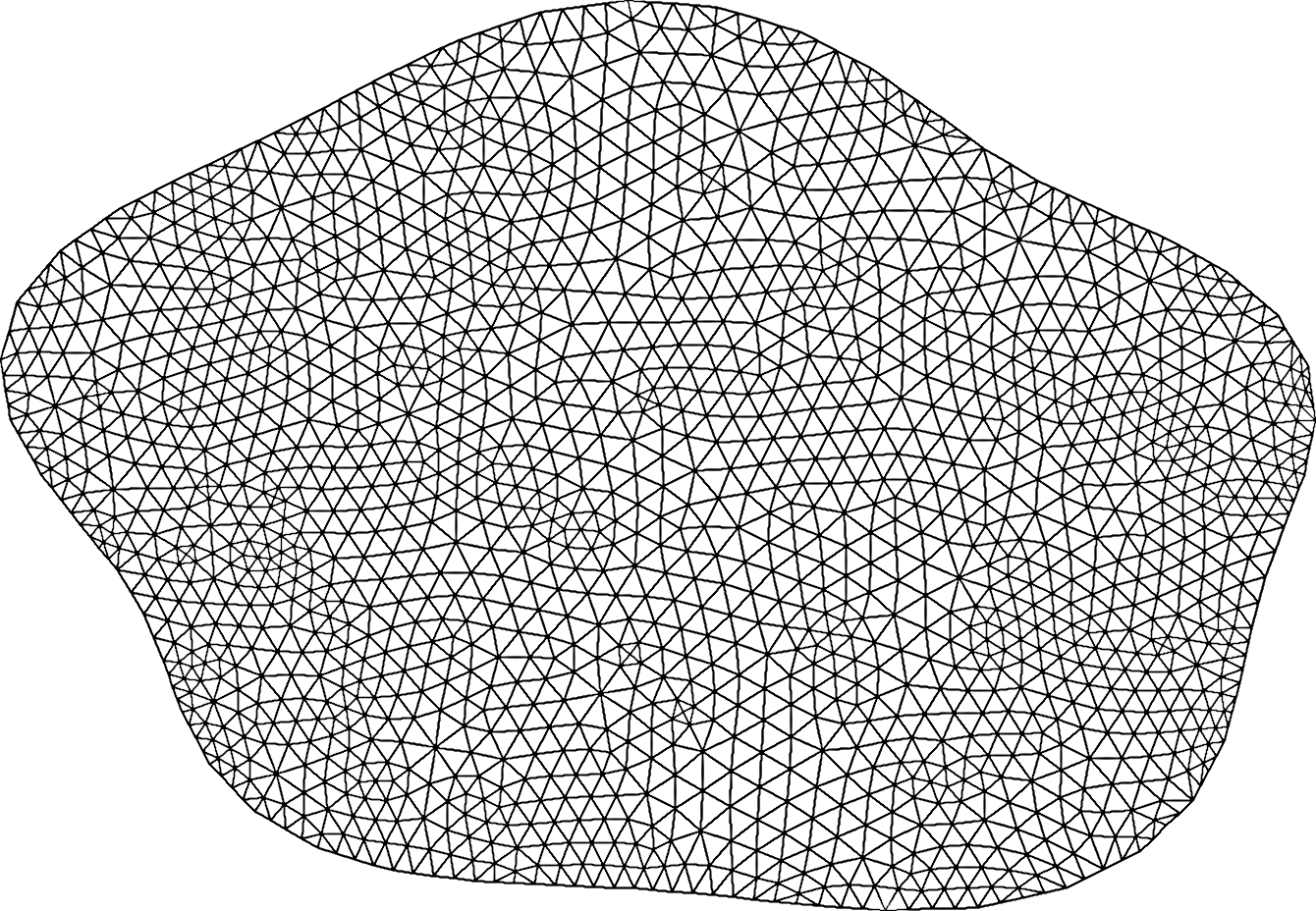} \\  $\mathcal{T}'$
\end{center} 
\end{minipage} 
\end{center}
\caption{Top row: Test domain $\Omega_0$ and granular permittivity distributions (I) and (II) utilized in the numerical experiments. Dark inclusions model vacuum cavites to be recovered. Distribution (II) includes additionally a surface layer structure, e.g.,  a dust or ice cover, which was also a target of inversion.  Bottom row: Finite element mesh $\mathcal{T}$ (left) and $\mathcal{T}'$ (right) inside $\Omega_0$. \label{omega_0}}
\end{figure}

\begin{figure}
\begin{center}
\begin{minipage}{3.9cm}
\begin{center}
\includegraphics[width=3.5cm]{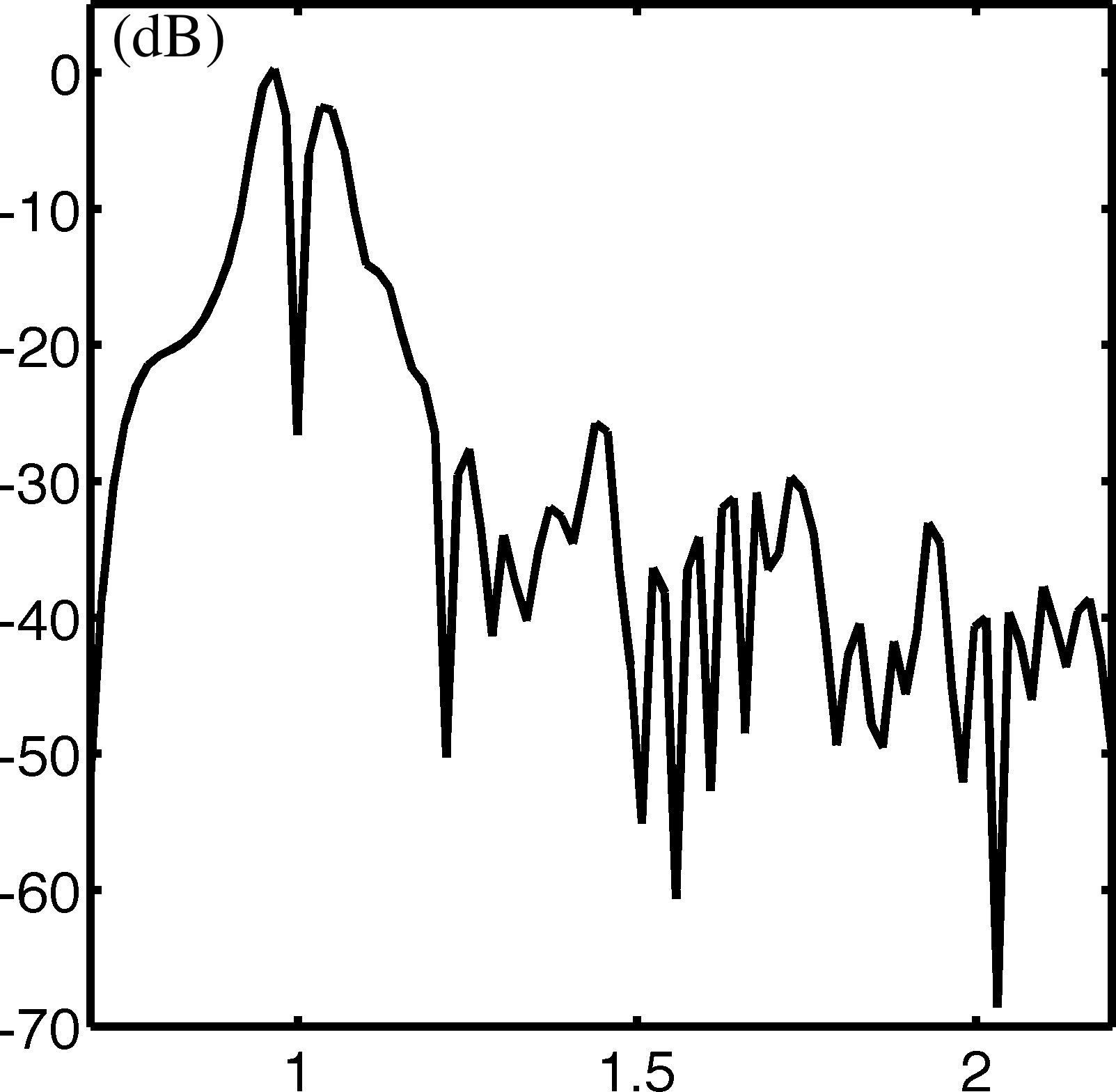} \\ without noise
\end{center}
\end{minipage} 
\begin{minipage}{3.9cm}
\begin{center} 
\includegraphics[width=3.5cm]{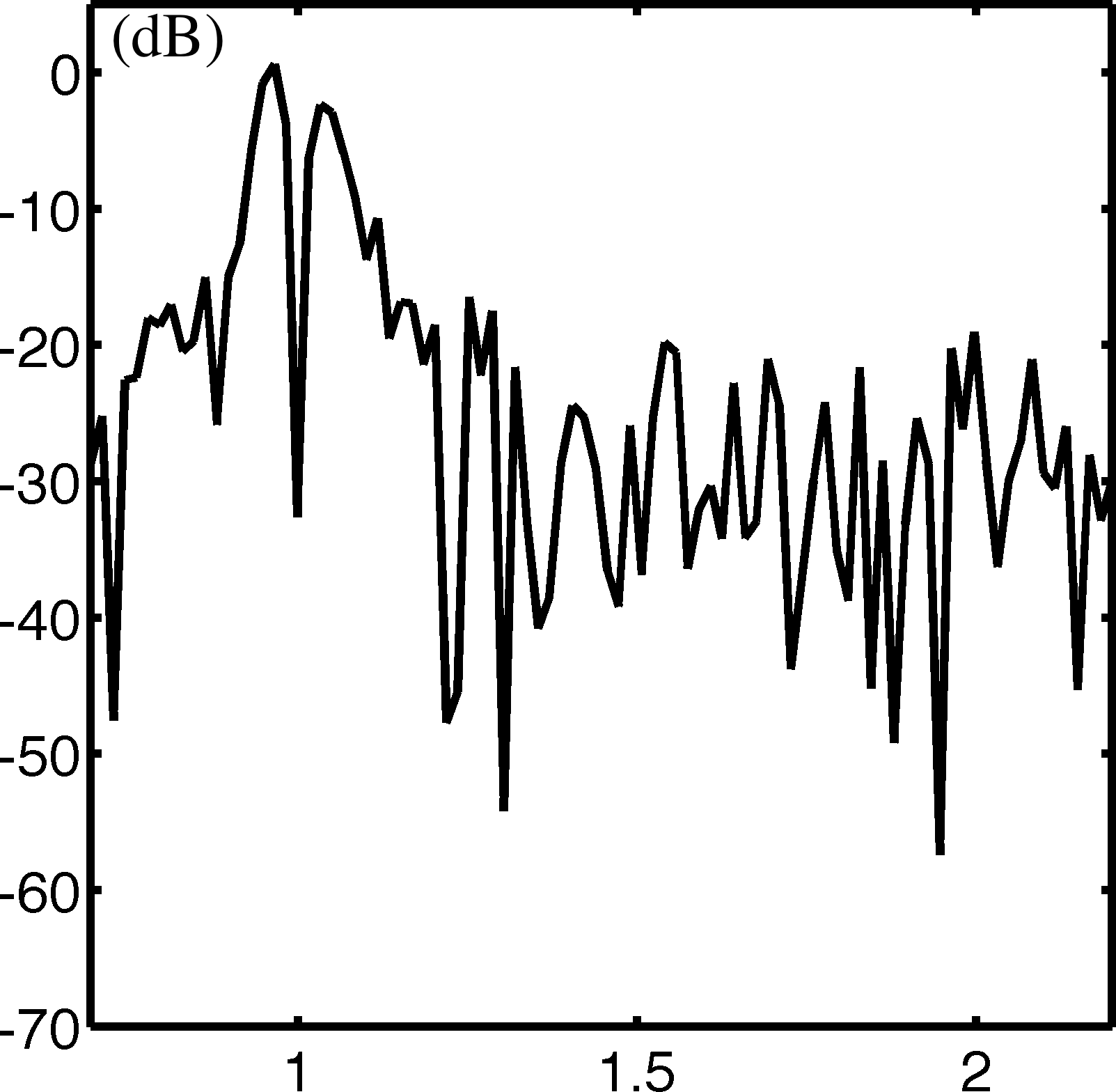} \\ with noise
\end{center} 
\end{minipage} 
\end{center}
\caption{A signal propagated through $\Omega_0$ without and with simulated measurement noise (left and right, respectively).  Similar to the CONSERT data \cite{kofman2015},  the peaks caused by the simulated noise stay mainly at least 20 dB below the main signal peak. \label{signal_fig} }
\end{figure}

\begin{figure*}
\begin{scriptsize}
\begin{center}
\begin{framed}
{ Single transmitter} \\ \vskip0.2cm
\begin{minipage}{7.6cm}
\begin{center}
\begin{framed}
{ Sparse (receiver spacing ${\pi}/{16}$)} \\ \vskip0.2cm
\begin{minipage}{3.2cm}
\begin{center}
\begin{framed}
{Low mixing  (transmitter spacing $\pi/8$)}\\ \vskip0.2cm
\includegraphics[width=2.3cm]{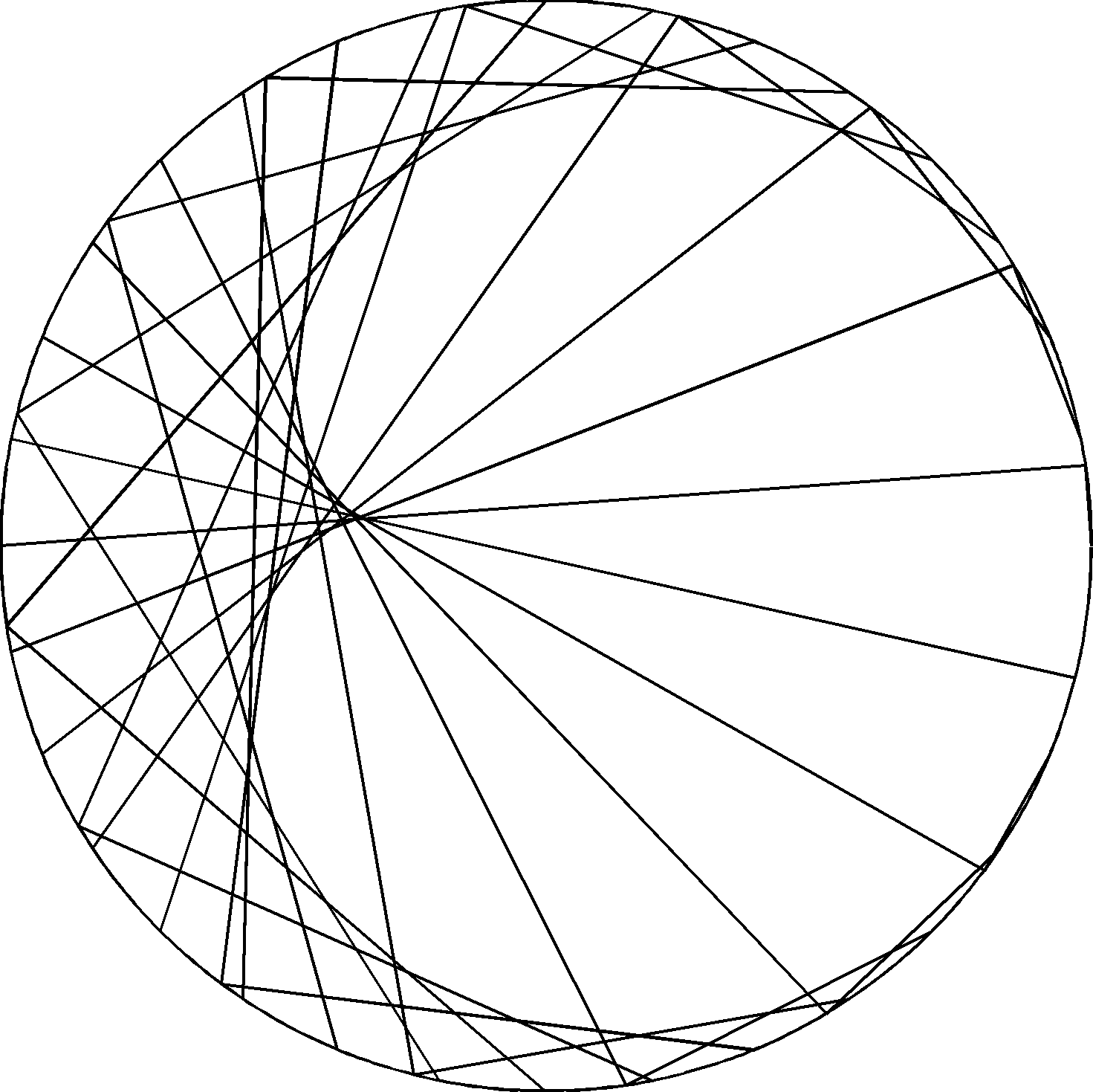} \\  (A) 
\end{framed}
\end{center}
\end{minipage} \hskip0.2cm
\begin{minipage}{3.2cm}
\begin{center} 
\begin{framed}
{ High mixing (transmitter spacing $15\pi/4$)} \\ \vskip0.2cm
\includegraphics[width=2.3cm]{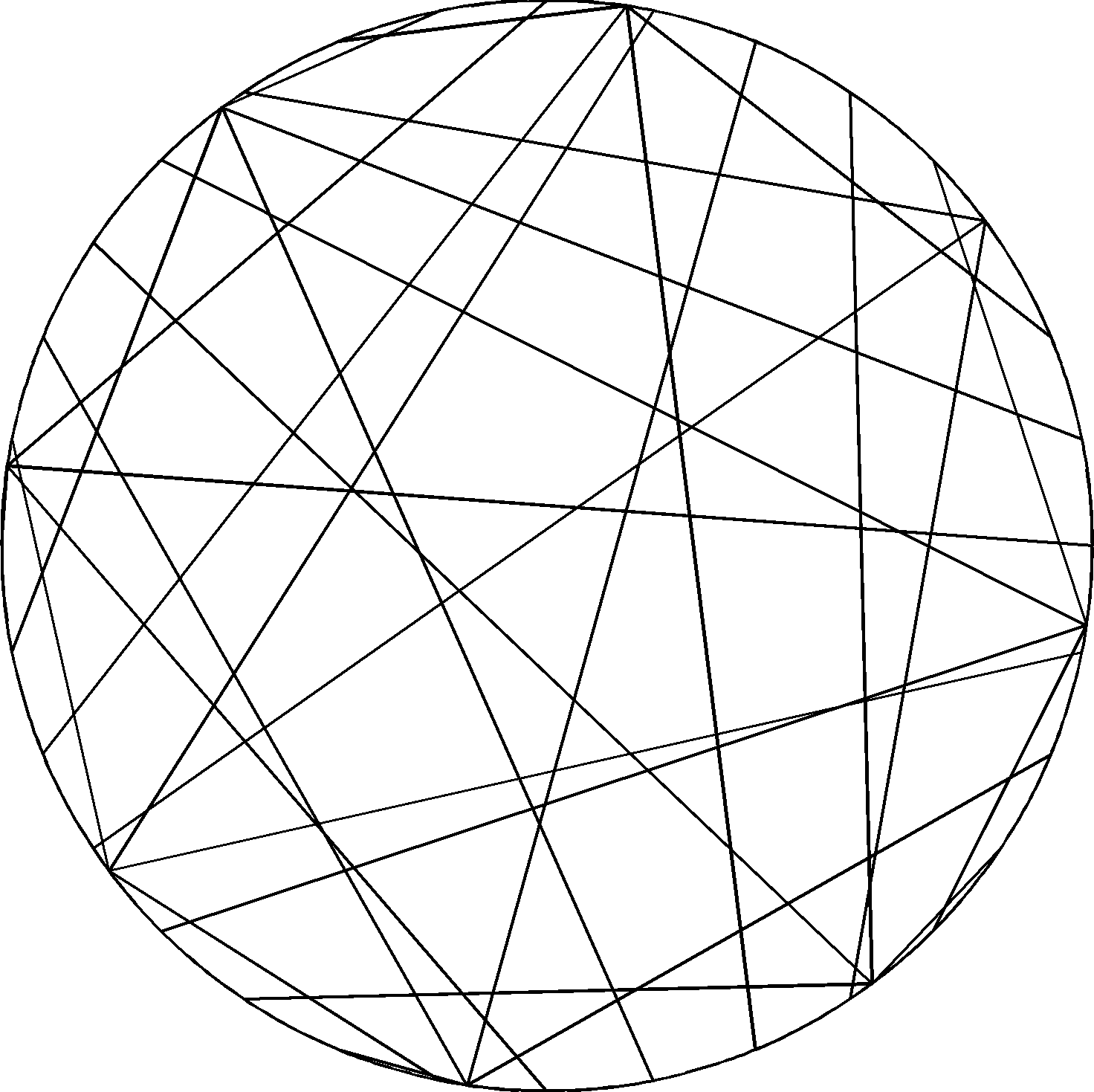} \\  (B) 
\end{framed}
\end{center} 
\end{minipage}
\end{framed}
\end{center}
\end{minipage} \hskip0.2cm 
\begin{minipage}{7.6cm}
\begin{center}
\begin{framed}
{ Dense (receiver spacing ${\pi}/{64}$)} \\ \vskip0.2cm
\begin{minipage}{3.2cm}
\begin{center}
\begin{framed}
{ Low mixing (transmitter spacing $\pi/32$)} \\ \vskip0.2cm
\includegraphics[width=2.3cm]{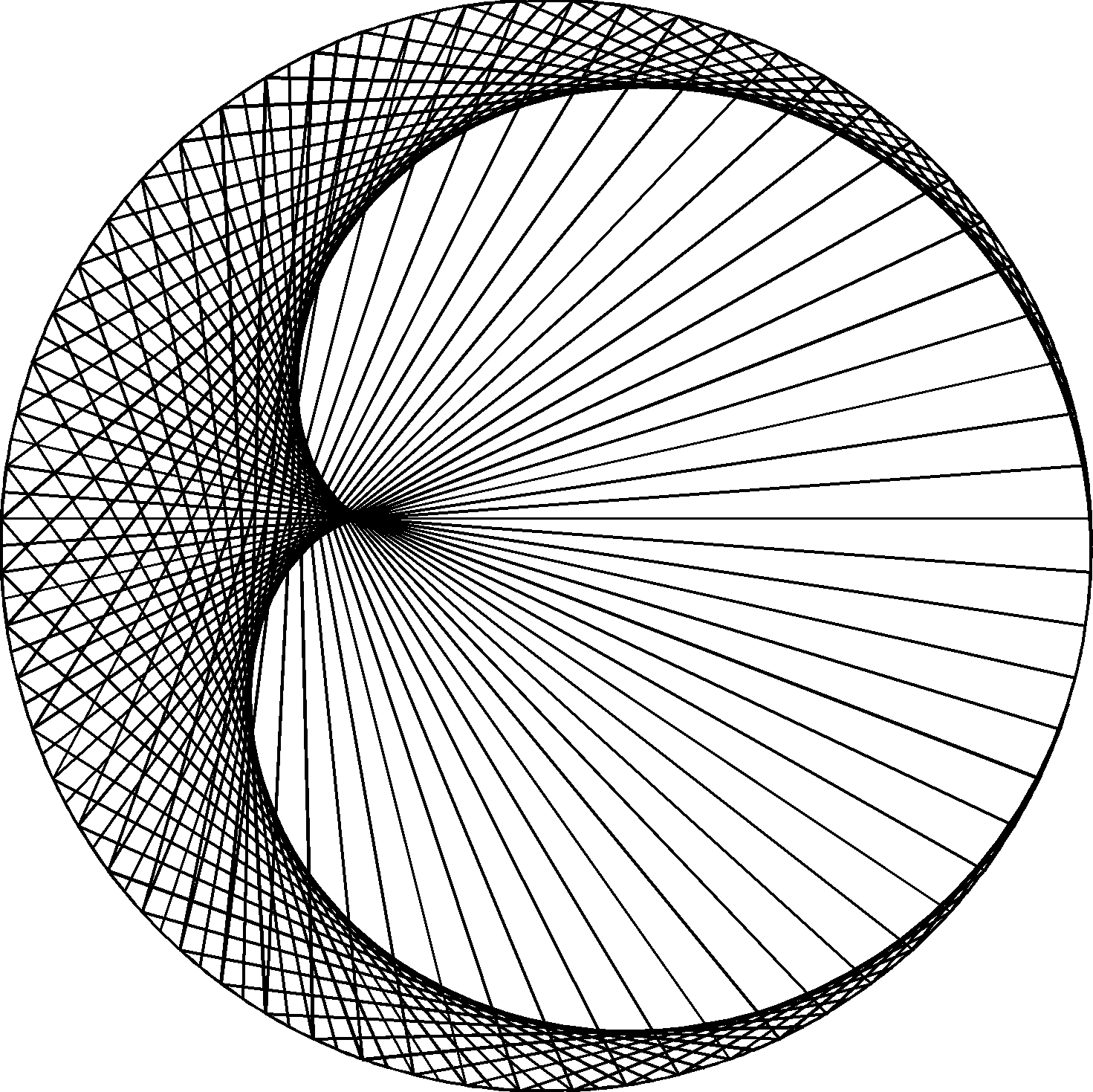} \\  (C) 
\end{framed}
\end{center}
\end{minipage} \hskip0.2cm 
\begin{minipage}{3.2cm}
\begin{center}
\begin{framed}
{ High mixing (transmitter spacing $15\pi/16$)} \\ \vskip0.2cm
\includegraphics[width=2.3cm]{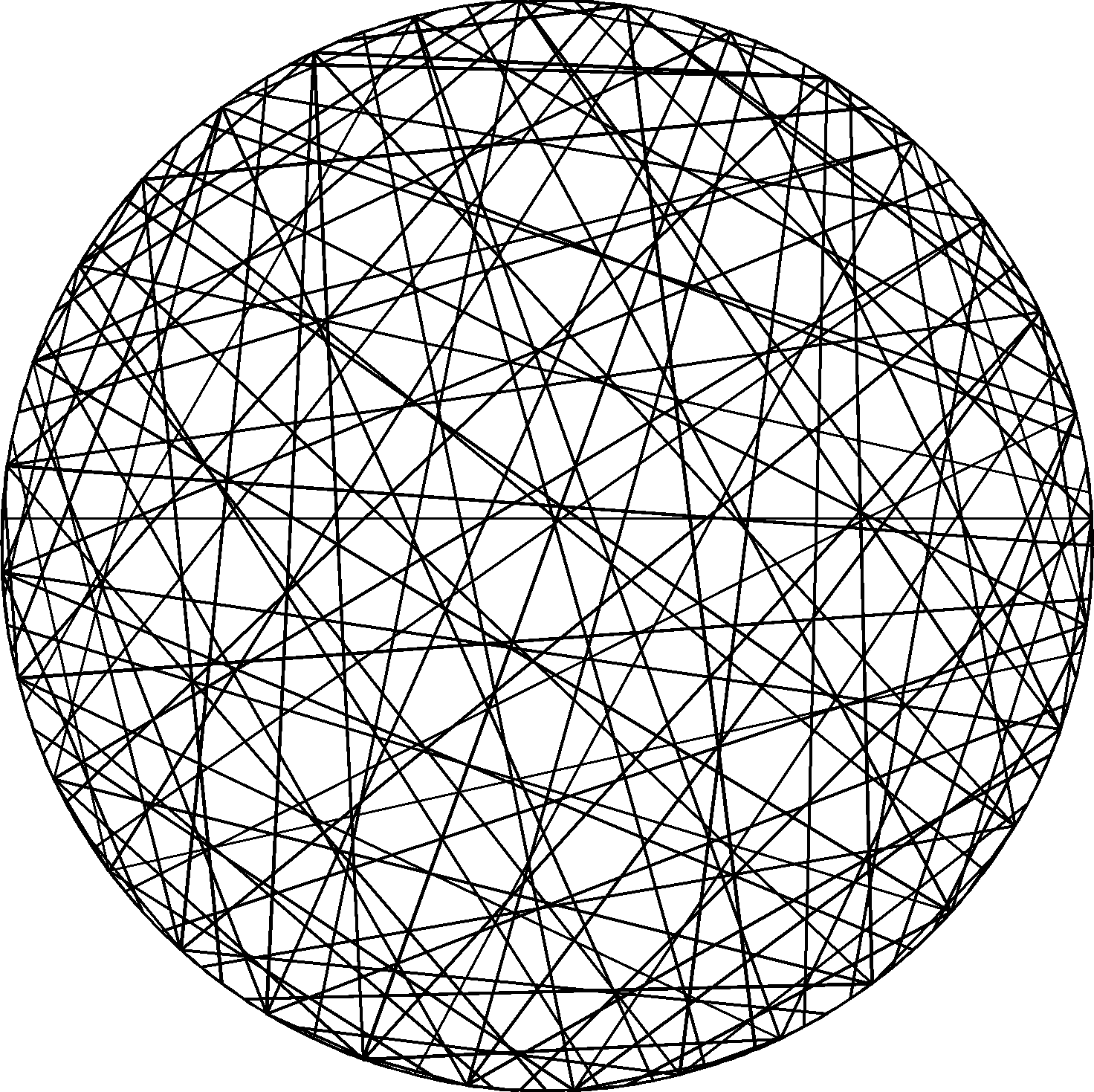} \\  (D)  
\end{framed}
\end{center}
\end{minipage} 
\end{framed}
\end{center} 
\end{minipage}
\end{framed} 
\begin{framed}
{ Three transmitters} \\ \vskip0.2cm
\begin{minipage}{7.6cm}
\begin{center}
\begin{framed}
{ Sparse (receiver spacing ${\pi}/{16}$)}  \\ \vskip0.2cm
\begin{minipage}{3.2cm}
\begin{center}
\begin{framed}
{ Low mixing  (transmitter spacing $\pi/8$)} \\ \vskip0.2cm
\includegraphics[width=2.3cm]{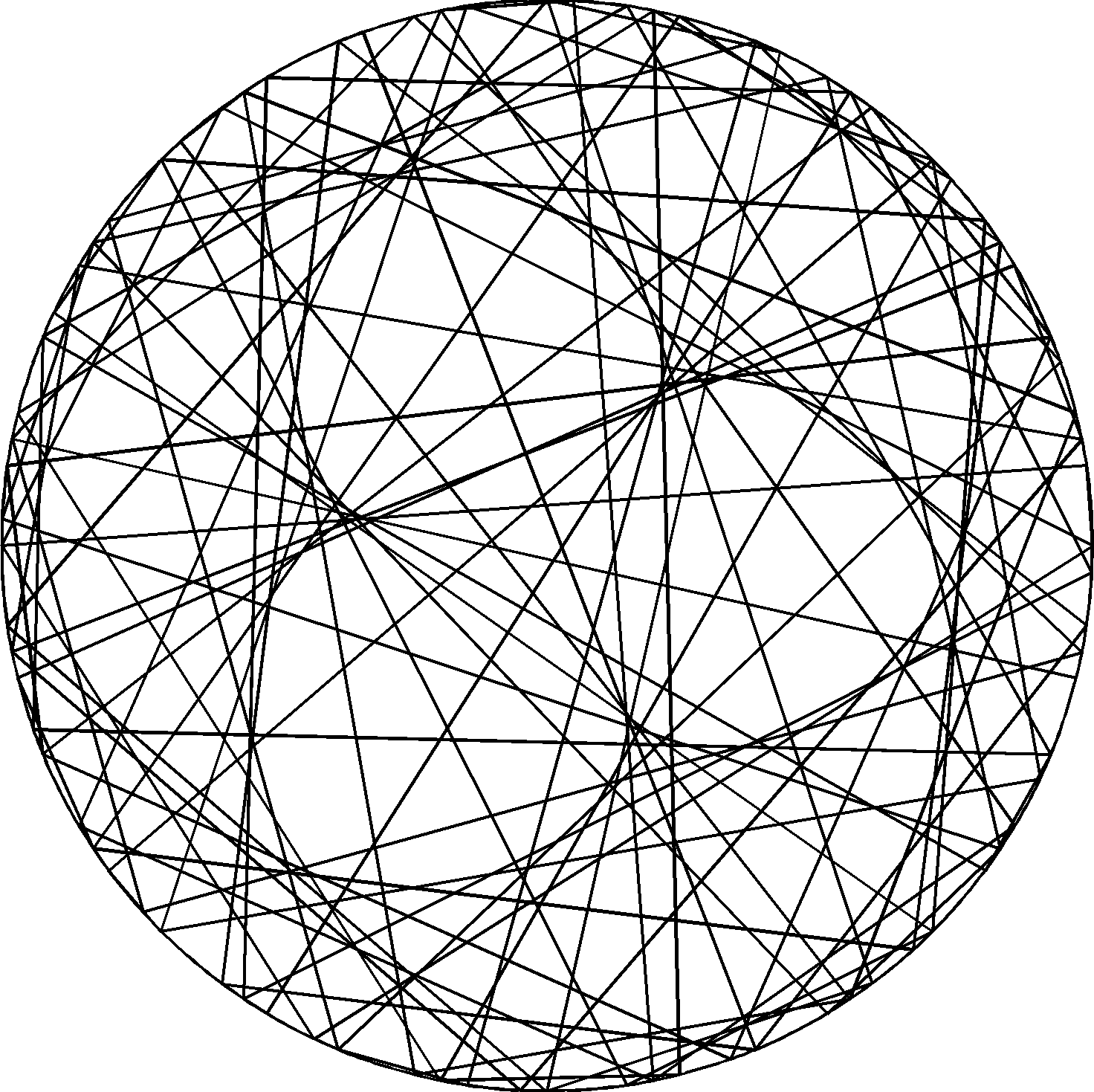} \\  (E) 
\end{framed}
\end{center}
\end{minipage} \hskip0.2cm
\begin{minipage}{3.2cm}
\begin{center} 
\begin{framed}
{ High mixing  (transmitter spacing $15\pi/4$)} \\ \vskip0.2cm
\includegraphics[width=2.3cm]{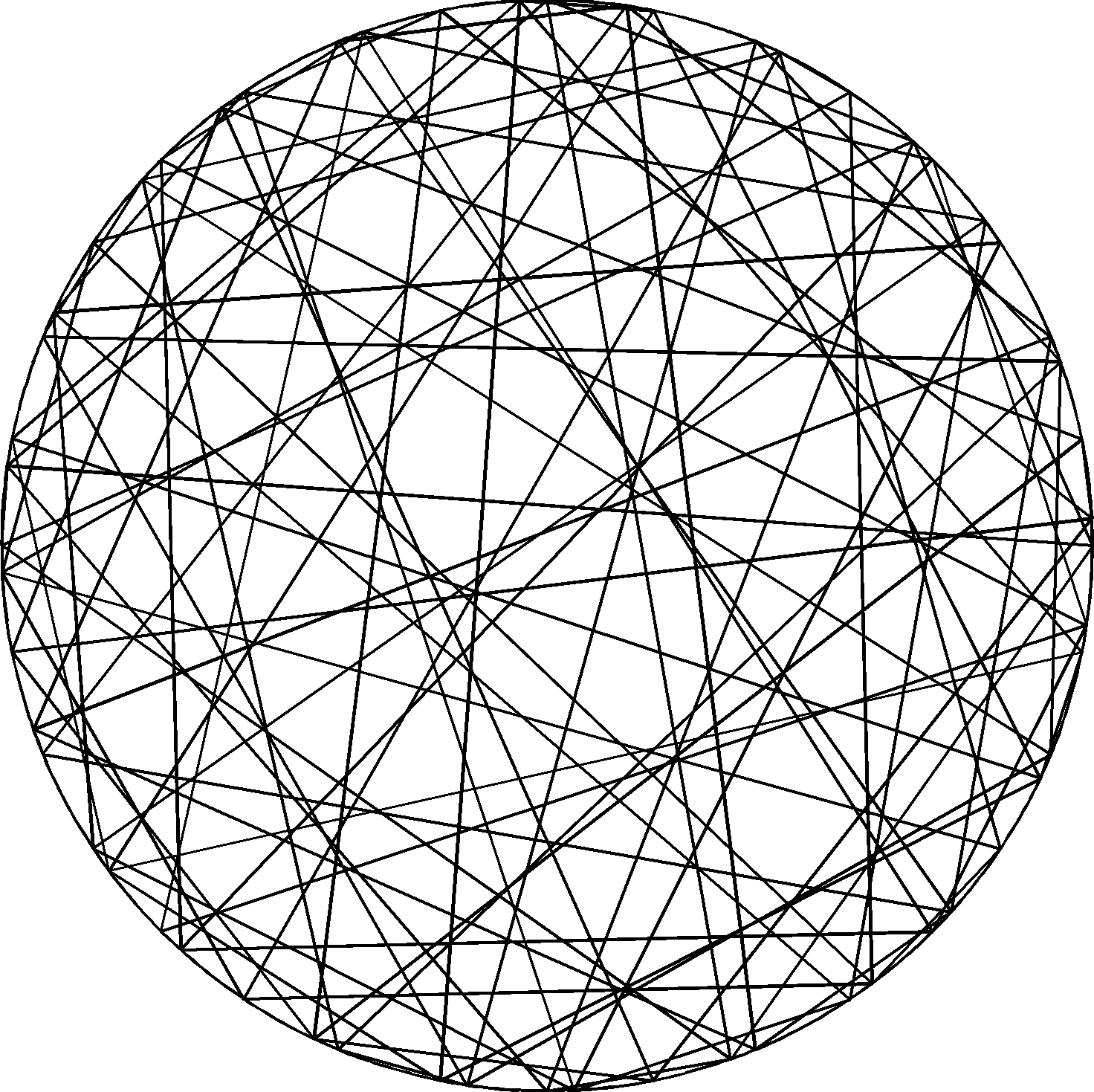} \\  (F) 
\end{framed}
\end{center} 
\end{minipage}
\end{framed}
\end{center}
\end{minipage} \hskip0.2cm
\begin{minipage}{7.6cm}
\begin{center}
\begin{framed}
{ Dense (receiver spacing ${\pi}/{64}$)} \\ \vskip0.3cm
\begin{minipage}{3.2cm}
\begin{center}
\begin{framed}
{ Low mixing (transmitter spacing $\pi/32$)} \\ \vskip0.2cm
\includegraphics[width=2.3cm]{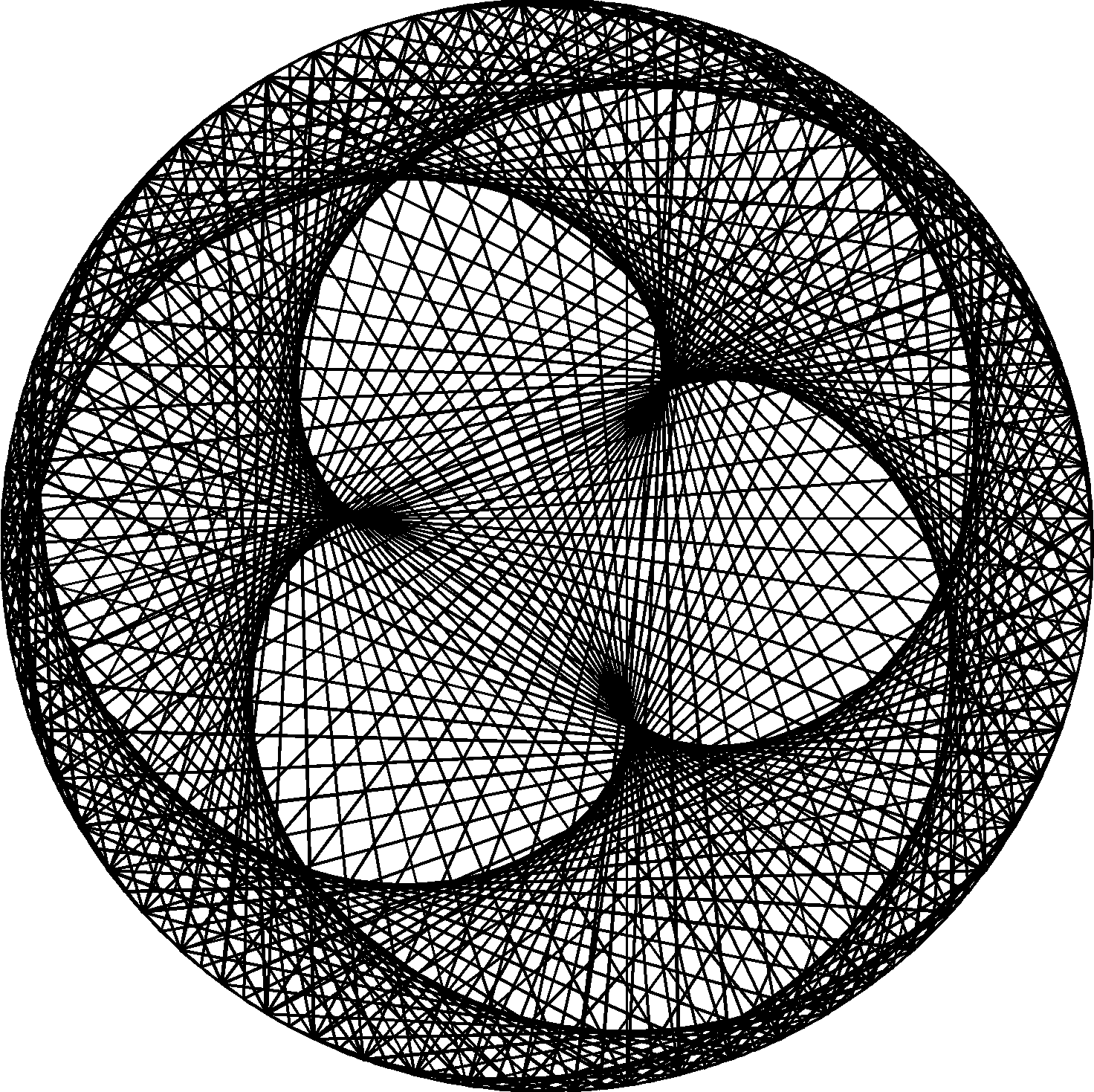} \\  (G) 
\end{framed}
\end{center}
\end{minipage} \hskip0.2cm
\begin{minipage}{3.2cm}
\begin{center}
\begin{framed}
{ High mixing (transmitter spacing $15\pi/16$)} \\ \vskip0.2cm
\includegraphics[width=2.3cm]{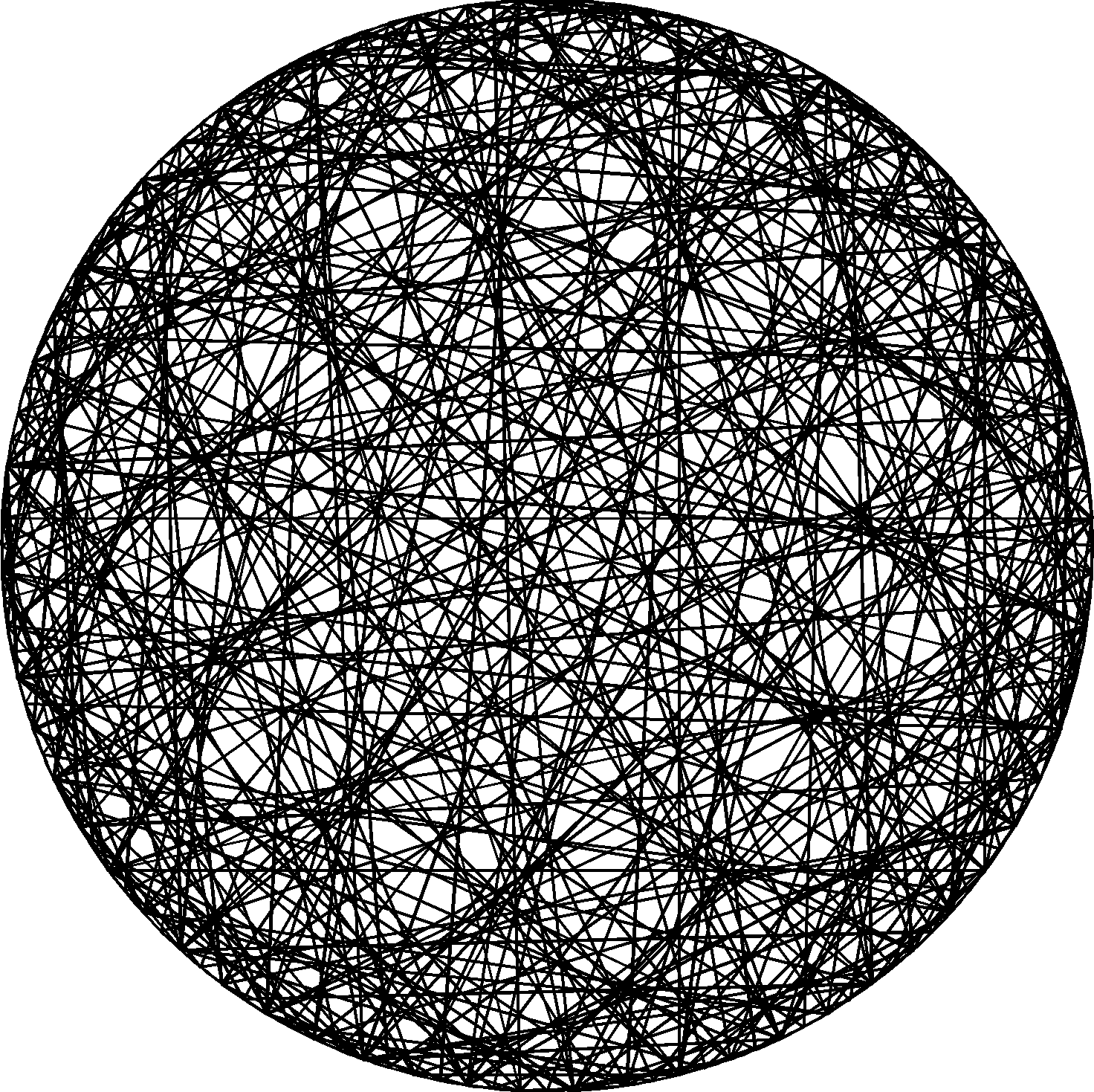} \\  (H) 
\end{framed}
\end{center}
\end{minipage} 
\end{framed}
\end{center}
\end{minipage}   \end{framed} 
\end{center}
\end{scriptsize}
\caption{Signal configurations (A)--(H) sketched with staight line segments connecting the transmitter and receiver positions. \label{signal_configurations}}
\end{figure*}

Numerical experiments were conducted in the square $\Omega = [-1,1] \times [-1,1]$. The boundary curve of $\Omega_0$  (Figure \ref{omega_0}) was identical to that of \cite{pursiainen2014}.  Similar to \cite{pursiainen2014}, the spatial scaling factor $s$ (Section \ref{section:forward_model})  was chosen to be $500$ m, leading to $90$--$135$ m (0.18--0.27) diameter of $\Omega_0$ and $30$--$45$ m (0.060--0.089) maximum diameter of the inclusions. The  data was asumed to have been transmitted and received  on a 160 m (0.32) diameter circle (synthetic circular orbit)  $\mathcal{C}$ centered at the origin.

\subsubsection{Permittivity}

Two different permittivity distributions (I) and (II) were explored. In (I), three elongated inclusions modeling vacuum cavities within $\Omega_0$ were to be recovered, and in (II), additionally a surface layer with thickness 10 \% of the diameter of $\Omega_0$, e.g., dust or ice cover,  was a target of reconstruction. The relative permittivity was assumed to be granular with the grain (finite element) size of  0.3--1.5 m (0.00060--0.0030) in other than the vacuum parts of the domain. In (I), the permittivity of each grain was drawn from a  flat distribution covering the interval $[2, 6]$. In (II), the choice was otherwise the same except that the interval $[1, 3]$ was used for the surface layer. The permittivity of the inclusions and the exterior of $\Omega_0$ was chosen to be one, i.e., that of the vacuum. 

The background permittivity of $\Omega_0$, i.e., the initial guess of the inversion procedure, was chosen to be four, matching, e.g., with granite, dunite or kaolinite \cite{davis1989,michel2015,herique2002}.  Vacuum background permittivity one was used for the remaining subdomain. 

\subsubsection{Conductivity}

In $\Omega_0$, the conductivity causing a signal energy loss was assumed to have the latent  distribution $\sigma = 5 \varepsilon_r$, that is, around $0.11$ m S/m (e.g.\ granite) or 0.03 dB/m   attenuation  for the background permittivity value $\varepsilon_r = 4$. For comparison, the range 0.001--0.02 dB/m has been suggested for a comet \cite{kofman1998}. The remaining part  $\Omega \setminus \Omega_0$ was assumed to be lossless, i.e.\ $\sigma = 0$. The {\em a priori} guess utilized in the forward simulation was $\sigma = 20$ in $\Omega_0$, and $\sigma = 0$, otherwise. 

\subsubsection{Signal}
\label{section:total_noise} 

The Blackman-Harris window \cite{irving2006,harris1978,nuttall1981} was used as the source function   $f(t, \vec{x} ) = \tilde{f}(t) \delta_{\vec{p}}(\vec{x})$, i.e., 
{\setlength\arraycolsep{2 pt} \begin{eqnarray}
\tilde{f}(t)  =  0.359 & - &  0.488 \cos \left ( {20 \pi t} \right) \nonumber \\  & + &  0.141 \cos \left ( {40 \pi t} \right)  - 0.012 \cos \left ( {60 \pi t}\right) 
\end{eqnarray}} for $t \leq 0.1$ (170 ns), and $\tilde{f}(t) = 0$, otherwise. This pulse is  centered around 10 MHz frequency  suitable for cavity detection \cite{binzel2005,irving2006,daniels2004,francke2009}. The data were gathered covering the time window from $t=0$ to $T=1.3$ (2.2 $\mu$s)  at $60$ MHz frequency, which was 1.7 relative to the Nyquist rate (NR), i.e.,  two times the density of sampling points divided by the bandwith of the signal.

Eight different signal path configurations (A)--(H) (Figure \ref{signal_configurations}, Table \ref{signal_configurations_table})   were tested. Of these, (A)--(D) were generated with two and (E)--(H) with four satellites.  In both cases, a single receiver recorded the signal transmitted separately by one or more satellites (Figure \ref{constellation_image}) orbiting in an evenly spaced  formation.  Both the receiver and the transmitters were placed on the circle $\mathcal{C}$. The set of receiver positions included 32 and 128 equispaced points in  (A), (B), (E), (F)   and (C), (D), (G), (H), respectively. For these groups, the spatial sampling rate (SR) (density of the receiver positions) was 0.53 and 2.1 relative to the NR, respectively. Mixing due to unequal orbital velocities was modeled by setting the angular spacing of the transmitting points to be 2 and 60 times that of the receiver positions in (A), (C), (E), (G) and (B), (D), (F), (H), respectively. Based on Figure \ref{signal_configurations}, the latter group can be observed to have an increased  variability in the signal directions compared to the former one. 

\subsubsection{Noise}

The noise term of  (\ref{lin_eq}) included  both forward errors and simulated measurement noise. Forward inaccuracies can be classified at least to modeling and simulation errors, of which the former can be associated with {\em a priori} uncertainty (e.g.\ initial guess) and, in particular, with the linearized approach, whereas the latter follows from the FDTD/FEM forward computations.  Measurement errors were simulated through additive zero-mean Gaussian white noise with the  standard deviation of 6 \% with respect to the main signal peak at the measurement location.  Similar to the CONSERT  data \cite{kofman2015},  the peaks caused by the simulated noise stayed mainly at least 20 dB below the main signal peak (Figure \ref{signal_fig}).

\subsubsection{Forward computations}

In order to avoid overly good fit (inverse crime) \cite{colton1998, kaipio2004}, exact data were computed using a different FE mesh (92569 nodes, 184336 triangular elements) than $\mathcal{T}$ of the forward simulation (78509 nodes, 156216 triangles). The temporal increment of leap-frog time integration was chosen to be $\Delta t = 2.5 \cdot 10^{-4}$  and  $\Delta t = 5 \cdot 10^{-4}$, respectively. The coarse nested mesh ${\mathcal{T}'}$ covering $\Omega_0$ consisted of 1647 nodes and 3104  triangles with each element covering four triangles of $\mathcal{T}$ following from the regular mesh refinement principle, in which each element edge is split into two equivalent halves.     

\subsubsection{Regularization}

Regularization parameters needed in the deconvolution and inversion procedure were chosen based on some preliminary tests in order  to find a balance between accuracy and numerical stability of the estimates. The regularized deconvolution formula (\ref{dc_eq}) was applied with $\nu = 10^{-3}$. In the inversion procedure (\ref{tv_iteration}), three iteration steps were performed with $\alpha = \beta = 0.01$. Each parameter was fixed to roughly in the middle of the logarithmic range of workable values, which for $\alpha$ and $\beta$  was approximately from $10^{-4}$ to $1$ and for $\nu$  
from $10^{-5}$ to $10^{-1}$. 

\subsubsection{Relative overlap}

The accuracy of the inversion results were examined through the relative overlapping area (ROA), i.e., the percentage \begin{equation} \hbox{ROA} = 100 \frac{\hbox{Area}( \mathcal{A}) }{ \hbox{Area}(\mathcal{S})}, \end{equation} where $\mathcal{A} = \mathcal{S} \cap \mathcal{R}$ is the overlap between the set $\mathcal{S}$ to be recovered and the set $\mathcal{R}$ in which a given reconstruction is smaller than a limit such that $\hbox{Area}(\mathcal{R}) = \hbox{Area}(\mathcal{S})$. The sets $\mathcal{A}$ and $\mathcal{S}$ were analyzed also visually.

\section{Results}

\begin{figure*}[t]
\begin{scriptsize}
\begin{center}
\begin{framed}
{ Single transmitter} \\ \vskip0.2cm
\begin{minipage}{7.6cm}
\begin{center}
\begin{framed}
{ Sparse (receiver spacing ${\pi}/{16}$)} \\ \vskip0.2cm
\begin{minipage}{3.2cm}
\begin{center}
\begin{framed}
{Low mixing  (transmitter spacing $\pi/8$)}\\ \vskip0.2cm
\includegraphics[width=2.4cm]{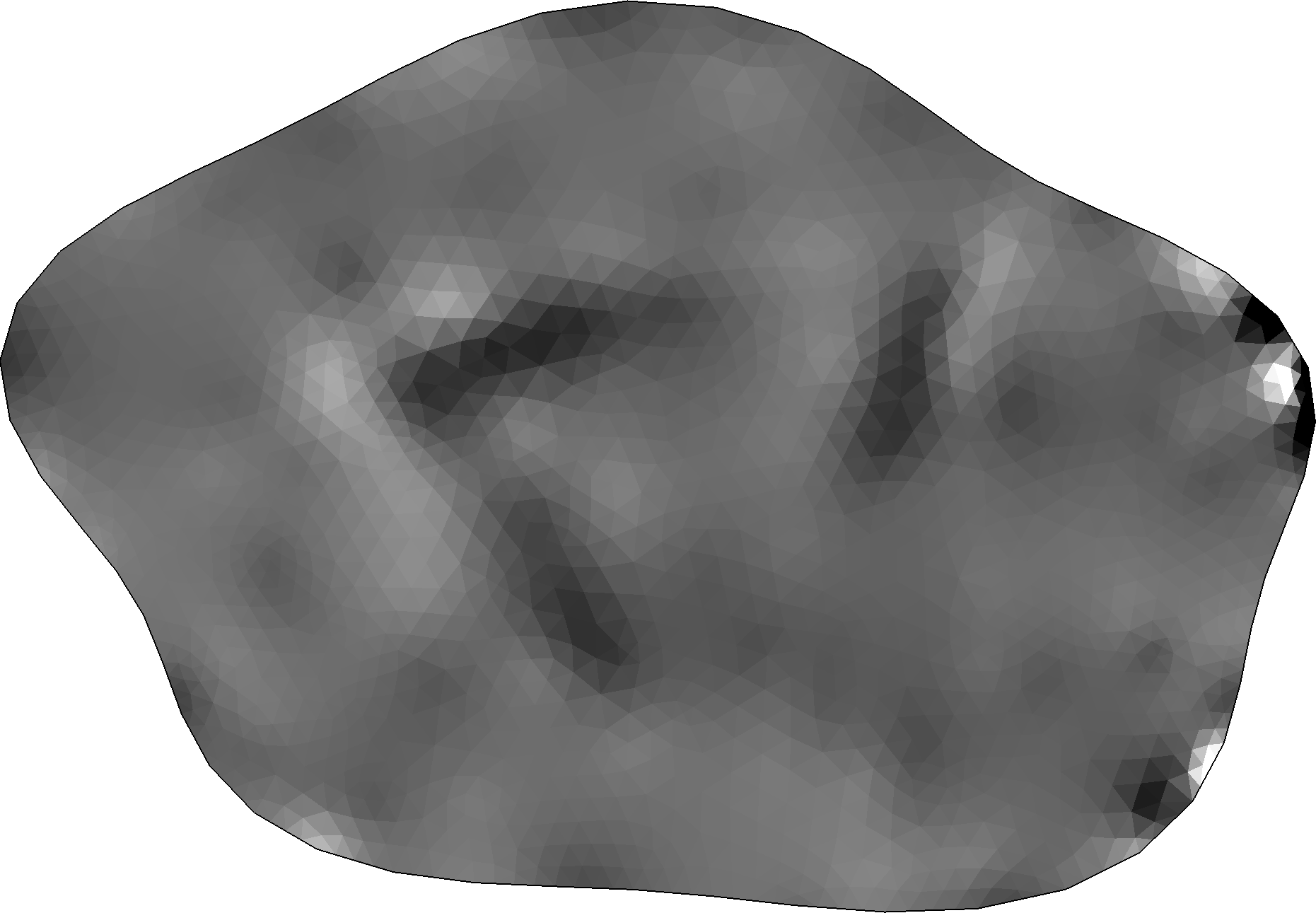} \\ \vskip0.2cm
\includegraphics[width=2.4cm]{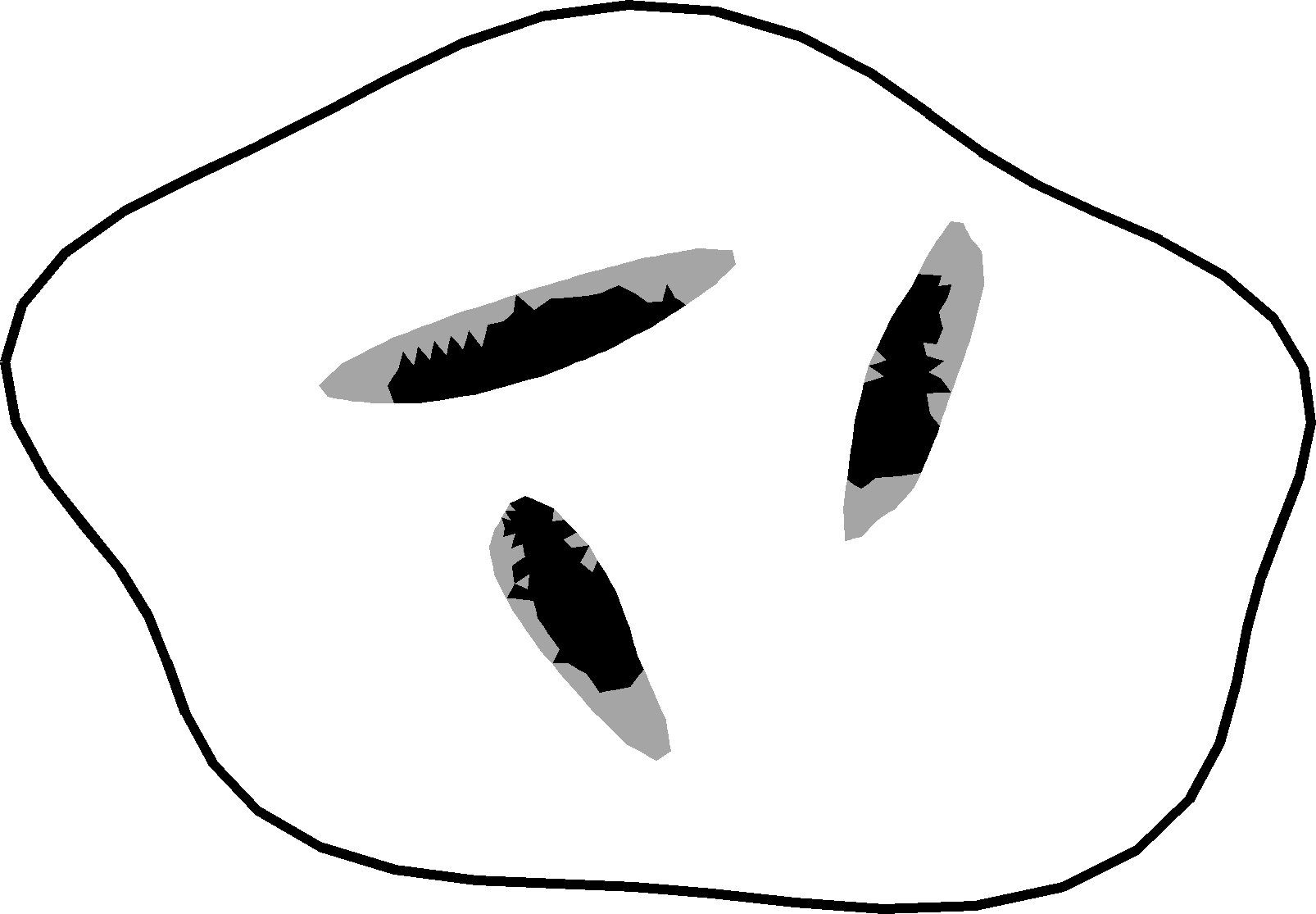} \\ (A)  \\ $\hbox{ROA} = 58 \%$
\end{framed}
\end{center}
\end{minipage} \hskip0.2cm
\begin{minipage}{3.2cm}
\begin{center} 
\begin{framed}
{ High mixing (transmitter spacing $15\pi/4$)} \\ \vskip0.2cm
\includegraphics[width=2.4cm]{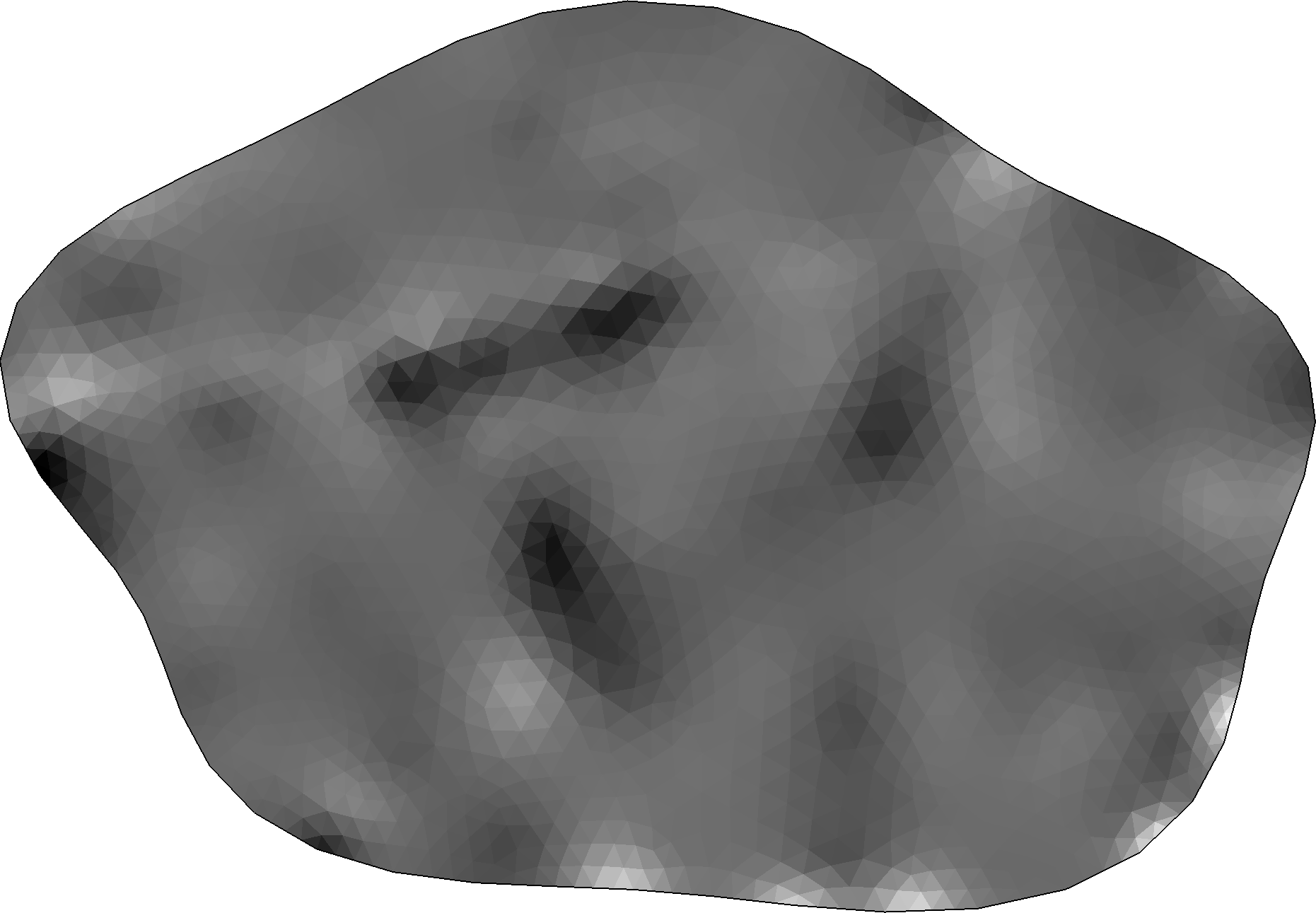} \\ \vskip0.2cm
\includegraphics[width=2.4cm]{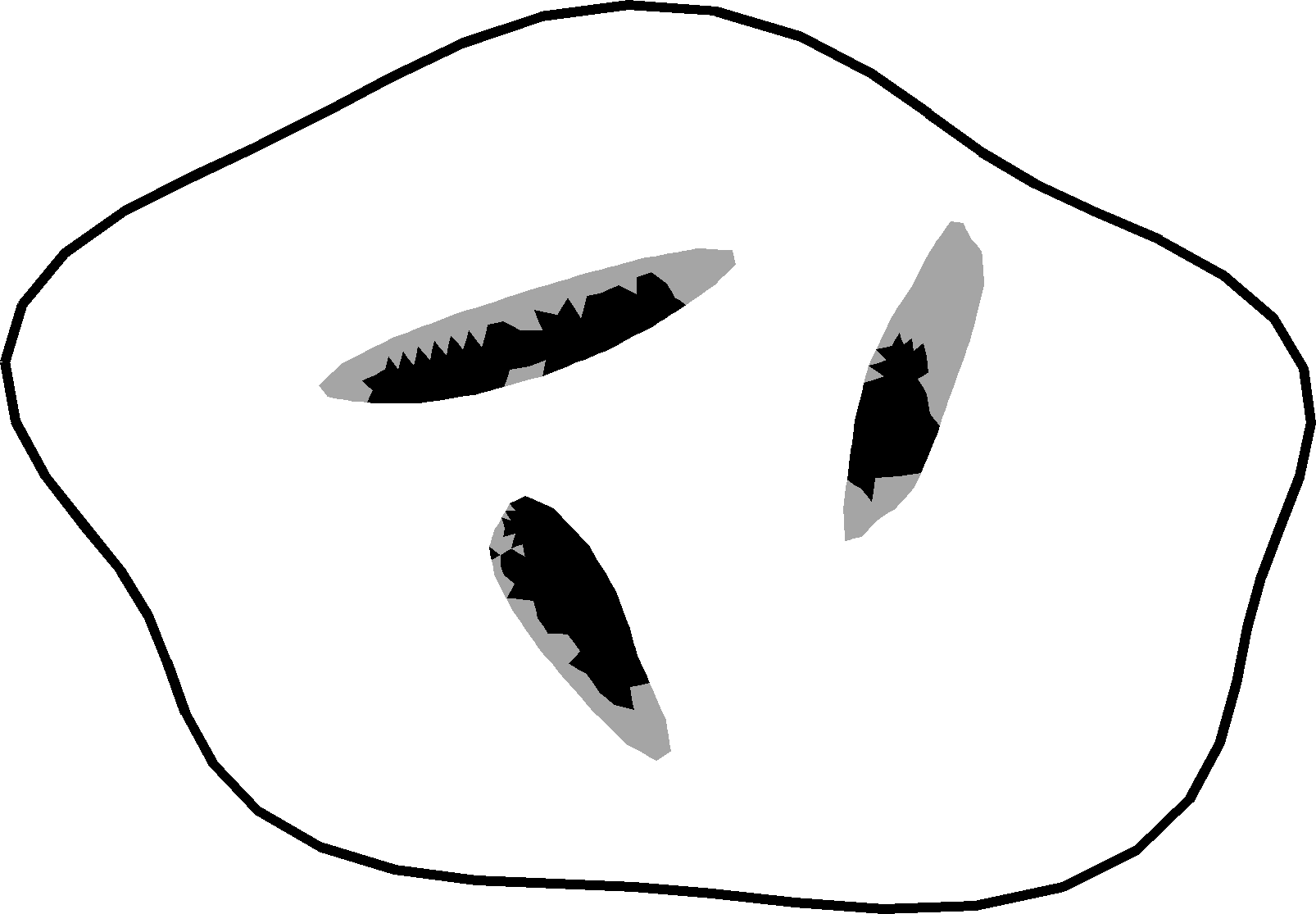} \\ (B) \\ $\hbox{ROA} = 55 \%$
\end{framed}
\end{center} 
\end{minipage}
\end{framed}
\end{center}
\end{minipage} \hskip0.2cm 
\begin{minipage}{7.6cm}
\begin{center}
\begin{framed}
{ Dense (receiver spacing ${\pi}/{64}$)} \\ \vskip0.2cm
\begin{minipage}{3.2cm}
\begin{center}
\begin{framed}
{ Low mixing (transmitter spacing $\pi/32$)} \\ \vskip0.2cm
\includegraphics[width=2.4cm]{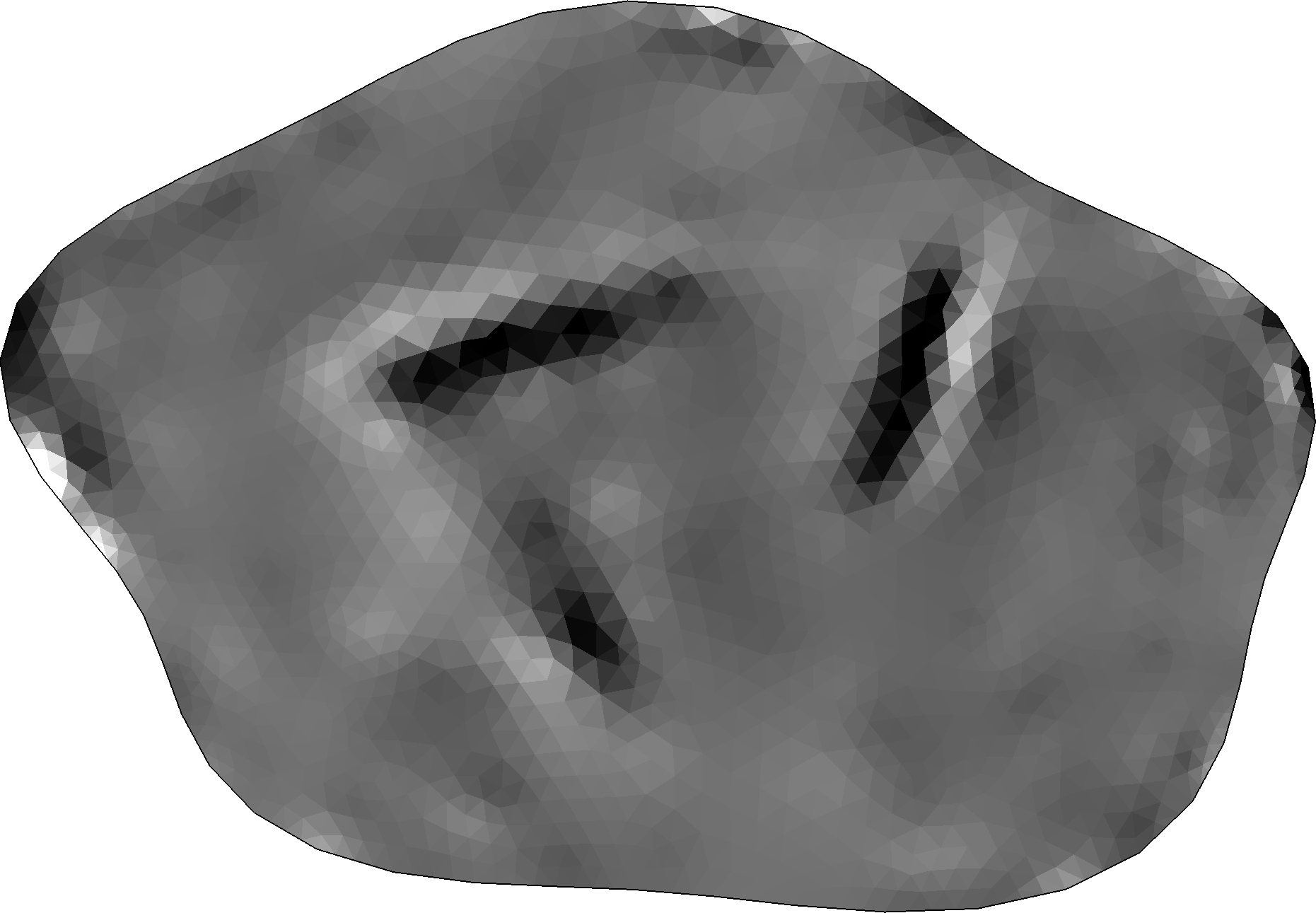} \\ \vskip0.2cm \includegraphics[width=2.4cm]{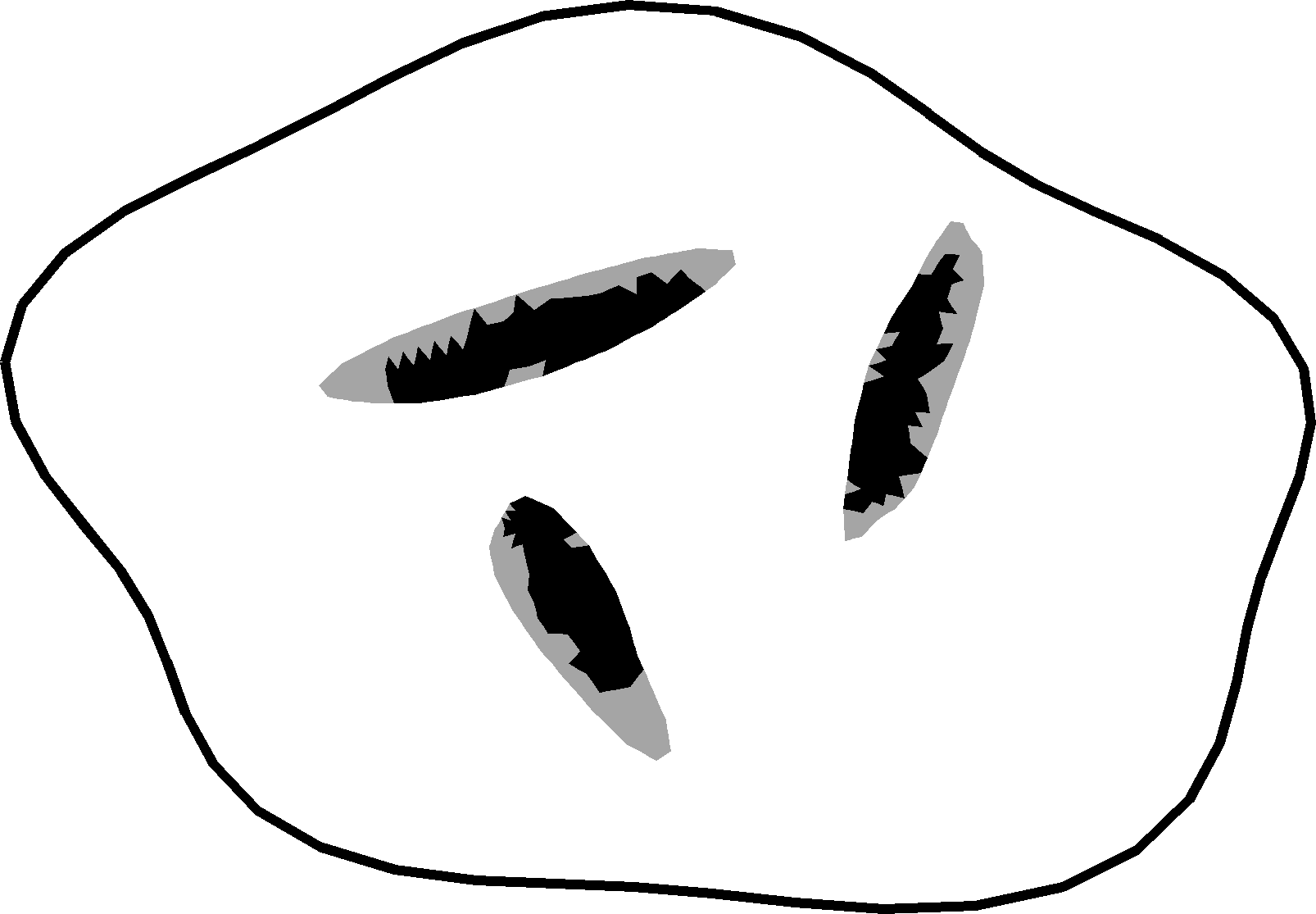} \\ (C) \\ $\hbox{ROA} = 60 \%$ 
\end{framed}
\end{center}
\end{minipage} \hskip0.2cm 
\begin{minipage}{3.2cm}
\begin{center}
\begin{framed}
{ High mixing (transmitter spacing $15\pi/16$)} \\ \vskip0.2cm
\includegraphics[width=2.4cm]{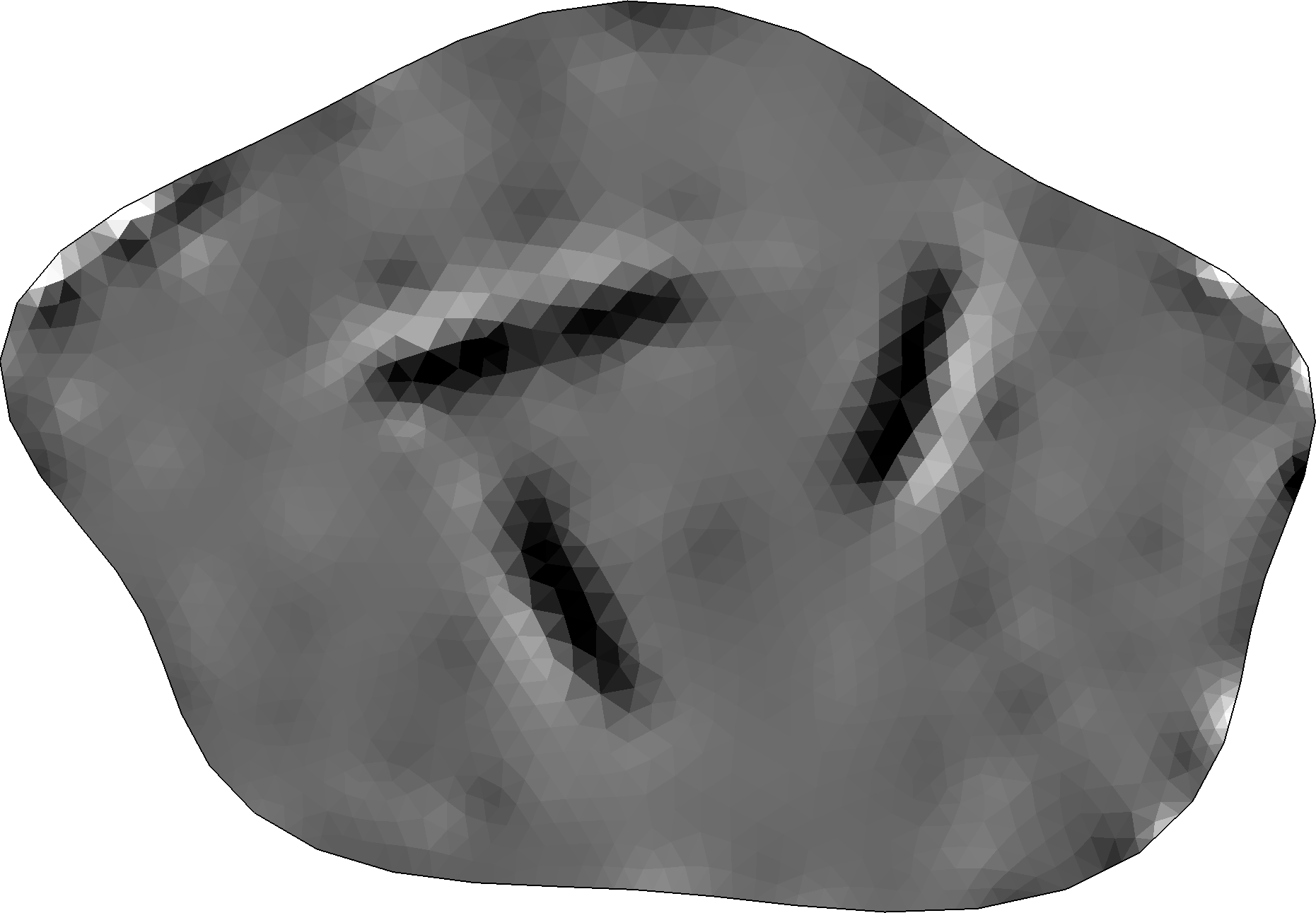} \\ \vskip0.2cm
\includegraphics[width=2.4cm]{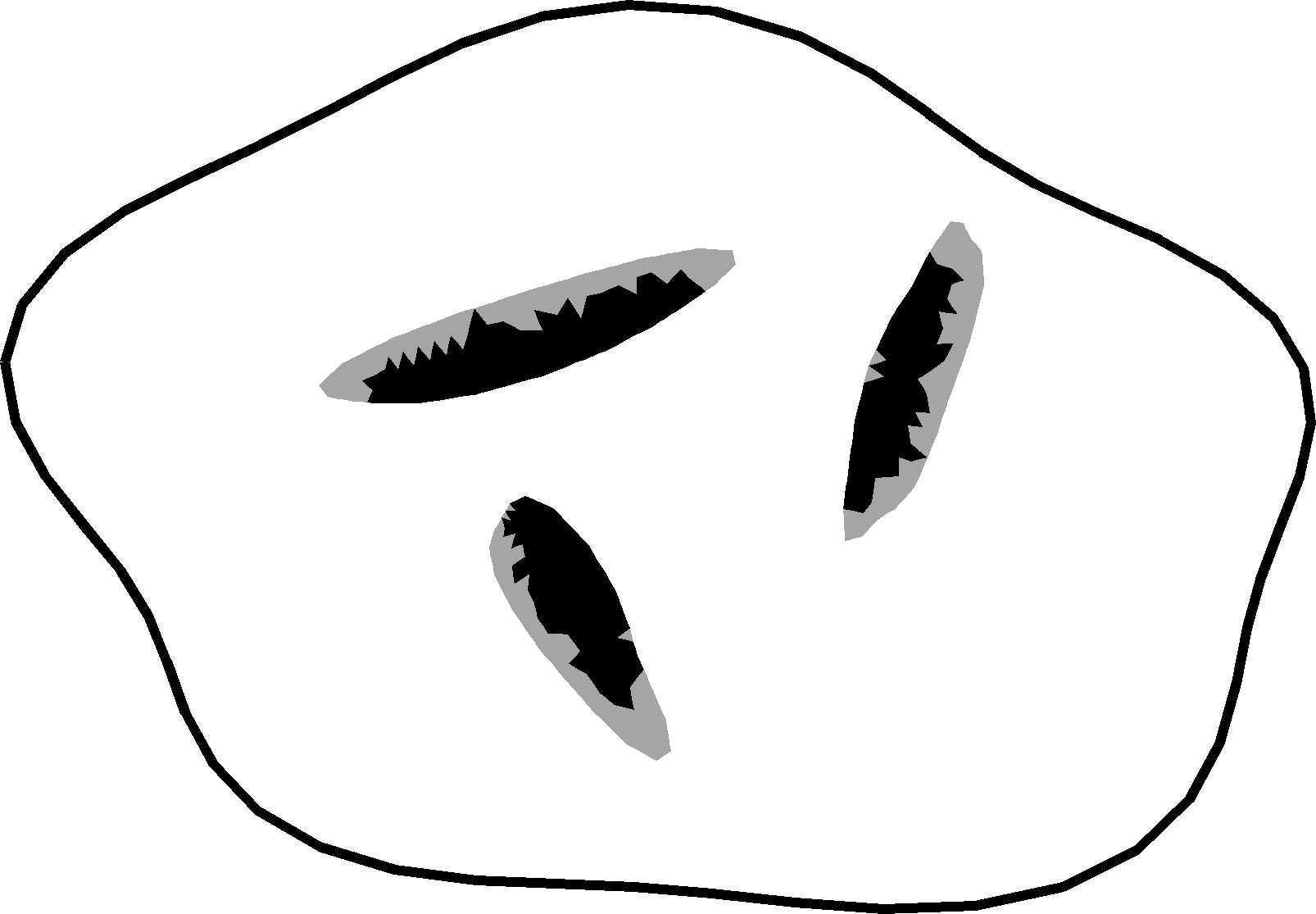} \\ (D) \\ $\hbox{ROA} = 61 \%$ 
\end{framed}
\end{center}
\end{minipage} 
\end{framed}
\end{center} 
\end{minipage}
\end{framed} 
\begin{framed}
{ Three transmitters} \\ \vskip0.2cm
\begin{minipage}{7.6cm}
\begin{center}
\begin{framed}
{ Sparse (receiver spacing ${\pi}/{16}$)}  \\ \vskip0.2cm
\begin{minipage}{3.2cm}
\begin{center}
\begin{framed}
{ Low mixing  (transmitter spacing $\pi/8$)} \\ \vskip0.2cm
\includegraphics[width=2.4cm]{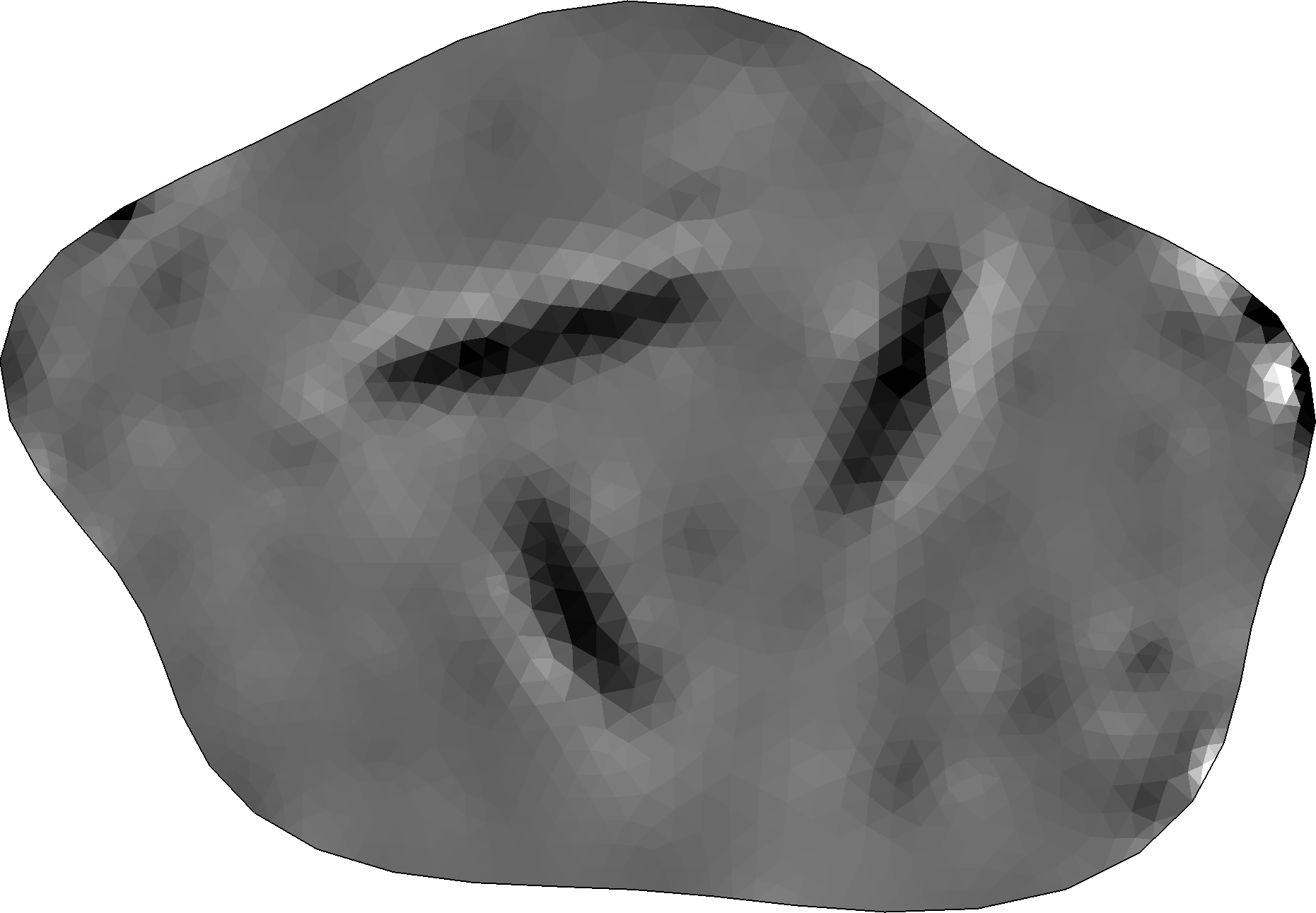} \\ \vskip0.2cm
\includegraphics[width=2.4cm]{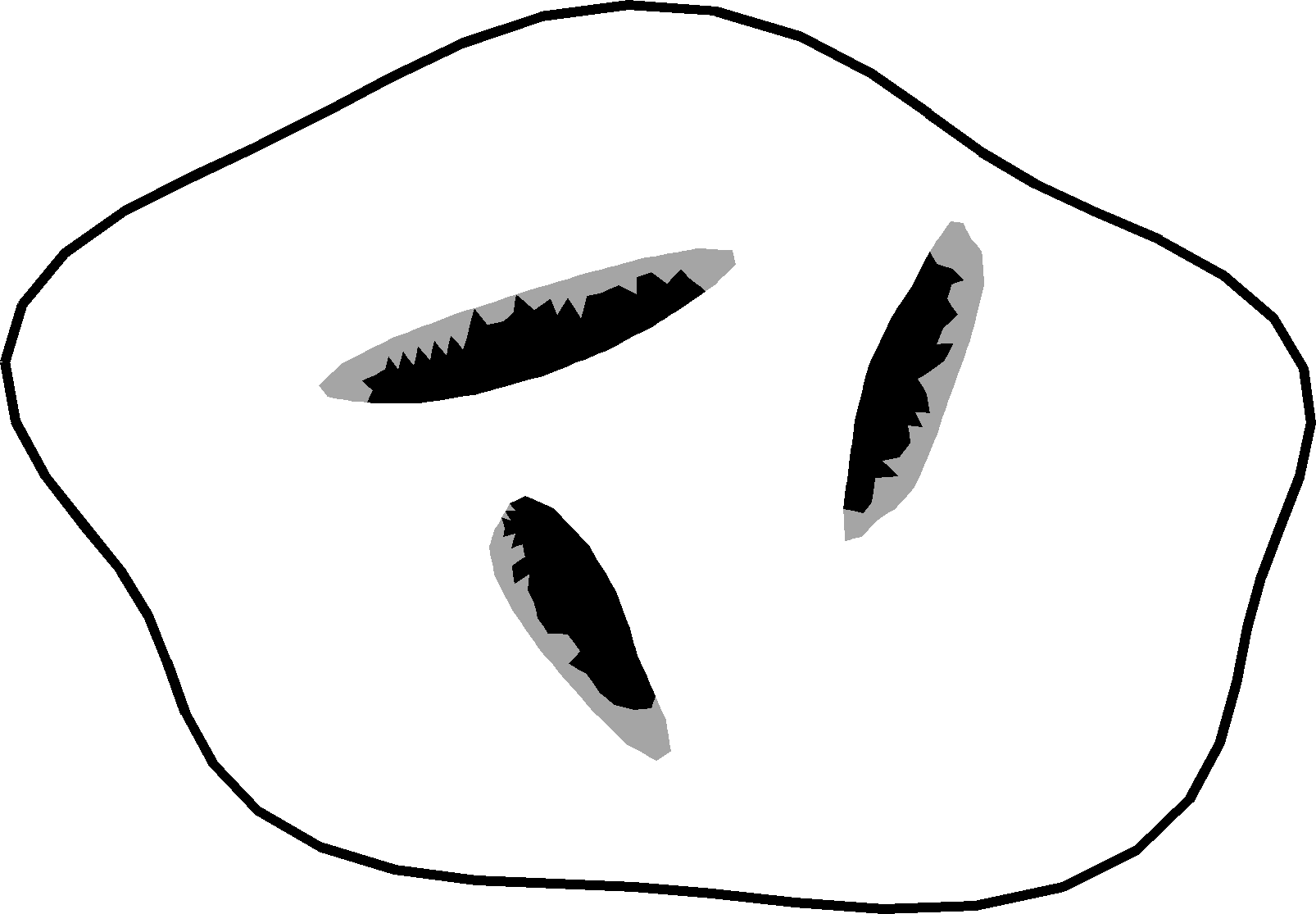} \\ (E)  \\ $\hbox{ROA} = 62 \%$
\end{framed}
\end{center}
\end{minipage} \hskip0.2cm
\begin{minipage}{3.2cm}
\begin{center} 
\begin{framed}
{ High mixing  (transmitter spacing $15\pi/4$)} \\ \vskip0.2cm
\includegraphics[width=2.4cm]{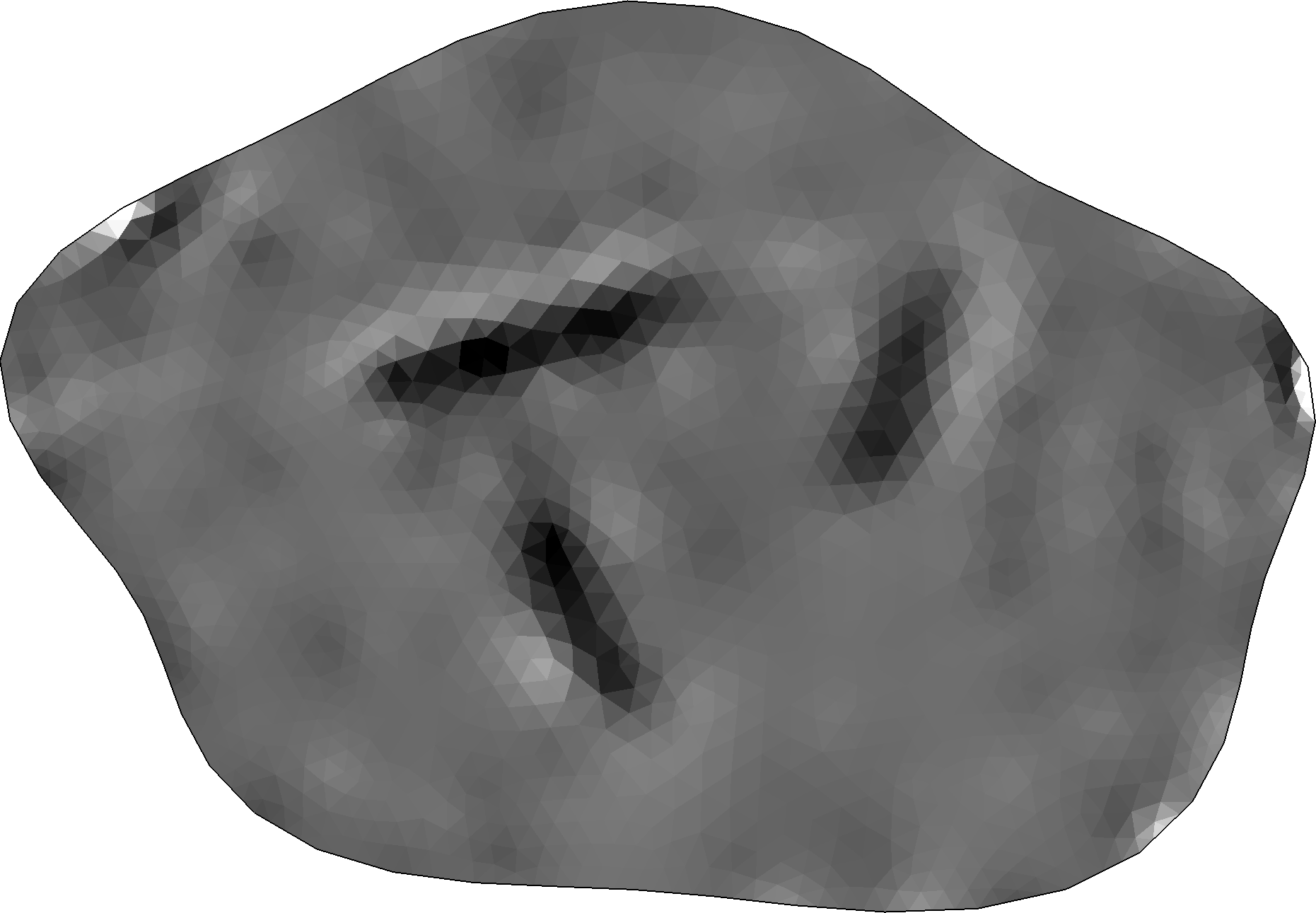} \\ \vskip0.2cm
\includegraphics[width=2.4cm]{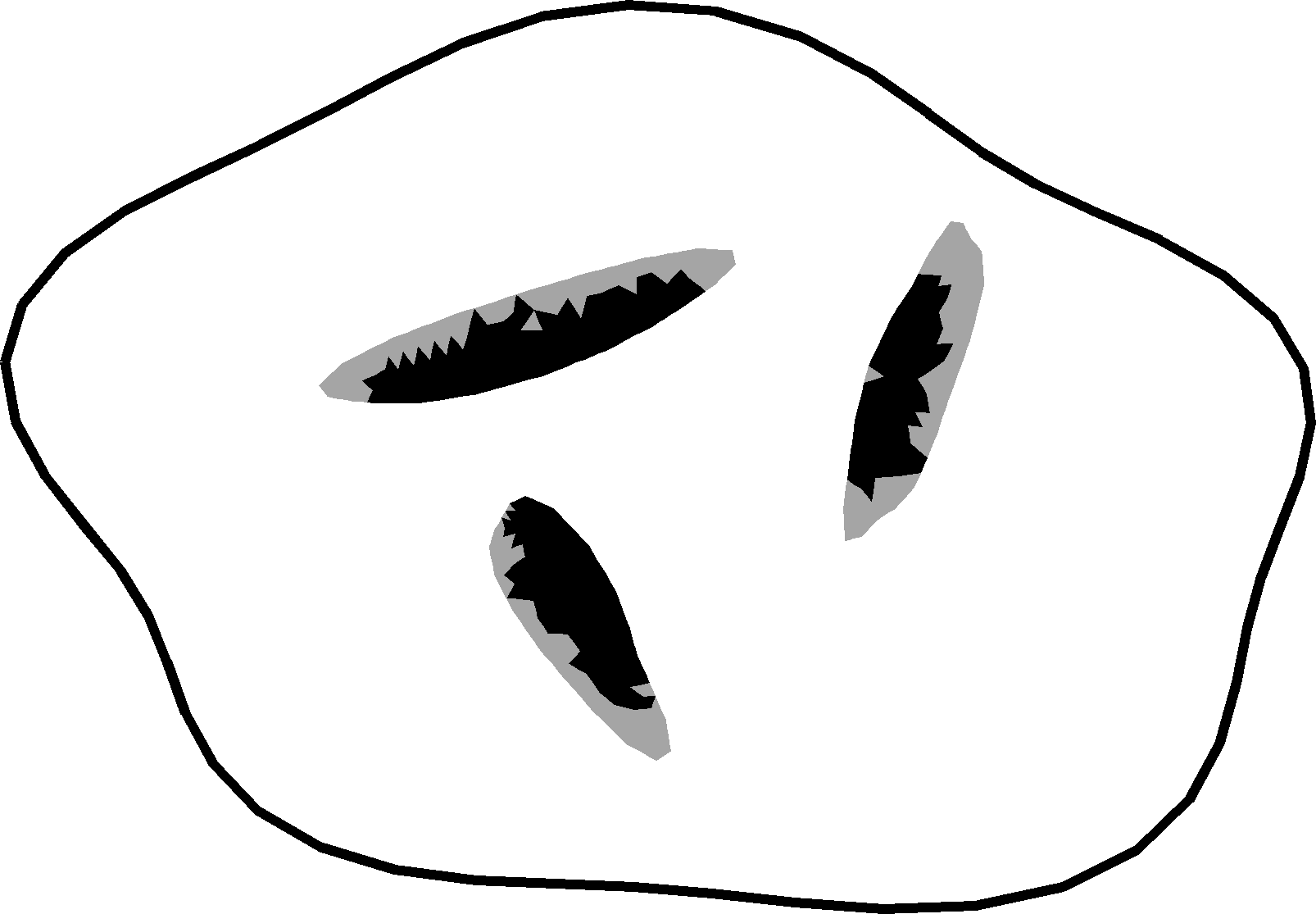} \\ (F) \\ $\hbox{ROA} = 61 \%$
\end{framed}
\end{center} 
\end{minipage}
\end{framed}
\end{center}
\end{minipage} \hskip0.2cm
\begin{minipage}{7.6cm}
\begin{center}
\begin{framed}
{ Dense (receiver spacing ${\pi}/{64}$)} \\ \vskip0.3cm
\begin{minipage}{3.2cm}
\begin{center}
\begin{framed}
{ Low mixing (transmitter spacing $\pi/32$)} \\ \vskip0.2cm
\includegraphics[width=2.4cm]{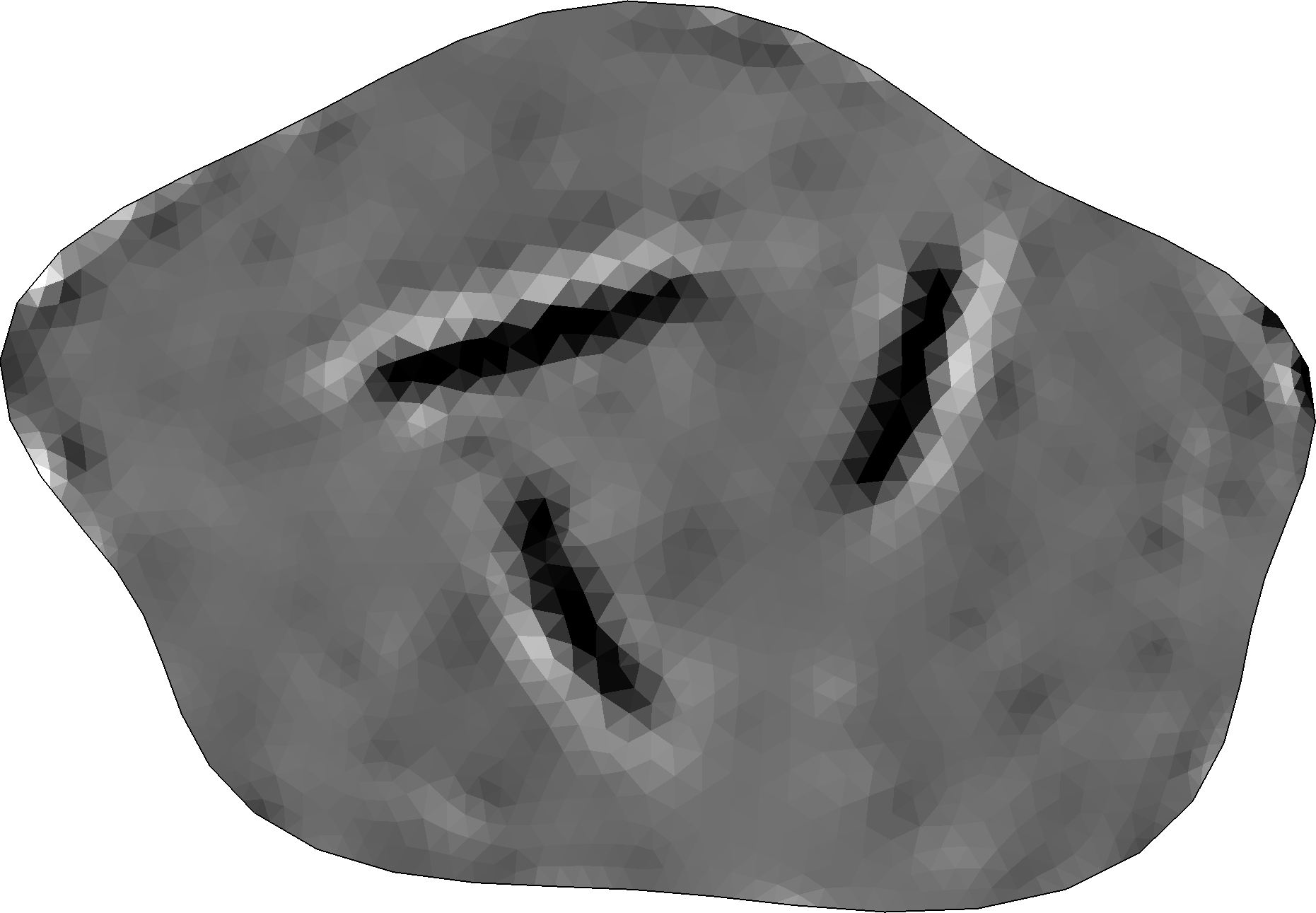} \\ \vskip0.2cm \includegraphics[width=2.4cm]{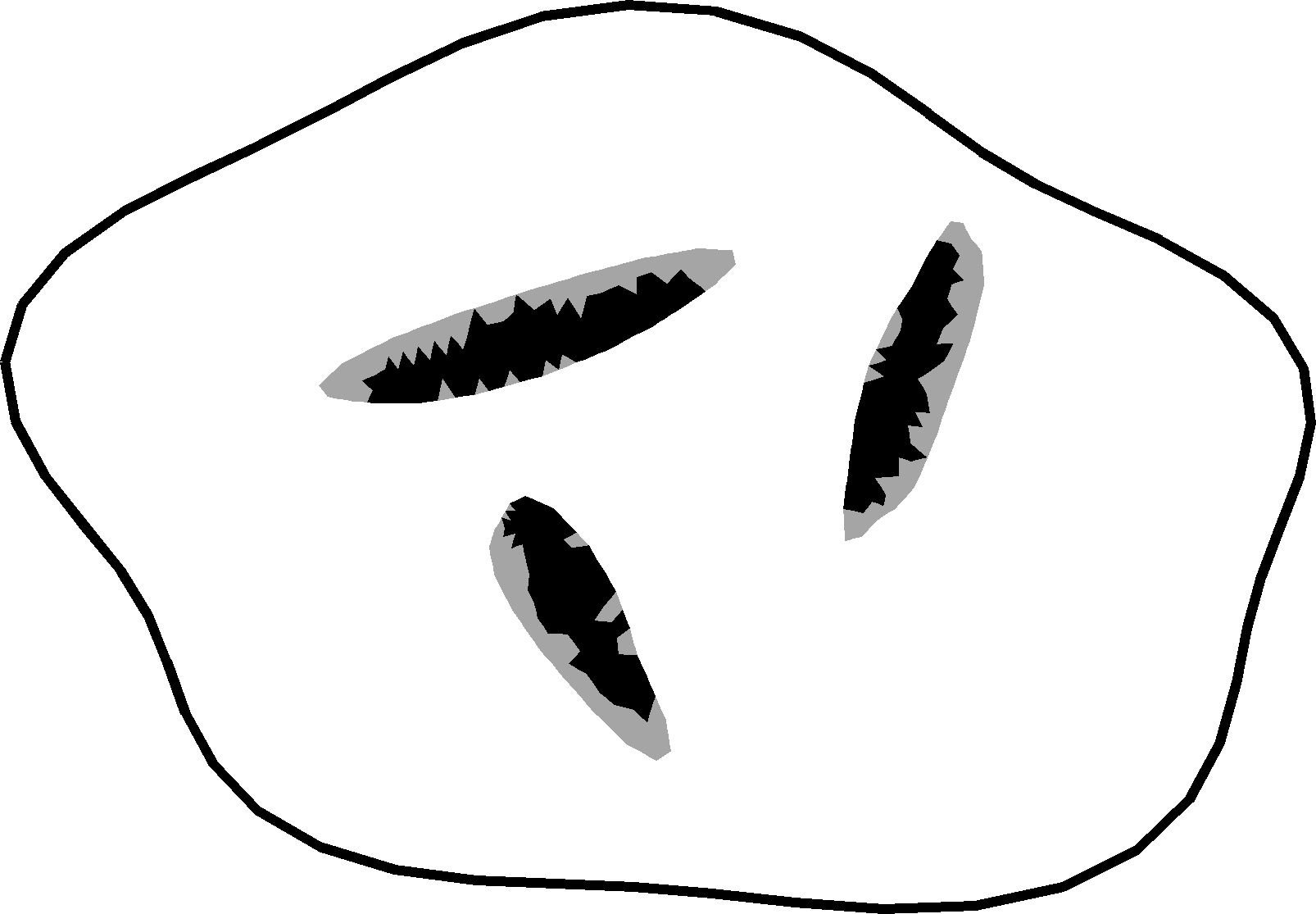} \\ (G) \\ $\hbox{ROA} = 61 \%$ 
\end{framed}
\end{center}
\end{minipage} \hskip0.2cm
\begin{minipage}{3.2cm}
\begin{center}
\begin{framed}
{ High mixing (transmitter spacing $15\pi/16$)} \\ \vskip0.2cm
\includegraphics[width=2.4cm]{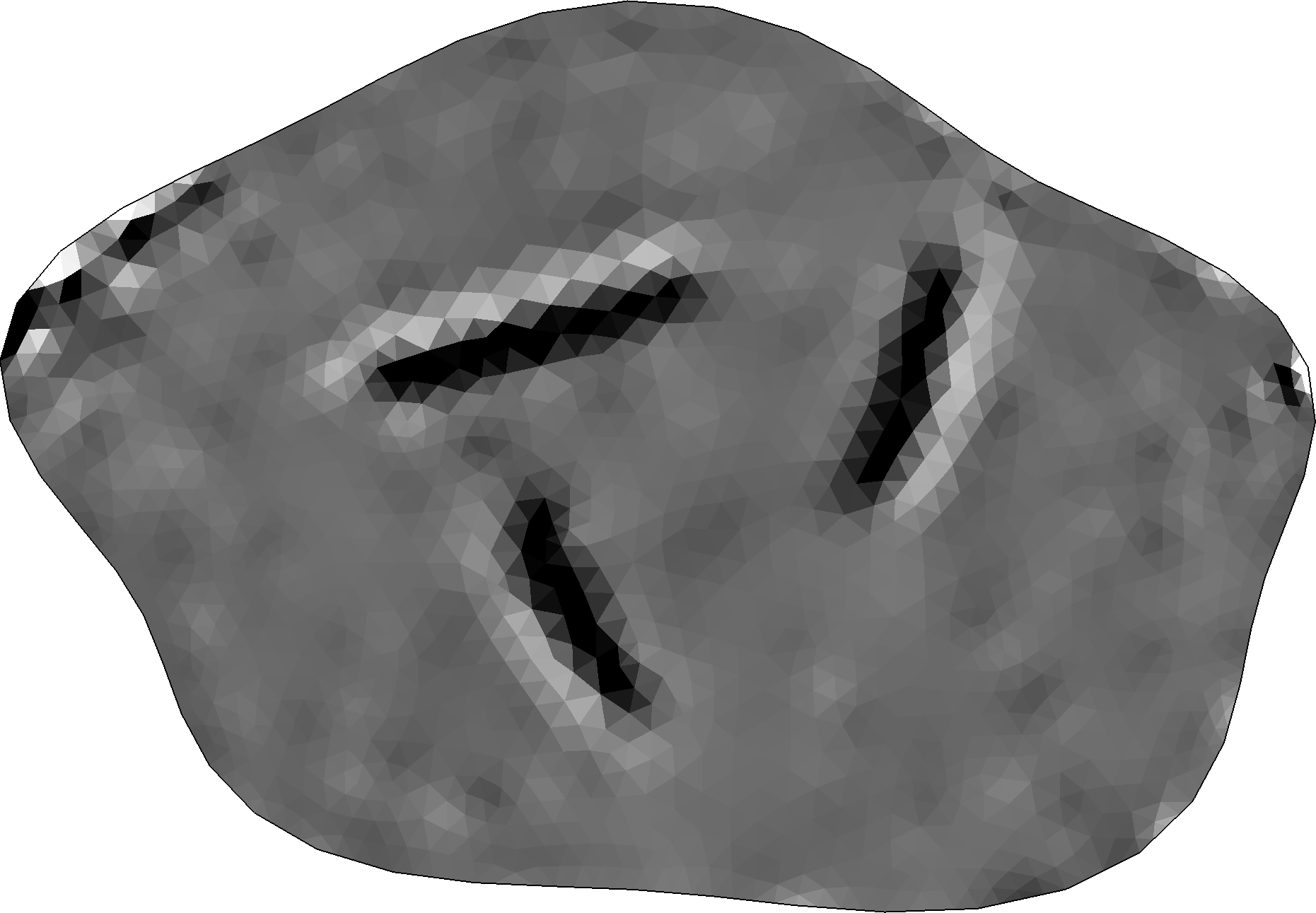} \\ \vskip0.2cm
\includegraphics[width=2.4cm]{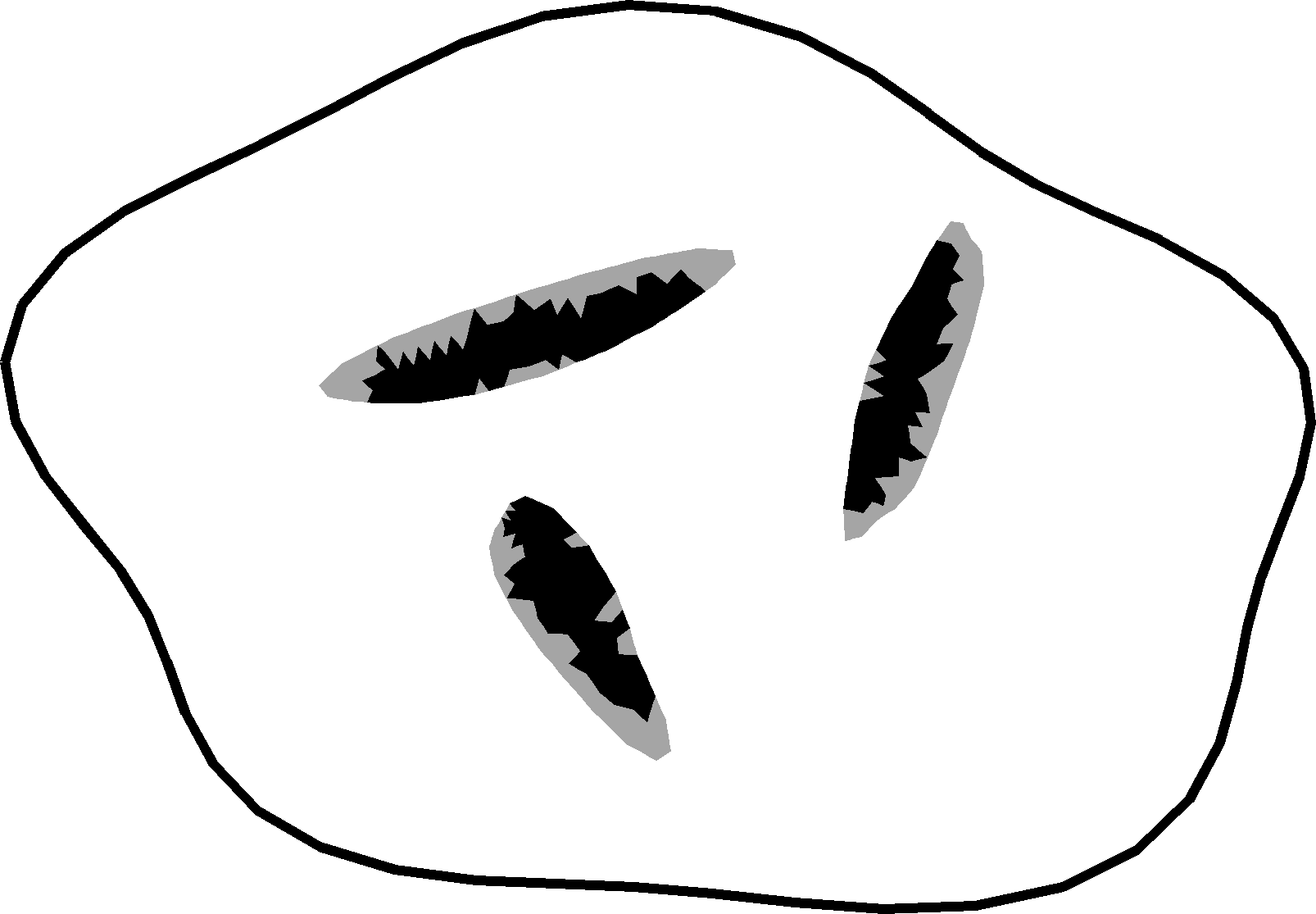} \\ (H) \\ $\hbox{ROA} = 62 \%$ 
\end{framed}
\end{center}
\end{minipage} 
\end{framed}
\end{center}
\end{minipage}   \end{framed} 
\end{center}
\end{scriptsize}
\caption{Reconstructions of permittivity distribution (I) for signal configurations (A)--(H).  In each case, the top image visualizes the actual reconstruction and the bottom one the sets $\mathcal{A}$ and $\mathcal{S}$. \label{results_1}}
\end{figure*}

\begin{figure*}[t]
\begin{scriptsize}
\begin{center}
\begin{framed}
{ Single transmitter} \\ \vskip0.2cm
\begin{minipage}{7.6cm}
\begin{center}
\begin{framed}
{ Sparse (receiver spacing ${\pi}/{16}$)} \\ \vskip0.2cm
\begin{minipage}{3.2cm}
\begin{center}
\begin{framed}
{Low mixing  (transmitter spacing $\pi/8$)}\\ \vskip0.2cm
\includegraphics[width=2.4cm]{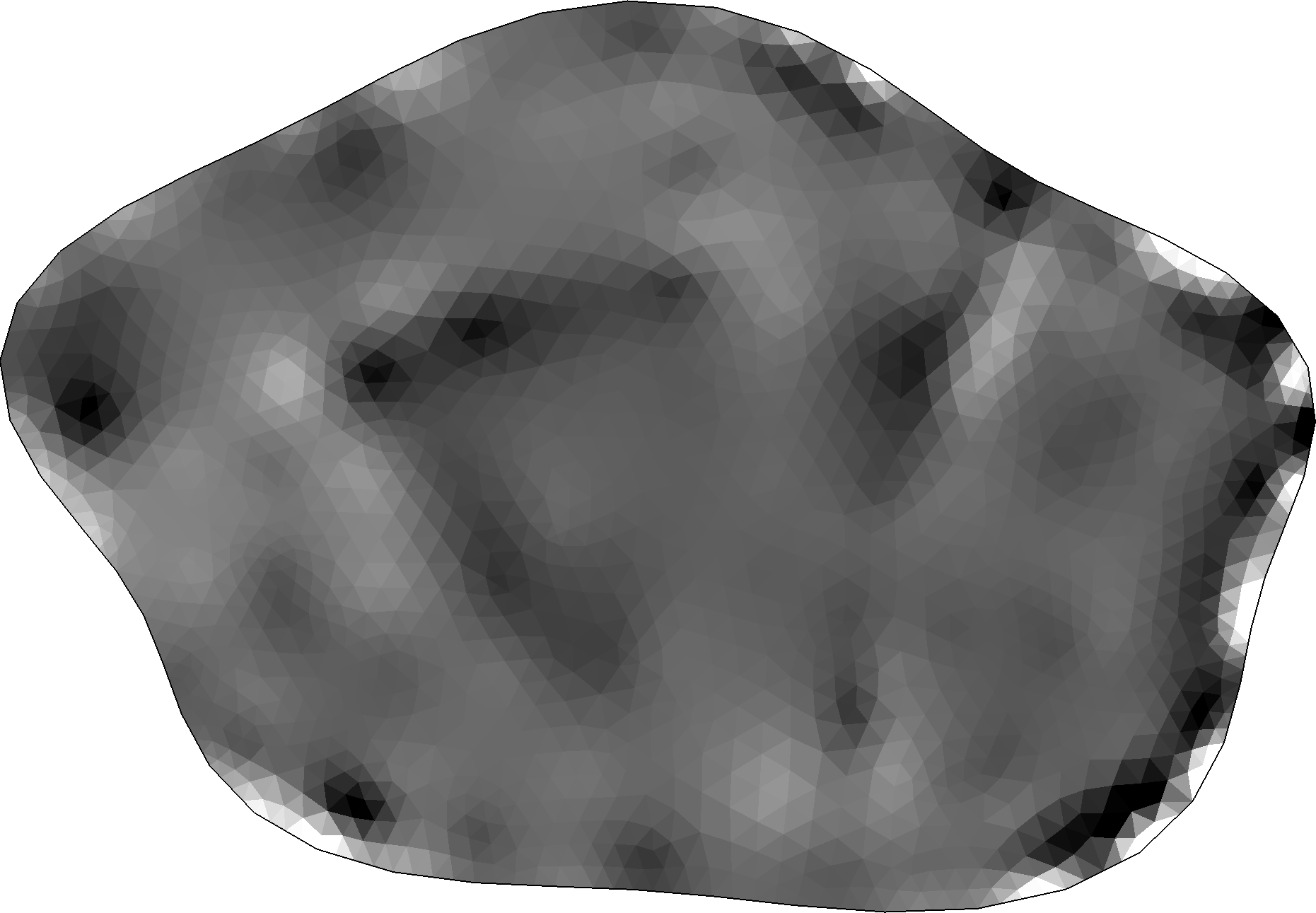} \\ \vskip0.2cm
\includegraphics[width=2.4cm]{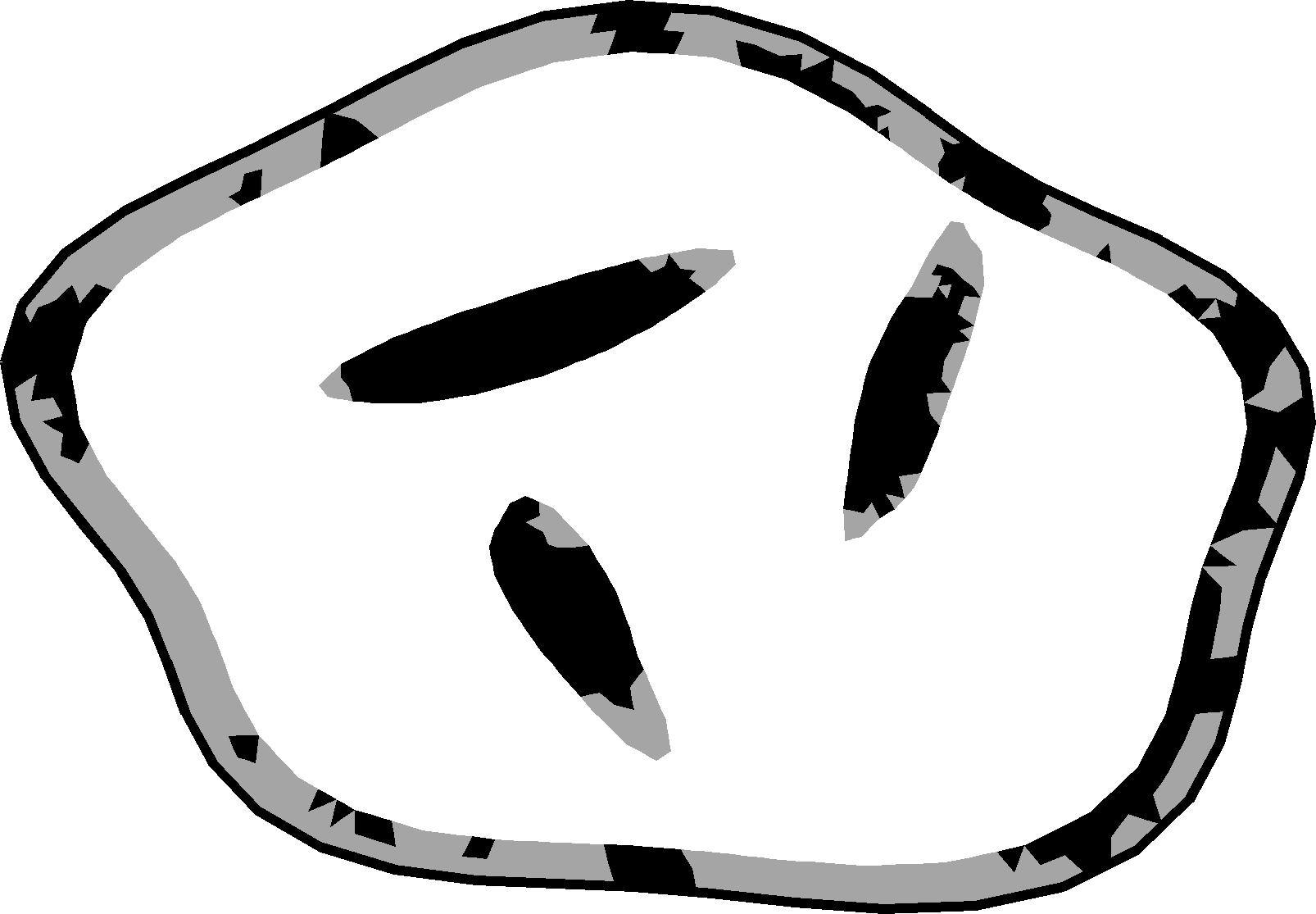} \\ (A)  \\ $\hbox{ROA} = 48 \%$
\end{framed}
\end{center}
\end{minipage} \hskip0.2cm
\begin{minipage}{3.2cm}
\begin{center} 
\begin{framed}
{ High mixing (transmitter spacing $15\pi/4$)} \\ \vskip0.2cm
\includegraphics[width=2.4cm]{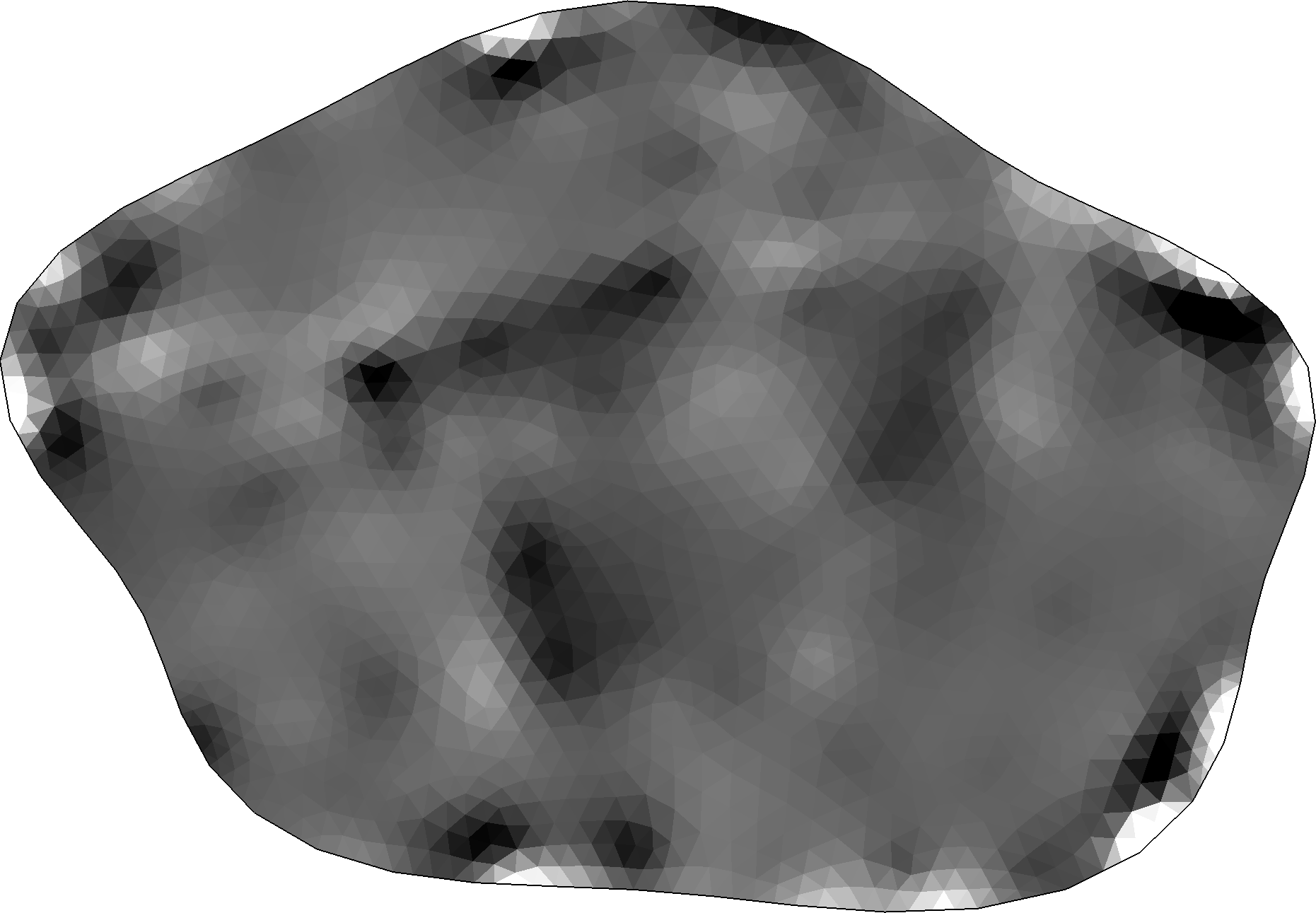} \\ \vskip0.2cm
\includegraphics[width=2.4cm]{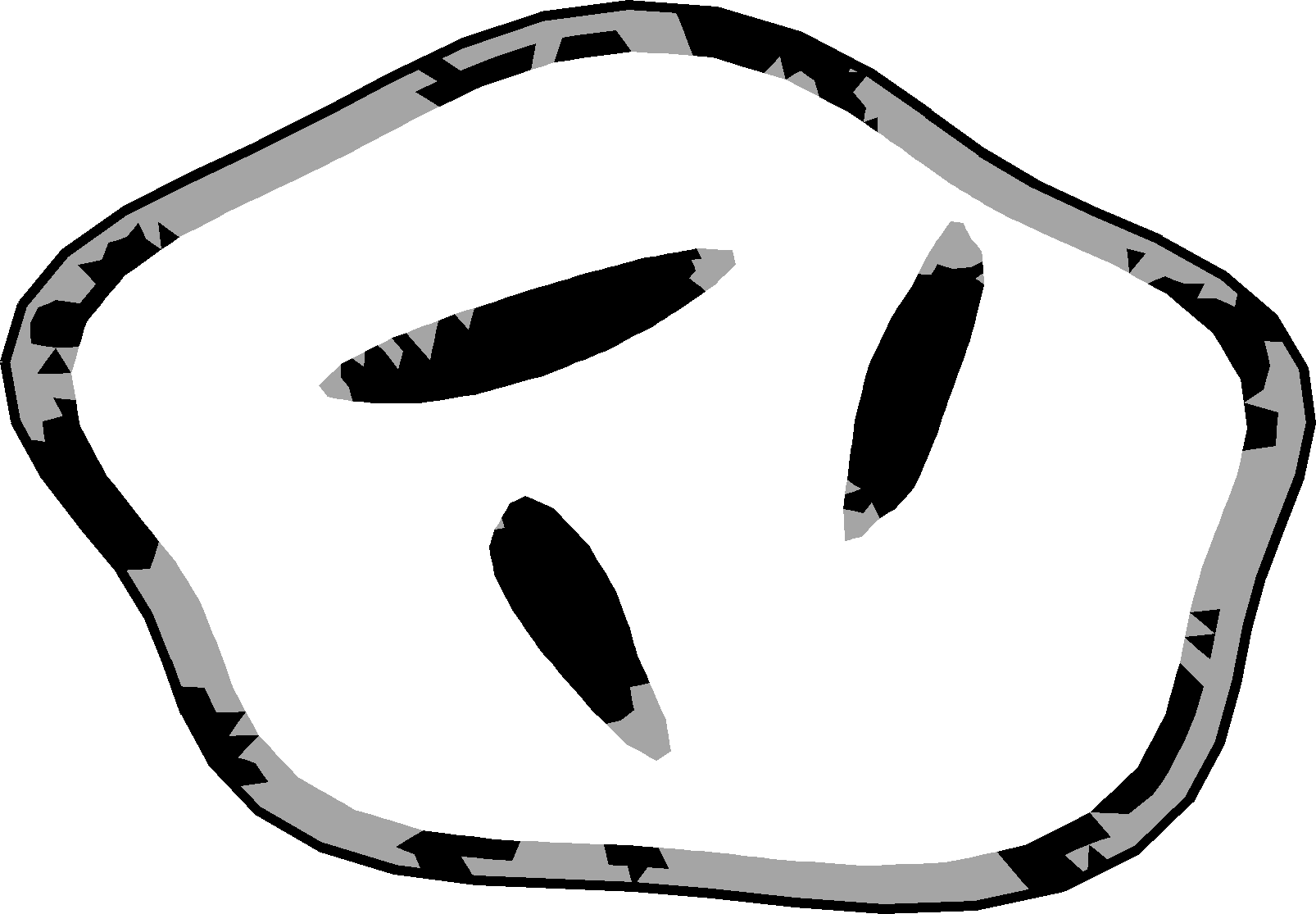} \\ (B) \\ $\hbox{ROA} = 49 \%$
\end{framed}
\end{center} 
\end{minipage}
\end{framed}
\end{center}
\end{minipage} \hskip0.2cm 
\begin{minipage}{7.6cm}
\begin{center}
\begin{framed}
{ Dense (receiver spacing ${\pi}/{64}$)} \\ \vskip0.2cm
\begin{minipage}{3.2cm}
\begin{center}
\begin{framed}
{ Low mixing (transmitter spacing $\pi/32$)} \\ \vskip0.2cm
\includegraphics[width=2.4cm]{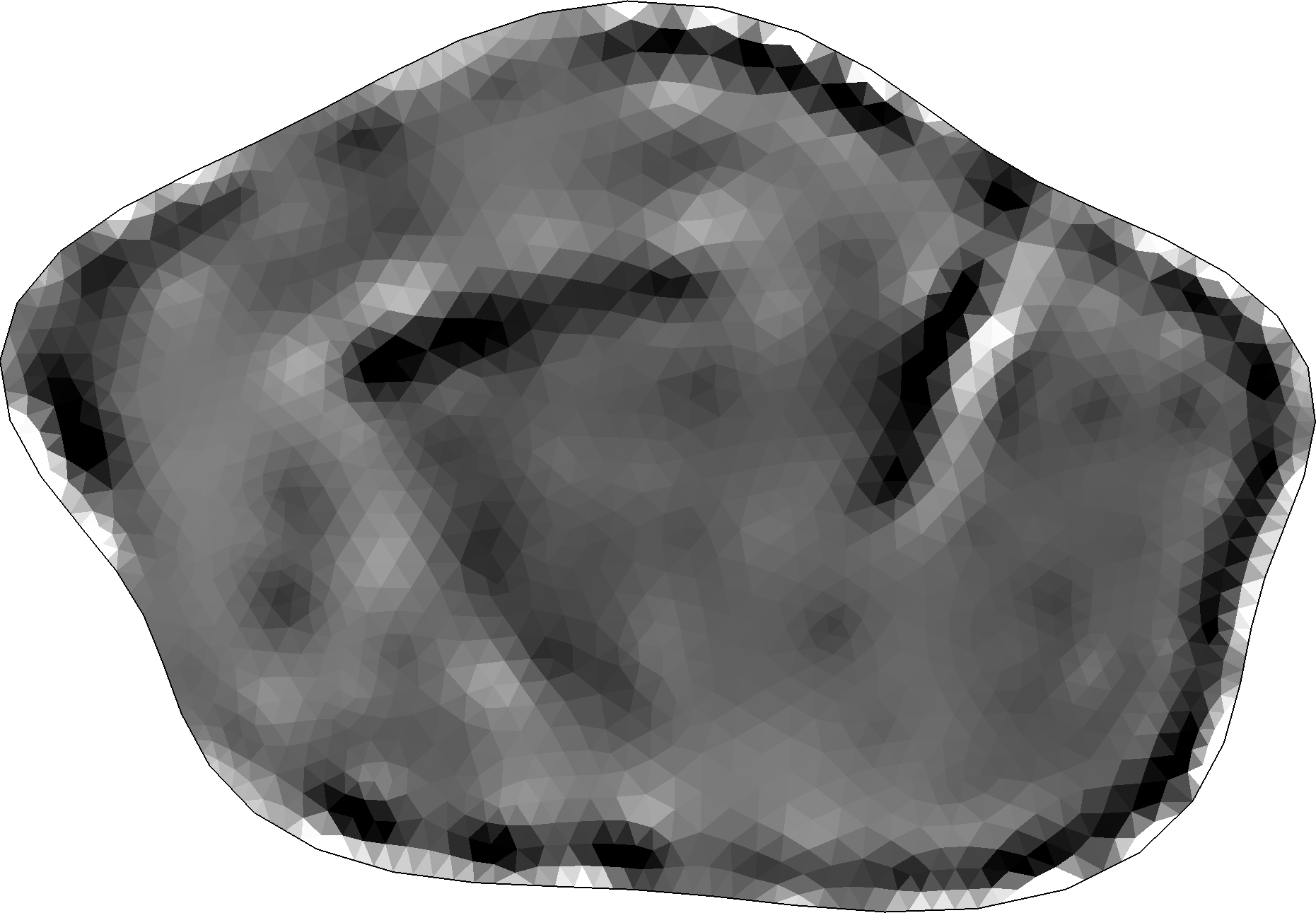} \\ \vskip0.2cm \includegraphics[width=2.4cm]{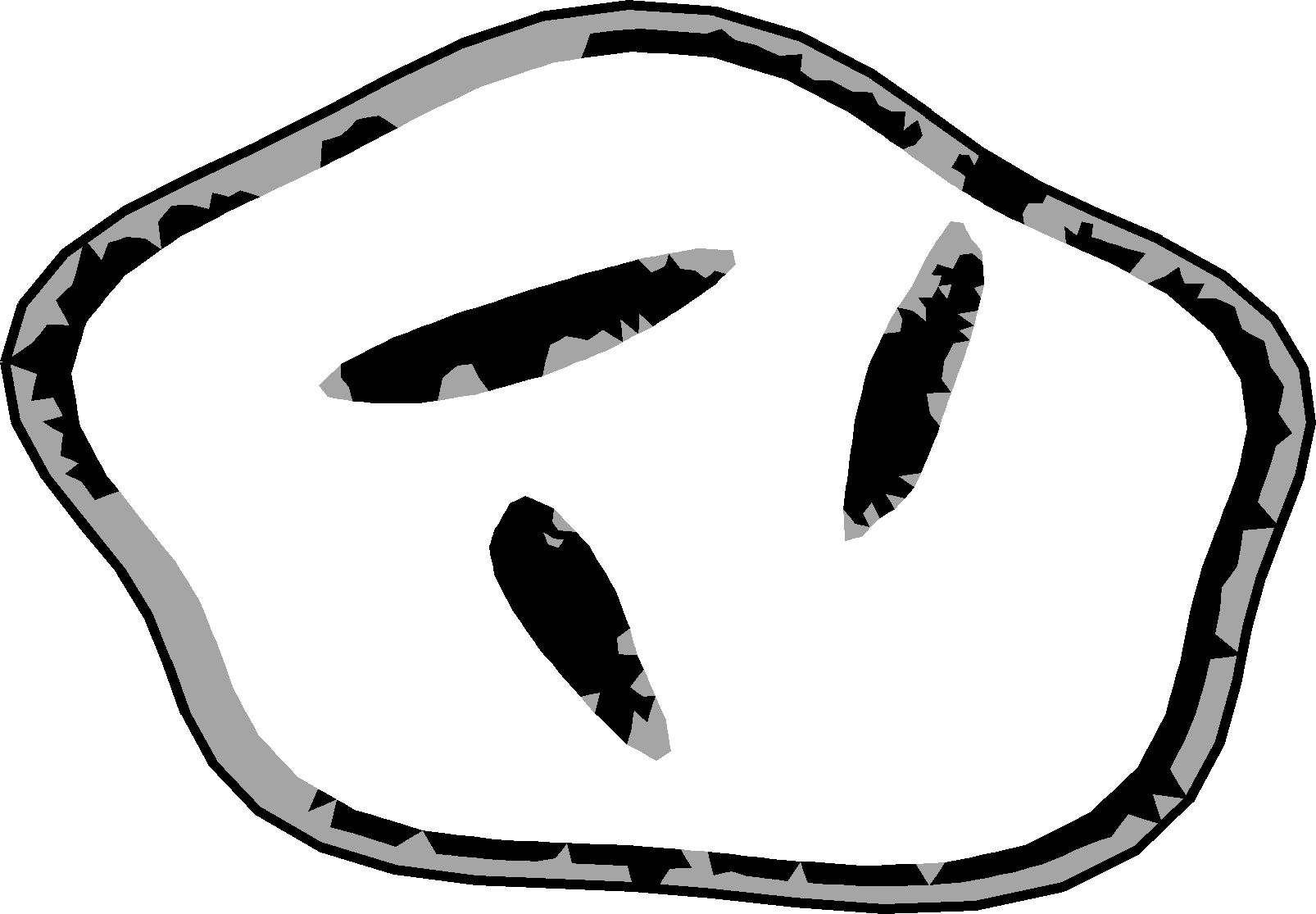} \\ (C) \\ $\hbox{ROA} = 52 \%$ 
\end{framed}
\end{center}
\end{minipage} \hskip0.2cm 
\begin{minipage}{3.2cm}
\begin{center}
\begin{framed}
{ High mixing (transmitter spacing $15\pi/16$)} \\ \vskip0.2cm
\includegraphics[width=2.4cm]{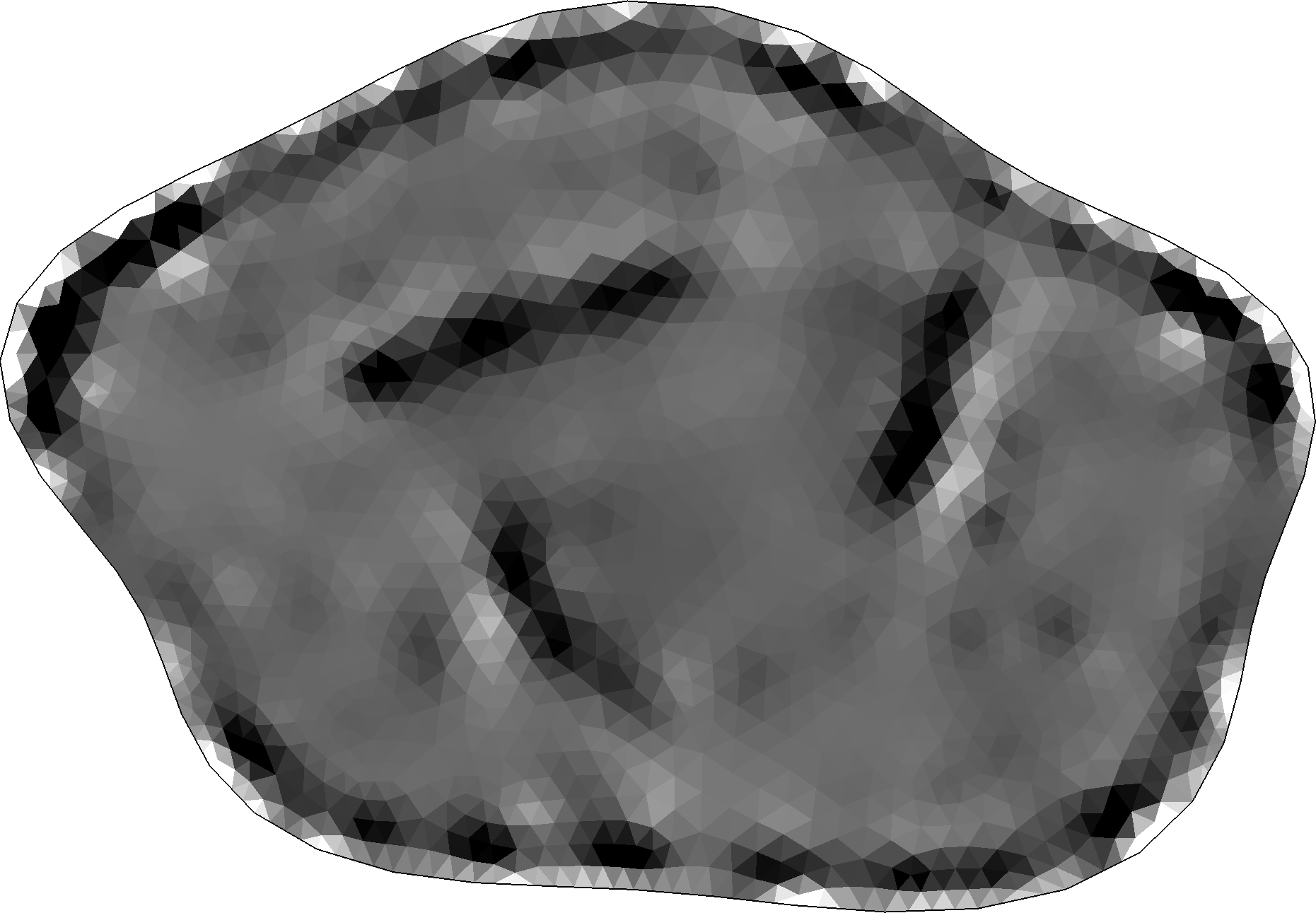} \\ \vskip0.2cm
\includegraphics[width=2.4cm]{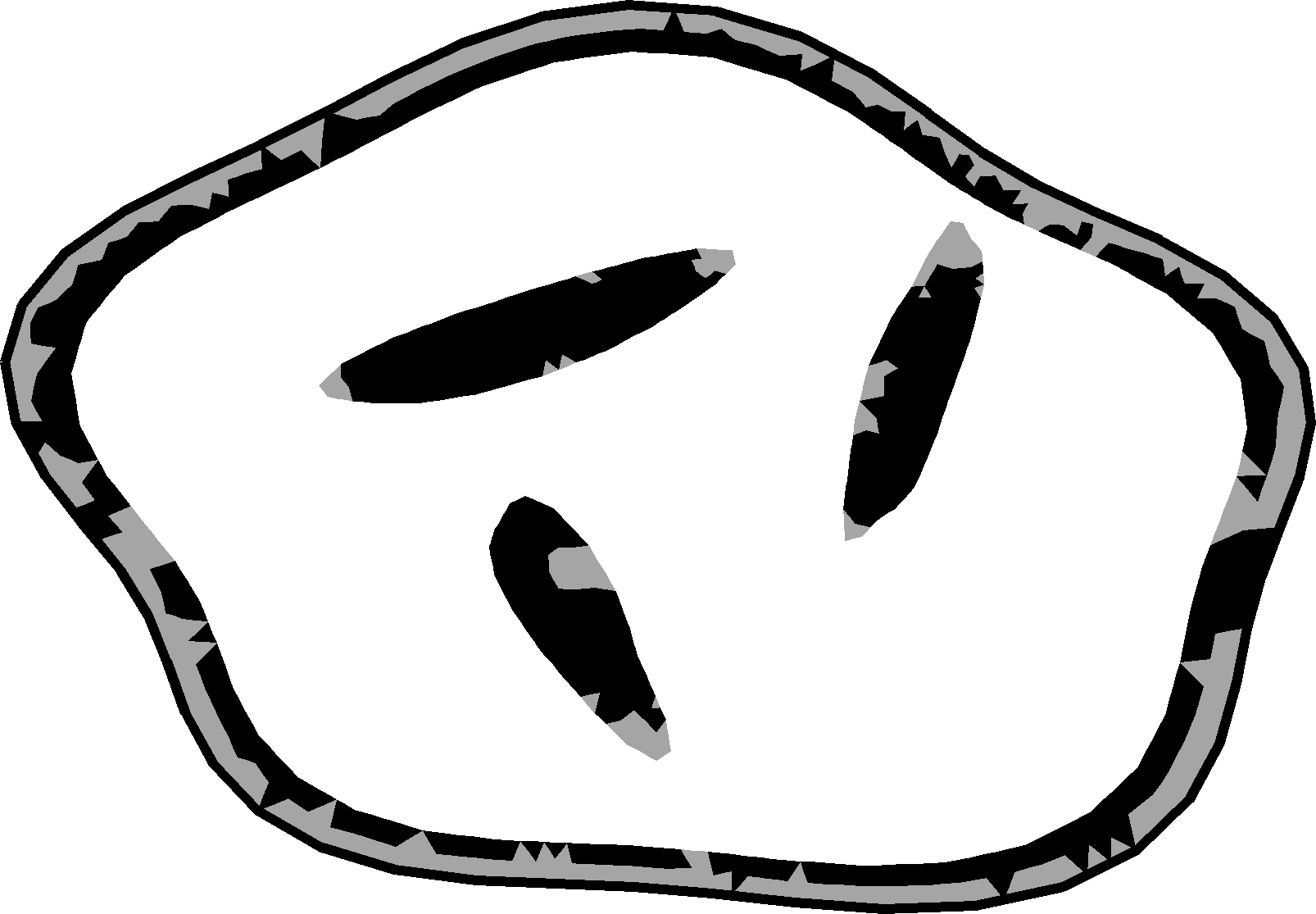} \\ (D) \\ $\hbox{ROA} = 61 \%$ 
\end{framed}
\end{center}
\end{minipage} 
\end{framed}
\end{center} 
\end{minipage}
\end{framed} 
\begin{framed}
{ Three transmitters} \\ \vskip0.2cm
\begin{minipage}{7.6cm}
\begin{center}
\begin{framed}
{ Sparse (receiver spacing ${\pi}/{16}$)}  \\ \vskip0.2cm
\begin{minipage}{3.2cm}
\begin{center}
\begin{framed}
{ Low mixing  (transmitter spacing $\pi/8$)} \\ \vskip0.2cm
\includegraphics[width=2.4cm]{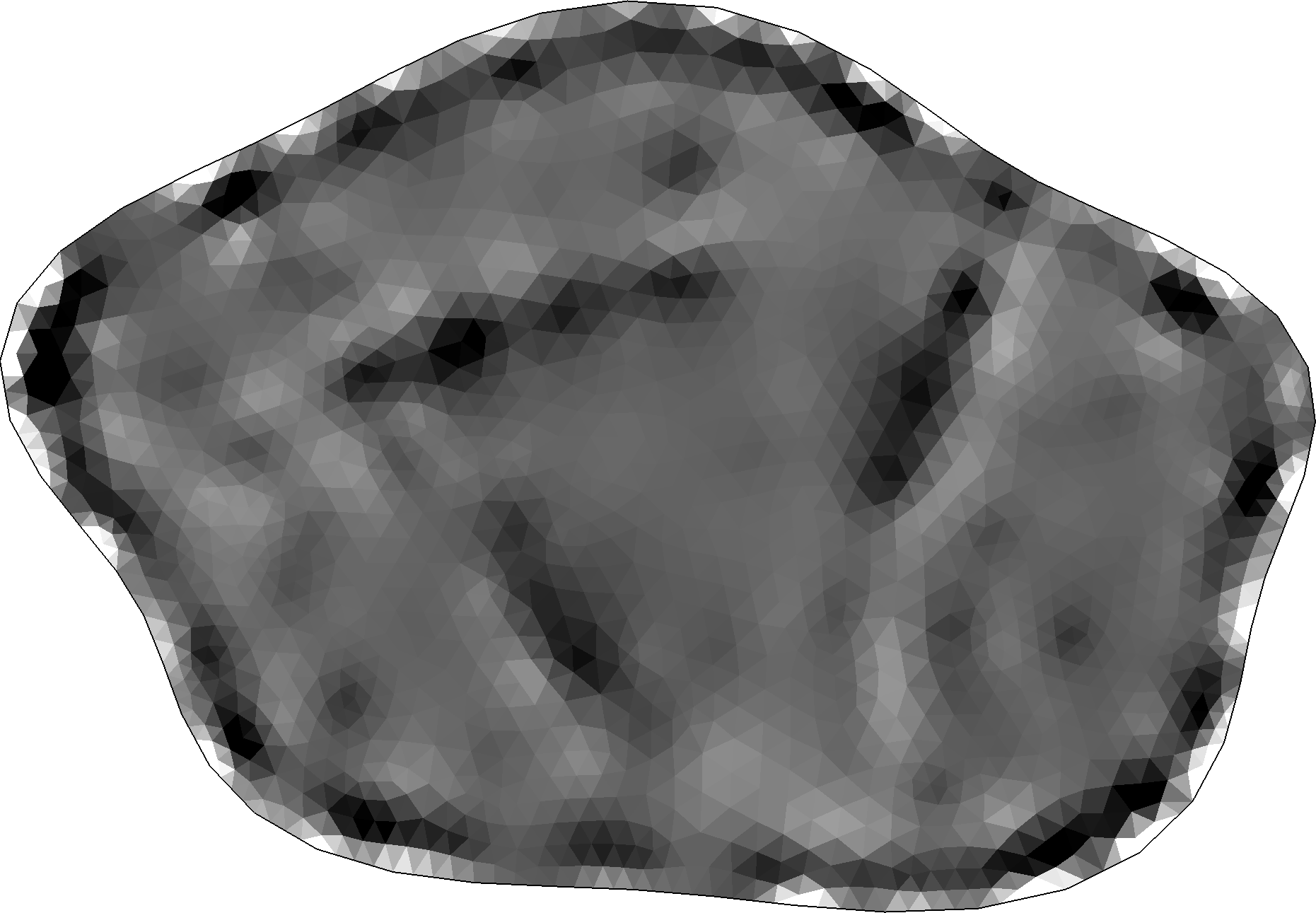} \\ \vskip0.2cm
\includegraphics[width=2.4cm]{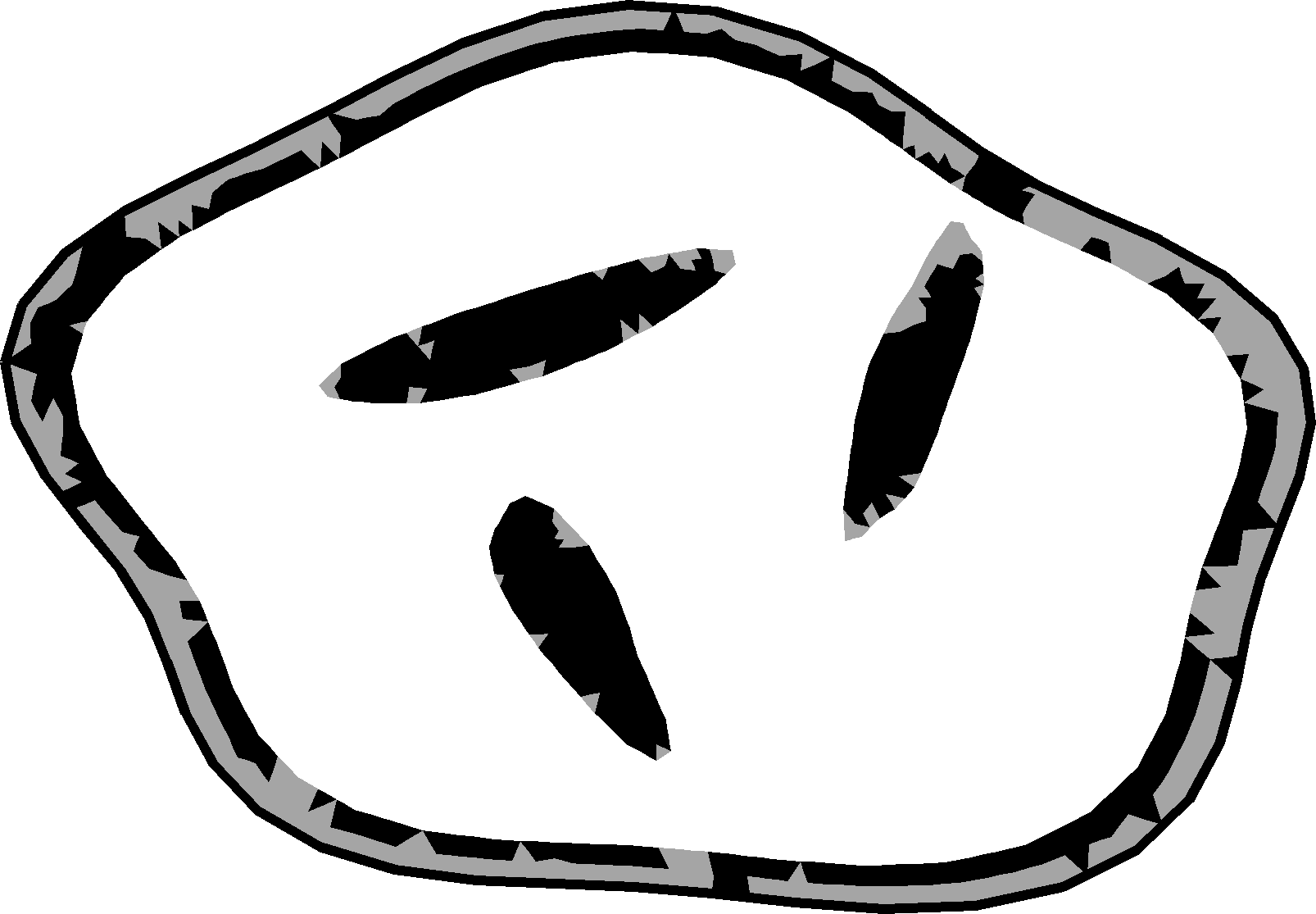} \\ (E)  \\ $\hbox{ROA} = 58 \%$
\end{framed}
\end{center}
\end{minipage} \hskip0.2cm
\begin{minipage}{3.2cm}
\begin{center} 
\begin{framed}
{ High mixing  (transmitter spacing $15\pi/4$)} \\ \vskip0.2cm
\includegraphics[width=2.4cm]{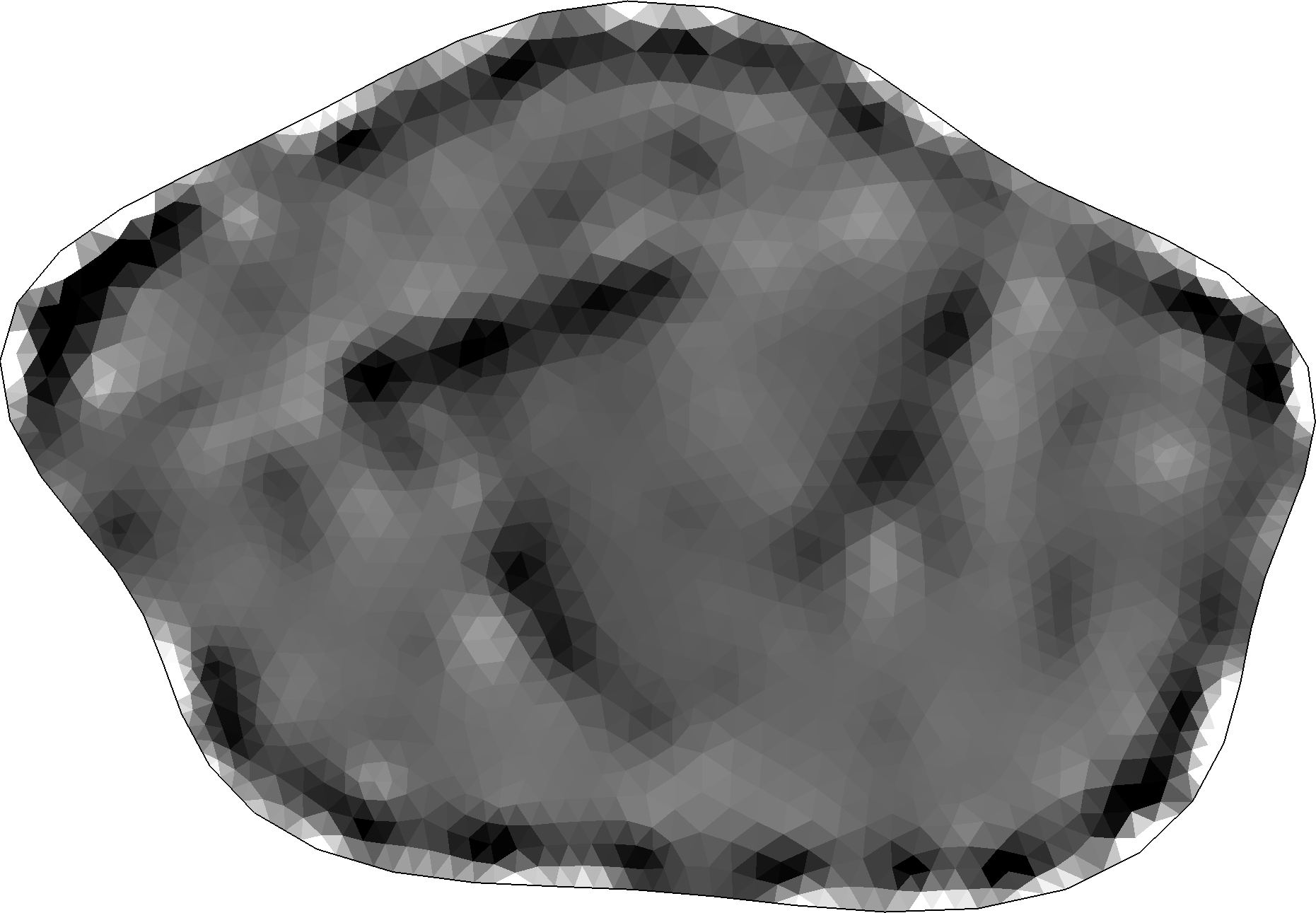} \\ \vskip0.2cm
\includegraphics[width=2.4cm]{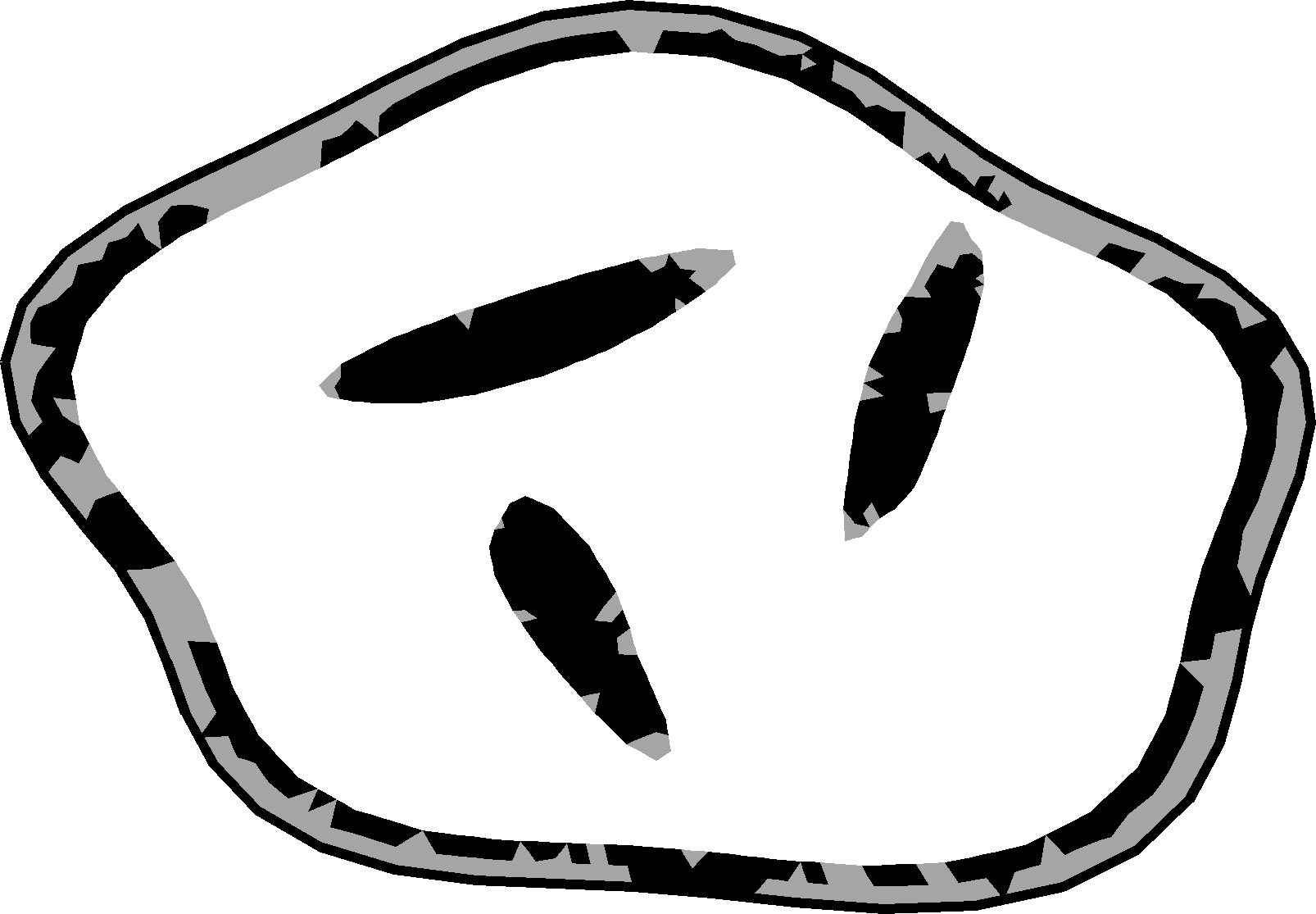} \\ (F) \\ $\hbox{ROA} = 58 \%$
\end{framed}
\end{center} 
\end{minipage}
\end{framed}
\end{center}
\end{minipage} \hskip0.2cm
\begin{minipage}{7.6cm}
\begin{center}
\begin{framed}
{ Dense (receiver spacing ${\pi}/{64}$)} \\ \vskip0.3cm
\begin{minipage}{3.2cm}
\begin{center}
\begin{framed}
{ Low mixing (transmitter spacing $\pi/32$)} \\ \vskip0.2cm
\includegraphics[width=2.4cm]{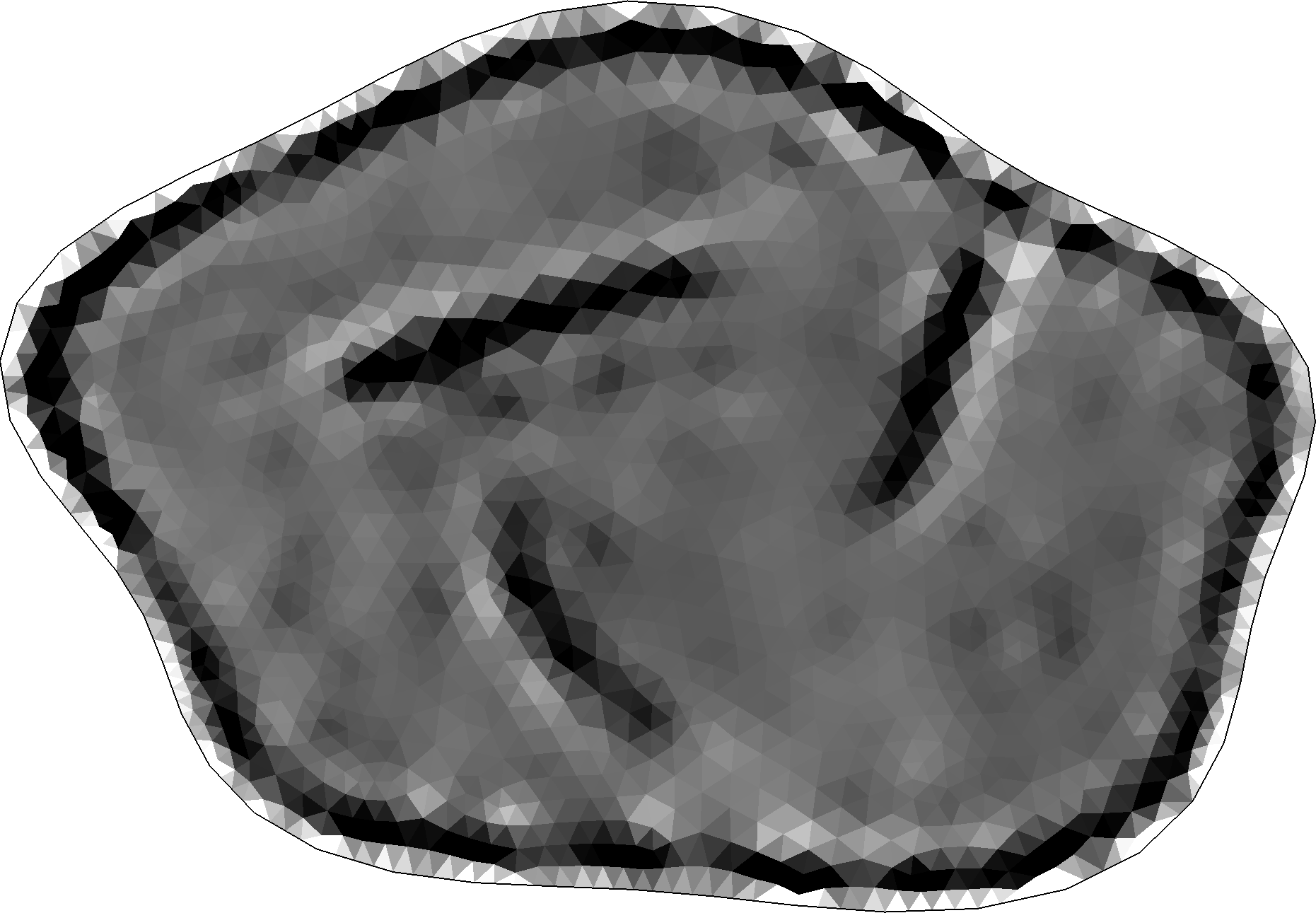} \\ \vskip0.2cm \includegraphics[width=2.4cm]{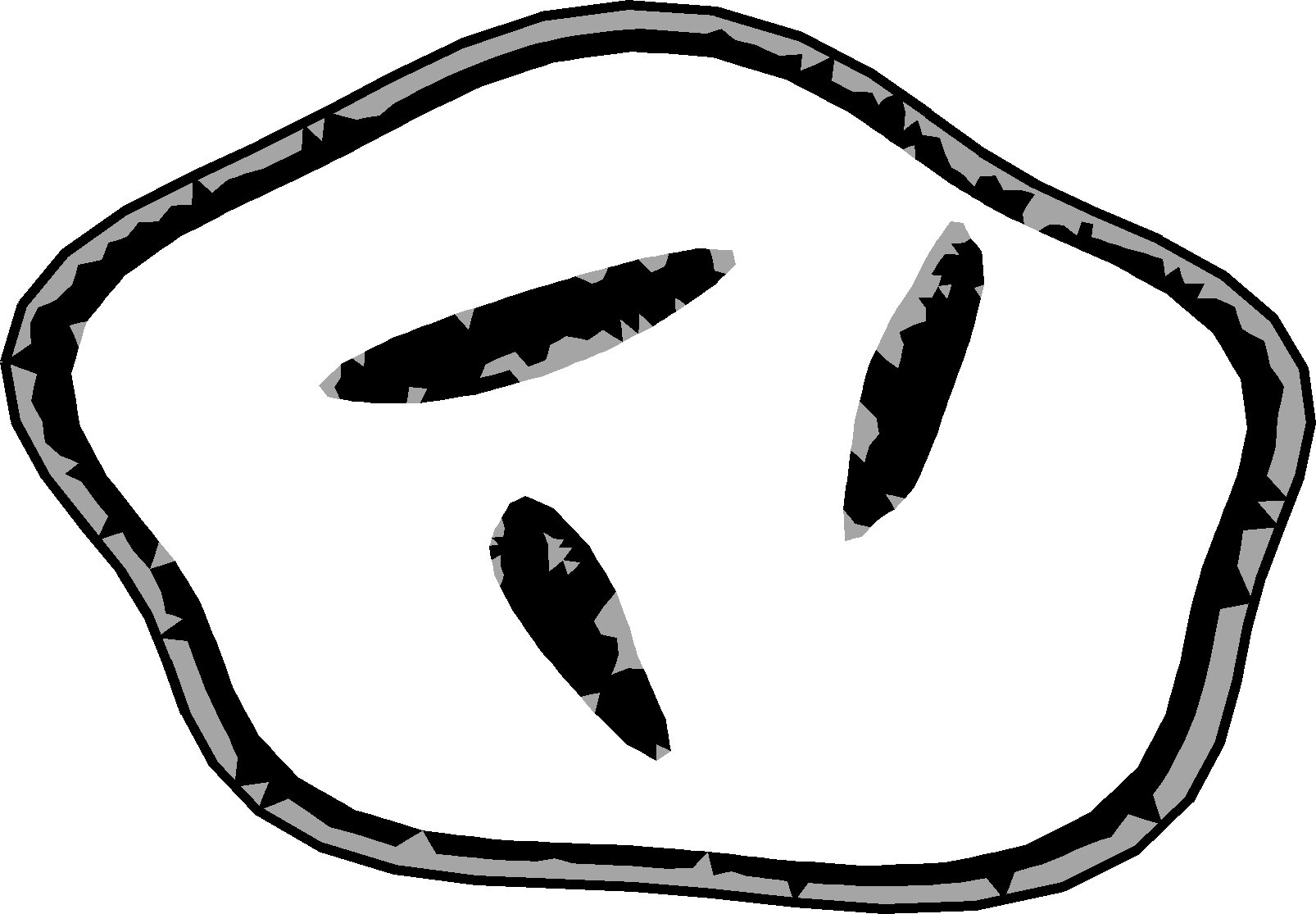} \\ (G) \\ $\hbox{ROA} = 61 \%$ 
\end{framed}
\end{center}
\end{minipage} \hskip0.2cm
\begin{minipage}{3.2cm}
\begin{center}
\begin{framed}
{ High mixing (transmitter spacing $15\pi/16$)} \\ \vskip0.2cm
\includegraphics[width=2.4cm]{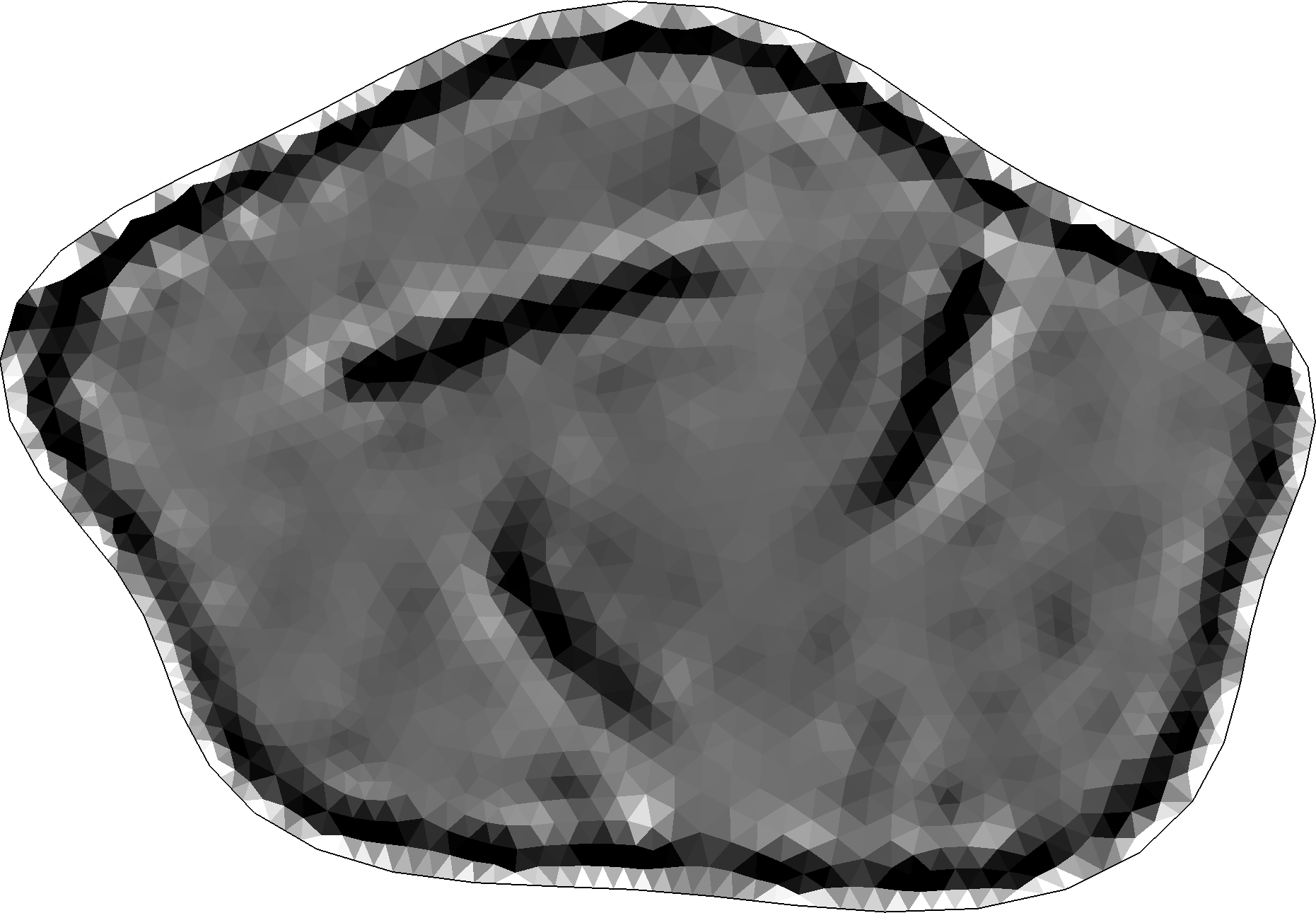} \\ \vskip0.2cm
\includegraphics[width=2.4cm]{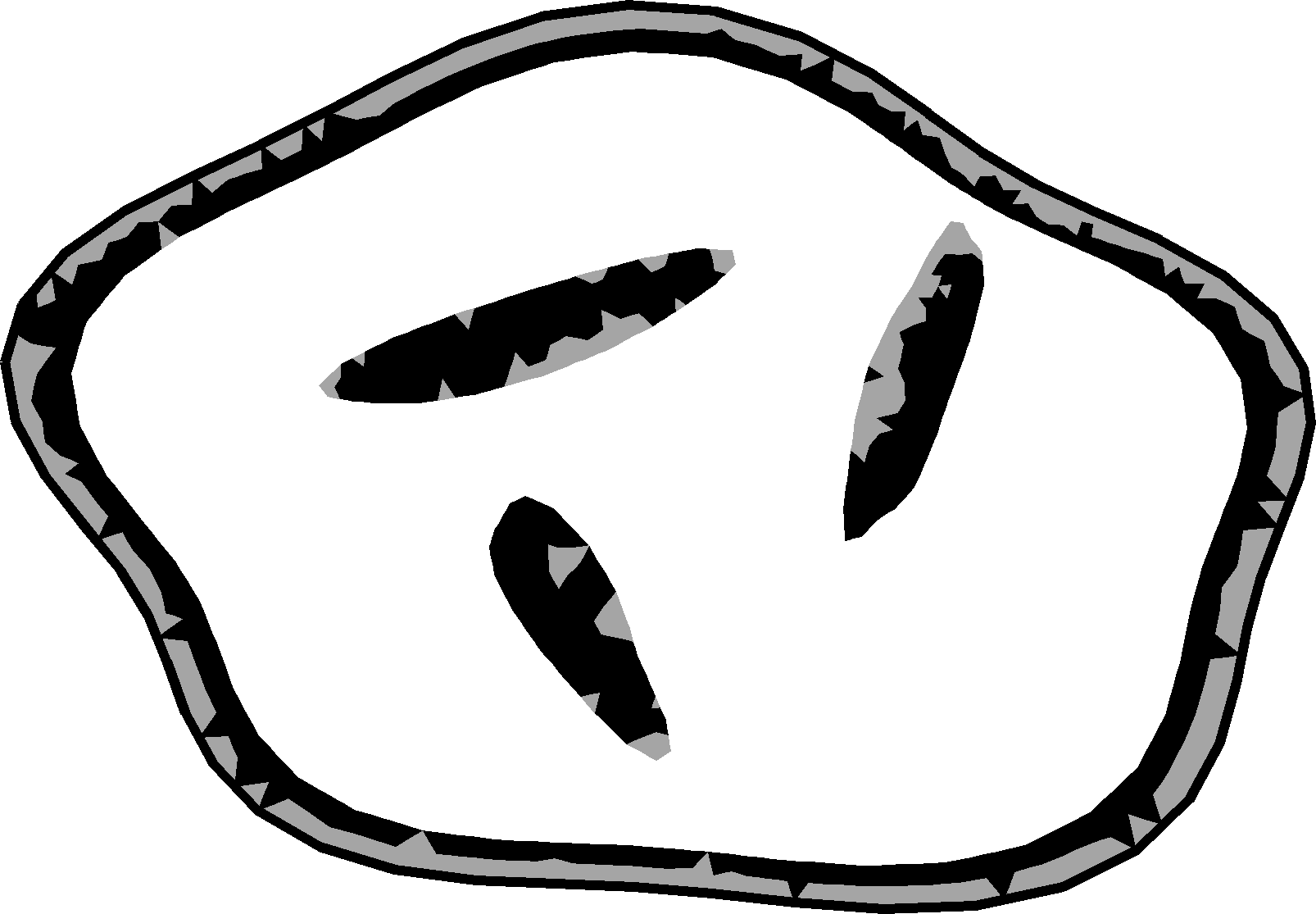} \\ (H) \\ $\hbox{ROA} = 62 \%$ 
\end{framed}
\end{center}
\end{minipage} 
\end{framed}
\end{center}
\end{minipage}   \end{framed} 
\end{center}
\end{scriptsize}
\caption{Reconstructions of permittivity distribution (II) for signal configurations (A)--(H). In each case, the top image visualizes the actual reconstruction and the bottom one the sets $\mathcal{A}$ and $\mathcal{S}$. \label{results_2}}
\end{figure*}

\begin{figure}
\begin{center}
\begin{minipage}{8cm}
\begin{center}
\includegraphics[width=7cm]{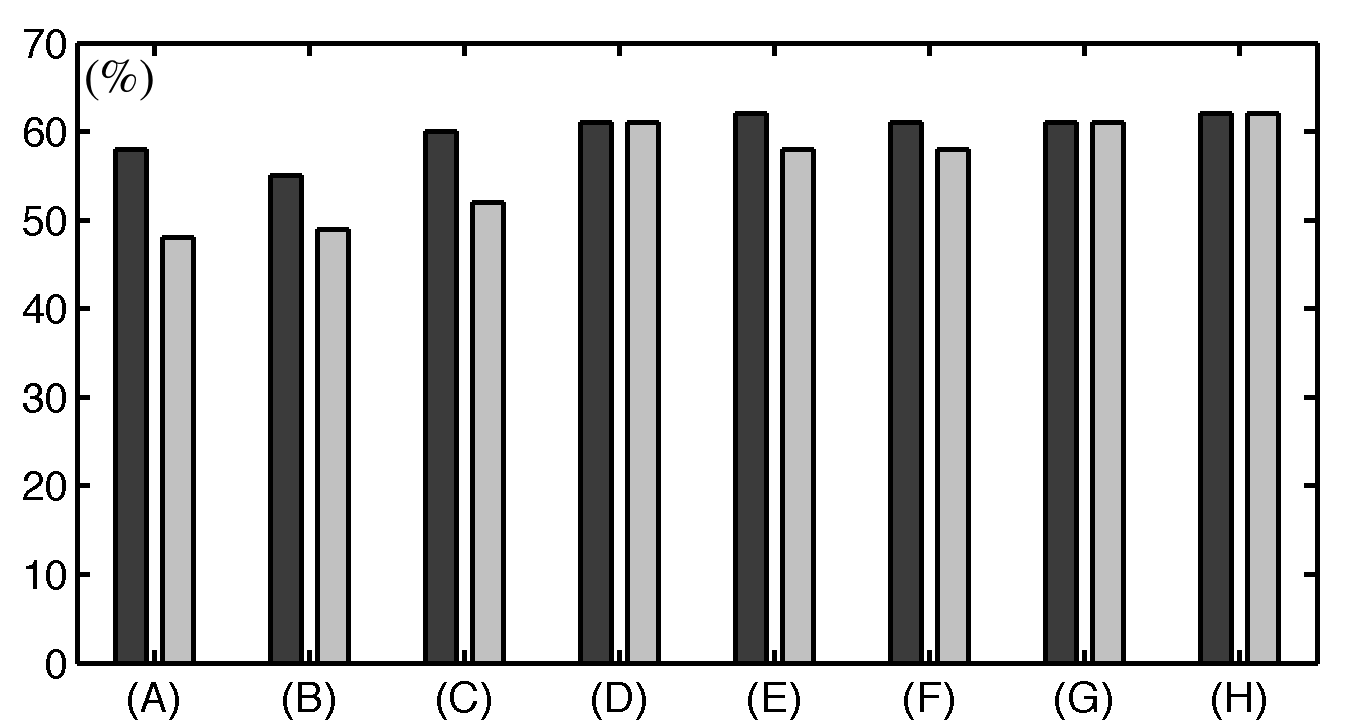} 
\end{center}
\end{minipage}  
\end{center}
\caption{ROA for signal configurations (A)--(H) and for permittivity distributions (I) (dark) and (II) (light grey). \label{results_3}}
\end{figure}

The results of the numerical experiments have been included in Figures  \ref{results_1}--\ref{results_3}.  Figures \ref{results_1} and \ref{results_2} visualize the reconstructions together with the sets $\mathcal{A}$ and  $\mathcal{S}$ for permittivity distributions (I) and (II), respectively. Figure \ref{results_3} includes a bar plot of the ROA results. 

The ROA was found to be systematically higher for (I) than (II).  For each individual signal configuration, the greater or equal ROA value was obtained with (I). The  range of variation was greater for distribution (II):  the values observed were from  55 to 62 \% and from 48 to 62 \% for (I) and (II), respectively. 
The dense signal configurations (C), (D), (G), (H) with 2.1 spatial sampling rate (SR) yielded a better reconstruction quality than the sparse patterns (A), (B), (E), (F) (0.53 spatial SR).  The ROA yielded by these groups was  60--62 \% and 55--61 \% for distribution (I) and 52--62 \% and 48--58 \% for (II), respectively. Furthermore, the difference between dense and sparse pattern under constant directional mixing was more pronounced in the case of one transmitter (A), (B), (C), (D) than in that of three  (E), (F), (G), (H). For the former group, up to 24 \%  relative difference in ROA was observed, whereas for the latter one it stayed below 7 \%.


The angular spacing of the transmitters had a notable effect on reconstruction quality in the case of single transmitter, where, e.g., the configurations (C) and (D) yielded the ROA of 52 and 61 \%, respectively, for distribution (II).  In the case of multiple transmitters, the effect of angular spacing was observed to be negligible. The number of transmitters and density of the data gathering points was found to be vital especially in the recovery of the surface layer of (II): the layer structure was virtually missing for sparse single-transmitter  configurations.

\section{Discussion}

This study concentrated on computational methods and techniques for radio tomography of SPOs \cite{mann2008, michel2015, kaasalainen2013}. We introduced and tested a full waveform tomography approach which allows   simulating  orbiter-to-orbiter measurements,  producing an inverse estimate for an  arbitrary permittivity distribution and speeding up of computations for 3D  imaging. The numerical experiments of this study focused, in particular,  on ({\bf 1}) the sparsity/density of the measurement  positions, ({\bf 2}) the number of transmitters in the satellite formation, ({\bf 3}) the directional variability of the signal paths, and ({\bf 4}) the recovery of (I) voids and (II) voids together with a surface layer, e.g.\  a  dust or ice cover.

The reconstruction quality was observed to increase along with the number of both transmitter and receiver  positions. For single transmitter signal configurations, spatial oversampling  (SR 2.1) relative to NR  yielded a significantly better quality compared to  undersampling (SR 0.53).  Reconstructing a surface layer  was observed  to be particularly difficult in the case of sparse data, when a single transmitter was used. Appropriate results were nevertheless obtained with three transmitters which  suggests that a multiple transmitter setup is advantageous for a sparse distribution of measurement points, supporting our previous findings for multiple sources (landers) and sparse data  \cite{pursiainen2014b, pursiainen2015}. In a real application context,  several different {\em in situ} constraints can lead to undersampling including the finite  time and energy resources, the restricted signal bandwith,  as well as the limited  control of a satellite formation \cite{miller2000, yeh2000,pongvthithum2005}.  One obvious challenge is the slow satellite movement due to the weakness of the gravity field.  Consequently, it is  very likely that the inversion methodology will need to rely on sparse undersampled data.  

In order to model the effect of orbiter movement on the reconstruction quality,  we tested two different values for the  angular spacing of the transmitter positions. Based on the results, this is a significant effect that  needs to be taken into  consideration in the mission design phase. Furthermore, it seems that the use of multiple transmitters can be advantageous in order to minimize this effect. It is also obvious that  generating synthetic \cite{benna2004} data based on realistic orbiter movement in three dimensions is an important topic that deserves future research. 

The multiresolution and incomplete Cholesky speedups \cite{braess2007, golub1989} were found to be well-suited for the current tomography context. For inversion in three dimensions, both of these techniques can be necessary extensions of our basic approach in order to achieve a reasonable computation time and compact memory usage. The incomplete Cholesky solver applied for the mass matrix ${\bf C}$ was found to yield fast and reliable results, when fixing the level of filling (number of nonzeros) to 20 \%  compared to that of the full Cholesky decomposition. Choosing a lower level was observed to deteriorate the reconstruction quality. The multiresolution idea, i.e., the redefinition of the differentiated potential with respect to a coarse FE mesh, was successfully implemented going up one degree in the hierarchy of the nested meshes. Based on our initial tests, a multi-level hierarchy can be a feasible solution in order to further compress the time and memory consumption.  The multiresolution speedup can be an essential tool to tackle the large and rapidly growing number of  elements, which is of the order $\propto \! \mathantt{h}^2$ and $\propto \! \mathantt{h}^3$ for two- and three-dimensional elements of diameter $\mathantt{h}$, respectively. Namely, the speedup gain is exponential within the hierarchy of nested meshes, being  comparable to the relative change in the number of elements $\propto \! 4^\mathantt{k}$ (2D) or $\propto \! 8^\mathantt{k}$ (3D) for a  coarsening of degree $\mathantt{k}$.

The inversion strategy of this paper is based on total variation regularization \cite{scherzer2008, stefan2008}. This technique was chosen as it, in principle, allows reconstructing an arbitrary permittivity distribution defined on the FE mesh.  The values of the regularization parameters $\alpha$ and $\beta$ were picked from the middle of the logarithmic interval of workable values which, based on our preliminary tests, led to approximately maximal overlap between the actual and reconstructed permittivities. 

Further analysis of the robustness of the inversion could be conducted in the future. Additionally,  implementing the present setting, e.g, within a Bayesian context  \cite{kaipio2004} might also be a potential future direction, to allow more advanced analysis of the noise and enhanced formalism for the {\em a priori} information.  Other future directions might concern realistic implementation of the present forward and inversion methodology.  To support this kind of future research, the development of advanced asteroid models with
various structures, such as cracks and porosity, would be essential. Features such as orbiter movement and  the  roughness of the body surface would  be vital from the application point of view.  Comparison between the current orbiter-to-orbiter measurement approach and, e.g., backscattering data would be important. 

\section{Appendix}

\subsection{Linearization}
\label{a_linearization}  
Differentiating  the leap-frog formulas (\ref{leap-frog1})--(\ref{leap-frog2}) with respect to $c_j$ at $\varepsilon_r = \varepsilon^{(bg)}_r$ yields:
{\setlength\arraycolsep{1 pt} \begin{eqnarray}
\label{lin1}
\frac{\partial {\bf q}^{(k)}_{\ell + \frac{1}{2}}}{\partial c_j } & = & \frac{\partial {\bf q}^{(k)}_{\ell - \frac{1}{2}}}{\partial c_j}  \! + \!  \Delta t {\bf  A}^{-1} \Big(  {{\bf B}^{(k)}} \frac{\partial {\bf  p}_{\ell}}{\partial c_j} \! - \! {\bf T}^{(k)} \frac{\partial {\bf q}^{(k)}_{\ell - \frac{1}{2}}}{\partial c_j} \Big), \\ 
 \frac{\partial {\bf p}_{\ell + 1} }{\partial c_j} & = & \frac{\partial {\bf p}_{\ell}}{\partial c_j}   \! + \!  \Delta t {\bf  C}^{-1} \Big(  - {\bf R} \frac{\partial {\bf p}_{\ell}}{\partial c_j}  - {\bf S} \frac{\partial {\bf p}_{\ell}}{\partial c_j}  \! - \!   \sum_{k = 1}^\mathantt{d}  {{\bf B}^{(k)}}^T \frac{\partial {\bf q}^{(k)}_{\ell \! + \! \frac{1}{2}}}{\partial c_j} \Big) \nonumber\\ & &  + \Delta t \, \frac{\partial {\bf  C}^{-1}}{\partial c_j} \Big( - \! {\bf R} {\bf p}_\ell \! - \!  {\bf S} {\bf p}_\ell \! - \! \sum_{k = 1}^\mathantt{d} {{\bf B}^{(k)}}^T {\bf q}^{(k)}_{\ell + \frac{1}{2}}  \Big). 
\label{lin2}
\end{eqnarray}}
The last row of the second equation follows from the classical product (derivative) rule, since ${\bf C}$   depends on $\varepsilon_r$. The formula for ${\partial {\bf C}^{-1}} / { \partial c_j}$ can be obtained via straightforward differentiation of ${\bf C} {\bf C}^{-1} = {\bf I}$. The product rule yields ${\bf C} {\partial {\bf C}^{-1}} / { \partial c_j}  + ({\partial {\bf C}}/{ \partial c_j}) {\bf C}^{-1} = {\bf 0}$, that is
\begin{equation} {\partial {\bf C}^{-1}} / { \partial c_j} = - {\bf C}^{-1} ({\partial {\bf C}}/{ \partial c_j}) {\bf C}^{-1}. 
\label{derivative}
\end{equation}  
As the permittivity perturbation is of the form $\varepsilon^{(p)}_r =  \sum_{j = 1}^M  {c}_j  \chi_j'$,  the entries of ${\bf C}$ can be expressed as
\begin{equation}  C_{i_1,i_2} = \int_{\Omega}  \varepsilon^{(bg)}_r \varphi_{i_1} \varphi_{i_2} \, \hbox{d} \Omega +  \sum_{j = 1}^M  {c}_j  \int_{\mathtt{T}'_j} \varphi_{i_1} \varphi_{i_2} \, \hbox{d} \Omega. \end{equation}
Hence, it follows that 
\begin{equation}  \Big( \frac{\partial {\bf C}} {\partial c_j} \Big)_{i_1,i_2} = \int_{\mathtt{T}'_j} \varphi_{i_1} \varphi_{i_2} \, \hbox{d} \Omega, \end{equation} if the $j$-th element includes nodes $i_1$ and $i_2$, otherwise $[\partial {\bf C} / \partial c_j]_{i_1,i_2}  = 0$. Substituting (\ref{derivative}) into  (\ref{lin2}) the last row of (\ref{lin2})  is of the form $\Delta t {\bf C}^{-1} \frac{\partial {\bf C}}{\partial c_j}  {\bf b}_\ell$, where 
\begin{equation} \quad {\bf b}_\ell =   {\bf C}^{-1} \Big( \! {\bf R} {\bf p}_\ell \! + \!  {\bf S} {\bf p}_\ell \! + \! \sum_{k = 1}^\mathantt{d} {{\bf B}^{(k)}}^T {\bf q}^{(k)}_{\ell + \frac{1}{2}}  \Big).
\end{equation}
Consequently, the system (\ref{lin1})--(\ref{lin2}) can be written as
{\setlength\arraycolsep{2 pt} \begin{eqnarray}
\label{aku}
{\frac{\partial {\bf q}^{(k)}_{\ell + \frac{1}{2}}}{\partial c_j}} & = & 
\frac{\partial {\bf q}^{(k)}_{\ell - \frac{1}{2}} }{\partial c_j}+  \Delta t  {\bf  A}^{-1} \Big(  {\bf B}^{(k)}  \frac{\partial {\bf p}_\ell}{\partial c_j}  - {\bf T}^{(k)} \frac{\partial {\bf q}^{(k)} _{\ell - \frac{1}{2}}}{\partial c_j}\Big),   \\ 
\frac{\partial {\bf p}_{\ell + 1}}{\partial c_j} & = & \frac{\partial {\bf p}_{\ell}}{\partial c_j} +  \Delta t {\bf  C}^{-1} \Big(  \frac{\partial {\bf C}}{\partial c_j} {\bf b}_\ell- {\bf R} \frac{\partial {\bf p}_\ell}{\partial c_j} - {\bf S} \frac{\partial {\bf p}_\ell}{\partial c_j}  \nonumber\\ & & \qquad \qquad \qquad \qquad  \qquad - \sum_{k = 1}^\mathantt{d} {{\bf B}^{(k)}}^T \frac{\partial {\bf q}^{(k)}_{\ell + \frac{1}{2}}}{\partial c_j}  \Big).
\label{iines} 
\end{eqnarray}}
This is similar to (\ref{leap-frog1})--(\ref{leap-frog2})  except from the source term $({\partial {\bf C}}/{\partial c_j}) {\bf b}_\ell$ that is specific, instead of a FE node, to the $j$-th element in the FE mesh $\mathcal{T}'$.  

In this paper, the solution of (\ref{aku})--(\ref{iines}) is found via the system  (\ref{linearized1})--(\ref{linearized2}) that has a point-specific source   ${\bf h}^{(i,j)} = ({\partial {\bf C}}/{\partial c_j}) {\bf b}_\ell^{(i)}$  with  $(b^{(i)}_\ell)_j=  (b_\ell)_i$, if $j = i$ and $(b^{(i)}_\ell)_j = 0$. This  is essential for our  deconvolution approach (Section \ref{deconvolution}) which relies on the reciprocity of the signal. Namely, the source term of (\ref{linearized1})--(\ref{linearized2}) is monopolar similar to that of (\ref{leap-frog1})--(\ref{leap-frog2}) and thus the reciprocity argument can be used. 

 Since ${\bf b}_\ell = \sum_{i = 1}^n {\bf b}_\ell^{(i)} $, the source of (\ref{iines}) can be expressed as the sum of point-specific sources $({\partial {\bf C}}/{\partial c_j}) {\bf b}_\ell = \sum_{i = 1}^n ({\partial {\bf C}}/{\partial c_j}) {\bf b}_\ell^{(i)} =  \sum_{i = 1}^n {\bf h}^{(i,j)}$. It follows that since the wave  equation is linear with respect to the source term, the solution of (\ref{aku})--(\ref{iines}) can be obtained via the sum \begin{equation} \label{monimutkainen} {\partial {\bf p}}/{\partial c_j}  = \sum_{i = 1}^n {\bf d}^{(i,j)}, \end{equation} where $ {\bf d}^{(i,j)}$ is a solution of (\ref{linearized1})--(\ref{linearized2}). Due to the sparse structure of $({\partial {\bf C}}/{\partial c_j})$  the vector ${\bf h}^{(i,j)}$ differs from zero only if the $i$-th node belongs to the element $\mathtt{T}'_j \in \mathcal{T}'$. For example, if the elements are triangular and the multiresolution speedup  is used, i.e., if ${\partial {\bf C}}/{\partial c_j}$ is defined w.r.t.\ $\mathcal{T}'$, the non-zero source terms are ${\bf h}^{(i_{1},j)}$ ${\bf h}^{(i_{2},j)}$, ${\bf h}^{(i_{3},j)}$, where $i_1, i_2, i_3$ denote the three nodes associated with $\mathtt{T}'_j$, and consequently, one can write 
\begin{equation}
\label{multires}
\frac{ {\partial {\bf p}}}{{\partial c_j}}  = {\bf d}^{(i_{1},j)} +  {\bf d}^{(i_{2},j)}  +  {\bf d}^{(i_{3},j)}.  
\end{equation}

The whole procedure of linearization, as implemented in this paper, can be summarized as follows 
\begin{enumerate}
\item {\em Forward simulation}: Transmit a wave from each transmitter and receiver position. Store the time-evolution at each receiver position and at each node of $\mathcal{T}'$. To compute the wave, use (\ref{leap-frog1})--(\ref{leap-frog2}) together with the incomplete Cholesky speedup (Section \ref{multiresolution}). 
\item {\em Linearization}: For each element $\mathtt{T}'_j$ of $\mathcal{T}'$, find the  non-zero source terms ${\bf h}^{(i,j)}$, $i = 1, 2, \ldots, n$ and compute the corresponding solution ${\bf d}^{(i,j)}$ of (\ref{linearized1})--(\ref{linearized2}) by applying the multiresolution speedup (\ref{multiresolution}) and deconvolution approach of Section  \ref{deconvolution}. Then, compute the sum $ {\partial {\bf p}}/{\partial c_j}  = \sum_{i = 1}^n {\bf d}^{(i,j)}$, and thus matrix ${\bf L}$  via  (\ref{multires}). 
\item {\em Inversion}: Invert the data as explained in Section \ref{inversion}.
\end{enumerate}

\subsection{Inversion}
\label{a_inversion}

In the algorithm (\ref{tv_iteration}), the matrix ${\bf D}$ is diagonalizable as a symmetric matrix. If $\beta>0$, then  ${\bf D}$ is positive definite and also invertible.   Consequently, one can define \begin{equation} \tilde{\bf x } = {{\bf D}} {\bf x}  \quad \hbox{and} \quad \tilde{\bf L}= {\bf L} {{\bf D}^{-1}}, \label{transform} \end{equation} which substituted into (\ref{tv_iteration})   yields the following iteration \cite{pursiainen2014, pursiainen2014b, kaipio2004} \begin{equation} \label{iter} \tilde{\bf x }_{\ell+1}  = (\tilde{\bf L}^T \tilde{\bf L}  + \alpha \tilde{\bf \Gamma}_{\ell} )^{-1} \tilde{\bf L}^T {\bf y}, \quad \! \! \tilde{\bf \Gamma}_{\ell} = \hbox{diag} ( |\tilde{\bf {x}}_{\ell}|)^{-1}, \quad \! \! \tilde{\bf \Gamma}_{0} = {\bf I} . \end{equation} This algorithm is closely related to alternating conditional  minimization of the function 
\begin{eqnarray}
G(\tilde{\bf x}, \tilde{\bf z}) & = &   \| \tilde{\bf L} \tilde{\bf x} - {\bf y} \|^2_2 + {\alpha}  \|  \hbox{diag} ( \tilde{\bf z})^{-1} \tilde{\bf x} \|^2_2  +  \sum_{j = 1}^M \tilde{z}_j \nonumber \\ 
& = &   \| \tilde{\bf L} \tilde{\bf x} - {\bf y} \|^2_2 + {\alpha} \sum_{j = 1}^M  \frac{\tilde{x}^2_j}{\tilde{z}_j} + {\alpha} \sum_{j = 1}^M \tilde{z}_j 
\end{eqnarray} 
in which $z_j>0$, for $i = 1, 2, \ldots, M$. Since $G(\tilde{\bf x}, \tilde{\bf z})$ is a quadratic function with respect to $\tilde{\bf x}$,  the conditional minimizer $\tilde{\bf x}^\ast  =\arg \min_{\tilde{\bf x}} G(\tilde{\bf x} \mid \tilde{\bf z})$ can be obtained through a least-squares approach of the form
\begin{equation}
\tilde{\bf x}^\ast =  (\tilde{\bf L}^T \tilde{\bf L}  + \alpha \tilde{\bf \Gamma}_{\tilde{\bf z}} )^{-1} \tilde{\bf L}^T {\bf y}, \quad \! \! \tilde{\bf \Gamma}_{\tilde{\bf z}} = \hbox{diag} ( \tilde{\bf z})^{-1} \quad \! \! \!   \hbox{and} \quad \! \!  \!   \tilde{\bf \Gamma}_0 = {\bf I}
\end{equation} 
At $\tilde{\bf z}^\ast  =\arg \min_{\tilde{\bf z}} G(\tilde{\bf z} \mid \tilde{\bf x})$, the gradient of $G( \tilde{\bf z} \mid \tilde{\bf x} )$ vanishes with respect to $\tilde{\bf z}$, i.e., 
\begin{equation}
\frac{\partial G( \tilde{\bf x}, \tilde{\bf z})}{\partial \tilde{z}_j} \Big|_{\tilde{\bf z}^\ast} =  - {\alpha} \frac{\tilde{x}^2_j}{(\tilde{z}^\ast_j)^2} + 1 = 0, \quad \hbox{i.e.} \quad \tilde{z}^\ast_j = |\tilde{x}_j|  \sqrt{\alpha}.
\end{equation} 
The global minimum can be sought via the following alternating iteration. 
\begin{enumerate}
\item Set $\tilde{\bf z}_{0} = (1, 1, \ldots, 1)$ and $\ell = 1$. For a desired number of iterations repeat the following two steps.
\item Find $\tilde{\bf x}_\ell = \arg \min_{\tilde{\bf x}} G(\tilde{\bf x}, \tilde{\bf z}_{\ell-1})$. 
\item Find  $\tilde{\bf z}_\ell = \arg \min_{\tilde{\bf z}} G(\tilde{\bf x}_\ell, \tilde{\bf z})$.
\end{enumerate}
In this algorithm, the sequence $\tilde{\bf x}_1, \tilde{\bf x}_2, \ldots$ is identical to that of (\ref{iter}) and ${\bf x}_\ell = {\bf D}^{-1} \tilde{\bf x}_\ell$ equals to the $\ell$-th iterate of (\ref{tv_iteration}). If for some $\ell < \infty$ the pair $({\bf \tilde{x}}_\ell, {\bf \tilde{z}}_\ell)$ is a global minimizer of  $G({\bf x}, {\bf z})$, then, since $(\tilde{z}_\ell)_j =  |(\tilde{x}_\ell)_j|$, $j = 1, 2, \ldots, M$,  it also minimizes the function \begin{eqnarray} \tilde{F}(\tilde{\bf x}) & = & G(\tilde{x}_1, \tilde{x}_2, \ldots , \tilde{x}_M, |\tilde{x}_j|, |\tilde{x}_j|, \ldots, |\tilde{x}_M|) \nonumber \\ & =&      \| \tilde{\bf L} \tilde{\bf x} - {\bf y} \|^2_2 + {\alpha} \sum_{j = 1}^M  \frac{\tilde{x}_j^2}{\tilde{z}_j} +\sum_{j = 1}^M \tilde{z}_j \nonumber \\ 
& = &  \| \tilde{\bf L} \tilde{\bf x} - {\bf y} \|^2_2 + {\alpha} \sum_{j = 1}^M  \frac{\tilde{x}_j^2}{|\tilde{x}_j| \sqrt{\alpha}} +\sum_{j = 1}^M |\tilde{x}_j| \sqrt{\alpha} \nonumber \\
& = &  \| \tilde{\bf L} \tilde{\bf x} - {\bf y} \|^2_2 + 2\sqrt{\alpha} \|\tilde{\bf x} \|_1.
\end{eqnarray}
Then, also ${\bf x}_\ell = {\bf D}^{-1} \tilde{\bf x}_\ell$ is the minimizer of $F({\bf x}) $ defined in (\ref{f_eq}), since $F({\bf x}) = \tilde{F}(\tilde{\bf x})$ with respect to the coordinate transform (\ref{transform}). The minimizer of $\tilde{F}(\tilde{\bf x})$ is also the 1-norm regularized solution of the linearized inverse problem.

\section*{Acknowledgements}
  
This work was supported by the Academy of Finland (Centre of Excellence in Inverse Problems Research).

\bibliography{references}

\begin{thebibliography}{10}
\providecommand{\url}[1]{#1}
\csname url@samestyle\endcsname
\providecommand{\newblock}{\relax}
\providecommand{\bibinfo}[2]{#2}
\providecommand{\BIBentrySTDinterwordspacing}{\spaceskip=0pt\relax}
\providecommand{\BIBentryALTinterwordstretchfactor}{4}
\providecommand{\BIBentryALTinterwordspacing}{\spaceskip=\fontdimen2\font plus
\BIBentryALTinterwordstretchfactor\fontdimen3\font minus
  \fontdimen4\font\relax}
\providecommand{\BIBforeignlanguage}[2]{{%
\expandafter\ifx\csname l@#1\endcsname\relax
\typeout{** WARNING: IEEEtran.bst: No hyphenation pattern has been}%
\typeout{** loaded for the language `#1'. Using the pattern for}%
\typeout{** the default language instead.}%
\else
\language=\csname l@#1\endcsname
\fi
#2}}
\providecommand{\BIBdecl}{\relax}
\BIBdecl

\bibitem{mann2008}
I.~Mann, A.~Nakamura, and T.~Mukai, \emph{Small Bodies in Planetary Systems},
  ser. Lecture Notes in Physics.\hskip 1em plus 0.5em minus 0.4em\relax
  Springer, 2008.

\bibitem{michel2015}
P.~Michel, F.~DeMeo, and W.~Bottke, \emph{Asteroids IV}, ser. Space Science
  Series.\hskip 1em plus 0.5em minus 0.4em\relax University of Arizona Press,
  2015.

\bibitem{kaasalainen2013}
M.~Kaasalainen and J.~Durech, ``Asteroid models for target selection and
  mission planning,'' in \emph{Asteroids: Prospective Energy and Material
  Resources}, V.~Badescu, Ed.\hskip 1em plus 0.5em minus 0.4em\relax
  Springer-Verlag GmbH, 2013.

\bibitem{kaasalainen1992}
M.~Kaasalainen, L.~Lamberg, and K.~Lumme, ``Interpretation of lightcurves of
  atmosphereless bodies.\ ii.\ practical aspects of inversion,'' \emph{A\&A},
  vol. 259, no.~1, pp. 333--340, 1992.

\bibitem{kaasalainen2006}
M.~Kaasalainen and L.~Lamberg, ``Inverse problems of generalized projection
  operators,'' \emph{Inverse Problems}, vol.~22, no.~3, pp. 749--769, 2006.

\bibitem{kaasalainen2011}
M.~Kaasalainen, ``Multimodal inverse problems: maximum compatibility estimate
  and shape reconstruction,'' \emph{Inverse Problems and Imaging}, vol.~5,
  no.~1, pp. 37--57, 2011.

\bibitem{kaipio2004}
J.~P. Kaipio and E.~Somersalo, \emph{Statistical and Computational Methods for
  Inverse Problems}.\hskip 1em plus 0.5em minus 0.4em\relax Berlin: Springer,
  2004.

\bibitem{nabighian1988}
M.~Nabighian, \emph{Electromagnetic Methods in Applied Geophysics: Volume 1,
  Theory}, ser. Electromagnetic Methods in Applied Geophysics: Applications
  Part A and Part B.\hskip 1em plus 0.5em minus 0.4em\relax Society of
  Exploration Geophysicists, 1988.

\bibitem{benna2002}
M.~Benna, J.-P. Barriot, and W.~Kofman, ``A priori information required for a
  two or three dimensional reconstruction of the internal structure of a comet
  nucleus ({CONSERT} experiment),'' \emph{Advances in Space Research}, vol.~29,
  no.~5, pp. 715 -- 724, 2002.

\bibitem{belton2004}
J.~Belton, \emph{Mitigation of Hazardous Comets and Asteroids}.\hskip 1em plus
  0.5em minus 0.4em\relax Cambridge University Press, 2004.

\bibitem{herique2002}
A.~Herique, J.~Gilchrist, W.~Kofman, and J.~Klinger,
  ``\BIBforeignlanguage{English}{Dielectric properties of comet analog
  refractory materials},'' \emph{\BIBforeignlanguage{English}{Planetary and
  Space Science}}, vol.~50, no.~9, pp. 857--863, AUG 2002.

\bibitem{kofman2015}
\BIBentryALTinterwordspacing
W.~Kofman, A.~Herique, Y.~Barbin, J.-P. Barriot, V.~Ciarletti, S.~Clifford,
  P.~Edenhofer, C.~Elachi, C.~Eyraud, J.-P. Goutail, E.~Heggy, L.~Jorda,
  J.~Lasue, A.-C. Levasseur-Regourd, E.~Nielsen, P.~Pasquero, F.~Preusker,
  P.~Puget, D.~Plettemeier, Y.~Rogez, H.~Sierks, C.~Statz, H.~Svedhem,
  I.~Williams, S.~Zine, and J.~Van~Zyl, ``Properties of the
  67p/churyumov-gerasimenko interior revealed by consert radar,''
  \emph{Science}, vol. 349, no. 6247, 2015. [Online]. Available:
  \url{http://www.sciencemag.org/content/349/6247/aab0639.abstract}
\BIBentrySTDinterwordspacing

\bibitem{kofman2007}
W.~Kofman, A.~Herique, J.-P. Goutail, T.~Hagfors, I.~P. Williams, E.~Nielsen,
  J.-P. Barriot, Y.~Barbin, C.~Elachi, P.~Edenhofer, A.-C. Levasseur-Regourd,
  D.~Plettemeier, G.~Picardi, R.~Seu, and V.~Svedhem, ``The comet nucleus
  sounding experiment by radiowave transmission ({CONSERT}): A short
  description of the instrument and of the commissioning stages,'' \emph{Space
  Science Reviews}, vol. 128, no. 1-4, pp. 413 -- 432, 2007.

\bibitem{kofman2004}
W.~Kofman, A.~Herique, and J.~Goutail, ``\BIBforeignlanguage{English}{{CONSERT}
  experiment: Description and performances in view of the new target},'' in
  \emph{\BIBforeignlanguage{English}{New {R}osetta Targets: Observations,
  Simulations and Instrument Performances}}, ser. Astrophysics and Space
  Science Library, L.~Colangeli, E.~Epifani, and P.~Palumbo, Eds., vol.
  311.\hskip 1em plus 0.5em minus 0.4em\relax Dordrecht, Netherlands: Springer,
  2004, Proceedings Paper, pp. 237--256.

\bibitem{benna2004}
M.~Benna, J.-P. Barriot, W.~Kofman, and Y.~Barbin, ``Generation of 3-d
  synthetic data for the modeling of the {CONSERT} experiment (the
  radiotomography of comet 67p/churyumov-gerasimenko),'' \emph{Antennas and
  Propagation, IEEE Transactions on}, vol.~52, no.~3, pp. 709 -- 716, march
  2004.

\bibitem{kofman1998}
W.~Kofman, Y.~Barbin, J.~Klinger, A.-C. Levasseur-Regourd, J.-P. Barriot,
  A.~Herique, T.~Hagfors, E.~Nielsen, and E.~Gr\, ``Comet nucleus sounding
  experiment by radiowave transmission,'' \emph{Advances in Space Research},
  vol.~21, no.~11, pp. 1589 -- 1598, 1998.

\bibitem{herique2011}
A.~{Herique}, A.~{Barucci}, J.~{Biele}, T.-M. {Ho}, W.~{Kofman}, C.~{Krause},
  P.~{Michel}, D.~{Plettemeier}, J.~Y. {Prado}, J.~C. {Souyris}, S.~{Zine}, and
  S.~{Ulamec}, ``{ASSERT : A Radar Tomography of Asteroids},'' in
  \emph{EPSC-DPS Joint Meeting 2011}, Oct. 2011, p. 924.

\bibitem{herique2011b}
------, ``{Radar Tomography of Asteroids},'' in \emph{EPSC-DPS Joint Meeting
  2011}, Oct. 2011, p. 920.

\bibitem{herique2010}
A.~{Herique}, D.~{Plettemeier}, W.~{Kofman}, S.~{Ulamec}, J.~{Biele},
  J.~{Goutail}, P.~{Beck}, J.~{Lassue}, A.~{Barucci}, and P.~{Michel},
  ``{{CONSERT} for Asteroid - radar tomography of Near Earth Asteroid},'' in
  \emph{EGU General Assembly 2010}, May 2010, p. 11011.

\bibitem{herique1999}
A.~{Herique}, W.~{Kofman}, T.~{Hagfors}, G.~{Caudal}, and J.-P. {Ayanides},
  ``{A characterization of a comet nucleus interior:inversion of simulated
  radio frequency data},'' \emph{\planss}, vol.~47, pp. 885--904, Jun. 1999.

\bibitem{landmann2010}
D.~Landmann, D.~Plettemeier, C.~Statz, F.~Hoffeins, U.~Markwardt, W.~Nagel,
  A.~Walther, A.~Herique, and W.~Kofman, ``Three-dimensional reconstruction of
  a comet nucleus by optimal control of maxwell's equations: A contribution to
  the experiment {CONSERT} onboard space craft {R}osetta,'' in \emph{Radar
  Conference, 2010 IEEE}, may 2010, pp. 1392 --1396.

\bibitem{nielsen2001}
E.~Nielsen, W.~Engelhardt, B.~Chares, L.~Bemmann, M.~Richards, F.~Backwinkel,
  D.~Plettemeier, P.~Edenhofer, Y.~Barbin, J.~Goutail, W.~Kofman, and
  L.~Svedhem, ``\BIBforeignlanguage{English}{Antennas for sounding of a
  cometary nucleus in the {R}osetta mission},'' in
  \emph{\BIBforeignlanguage{English}{Eleventh International Conference on
  Antennas and Propagation, Vols 1 and 2}}, ser. IEE Conference Publications,
  no. 480.\hskip 1em plus 0.5em minus 0.4em\relax 379 THORNALL ST, EDISON, NJ
  08837 USA: Inst Electrical Engineers Inspec INC, 2001, Proceedings Paper, pp.
  436--441.

\bibitem{barriot1999}
J.-P. Barriot, W.~Kofman, A.~Herique, S.~Leblanc, and A.~Portal, ``A two
  dimensional simulation of the {CONSERT} experiment (radio tomography of comet
  wirtanen),'' \emph{Advances in Space Research}, vol.~24, no.~9, pp. 1127 --
  1138, 1999.

\bibitem{pursiainen2013}
S.~Pursiainen and M.~Kaasalainen, ``Iterative alternating sequential ({IAS})
  method for radio tomography of asteroids in 3{D},'' \emph{Planetary and Space
  Science}, vol.~78, 2013.

\bibitem{pursiainen2014}
------, ``Detection of anomalies in radio tomography of asteroids: Source count
  and forward errors,'' \emph{Planetary and Space Science}, vol.~99, no.~0, pp.
  36 -- 47, 2014.

\bibitem{pursiainen2014b}
------, ``Sparse source travel-time tomography of a laboratory target: accuracy
  and robustness of anomaly detection,'' p. 114016, 2014.

\bibitem{daniels2004}
D.~J. Daniels, \emph{Ground Penetrating Radar (2nd Edition)}.\hskip 1em plus
  0.5em minus 0.4em\relax Stevenage: Institution of Engineering and Technology,
  2004.

\bibitem{agrawal2014}
V.~Agrawal, \emph{Satellite Technology: Principles and Applications}.\hskip 1em
  plus 0.5em minus 0.4em\relax Wiley, 2014.

\bibitem{schneider2012}
\BIBentryALTinterwordspacing
J.~B. Schneider, \emph{Understanding the {FDTD} Method}.\hskip 1em plus 0.5em
  minus 0.4em\relax John B. Schneider, 2012. [Online]. Available:
  \url{http://www.eecs.wsu.edu/~schneidj/ufdtd/}
\BIBentrySTDinterwordspacing

\bibitem{braess2007}
D.~Braess, \emph{Finite Elements: Theory, Fast Solvers, and Applications in
  Solid Mechanics}.\hskip 1em plus 0.5em minus 0.4em\relax Cambridge University
  Press, 2007.

\bibitem{scherzer2008}
O.~Scherzer, M.~Grasmair, H.~Grossauer, M.~Haltmeier, and F.~Lenzen,
  \emph{Variational Methods in Imaging}, ser. Applied Mathematical
  Sciences.\hskip 1em plus 0.5em minus 0.4em\relax Springer New York, 2008.

\bibitem{stefan2008}
W.~Stefan, \emph{Total Variation Regularization for Linear Ill-posed Inverse
  Problems: Extensions and Applications}.\hskip 1em plus 0.5em minus
  0.4em\relax Arizona State University, 2008.

\bibitem{golub1989}
G.~Golub and C.~van Loan, \emph{Matrix Computations}.\hskip 1em plus 0.5em
  minus 0.4em\relax Baltimore: The John Hopkins University Press, 1989.

\bibitem{evans1998}
L.~Evans, \emph{Partial Differential Equations}, ser. Graduate studies in
  mathematics.\hskip 1em plus 0.5em minus 0.4em\relax American Mathematical
  Society, 1998.

\bibitem{bossavit1999}
A.~Bossavit and L.~Kettunen, ``Yee-like schemes on a tetrahedral mesh, with
  diagonal lumping,'' \emph{International Journal of Numerical Modelling:
  Electronic Networks, Devices and Fields}, vol.~12, no. 1-2, pp. 129--142,
  1999.

\bibitem{yee1966}
K.~Yee, ``Numerical solution of initial boundary value problems involving
  maxwell's equations in isotropic media,'' \emph{Antennas and Propagation,
  IEEE Transactions on}, vol.~14, no.~3, pp. 302--307, 1966.

\bibitem{altman1991}
C.~Altman and K.~Suchy, \emph{Reciprocity, Spatial Mapping and Time Reversal in
  Electromagnetics}, ser. Developments in Electromagnetic Theory and
  Applications.\hskip 1em plus 0.5em minus 0.4em\relax Springer Netherlands,
  1991.

\bibitem{davis1989}
J.~L. Davis and A.~P. Annan, ``Ground penetrating radar for high resolution
  mapping of soil and rock stratigraphy,'' \emph{Geoscience Canada}, vol.~37,
  no.~5, pp. 531--551, 1989.

\bibitem{irving2006}
J.~Irving and R.~Knight, ``Numerical modeling of ground-penetrating radar in
  2-{D} using {MATLAB},'' \emph{Computers \& Geosciences}, vol.~32, no.~9, pp.
  1247 -- 1258, 2006.

\bibitem{harris1978}
F.~J. Harris, ``On the use of windows for harmonic analysis with the discrete
  fourier transform,'' \emph{Proceedings of the IEEE}, vol.~66, no.~1, pp.
  51--83, 1978.

\bibitem{nuttall1981}
A.~H. Nuttall, ``Some windows with very good sidelobe behavior,'' \emph{IEEE
  Transactions on Acoustics, Speech, Signal Processing}, vol. ASSP-29, no.~1,
  pp. 84--91, 1981.

\bibitem{binzel2005}
R.~P. Binzel and W.~Kofman, ``Internal structure of near-earth objects,''
  \emph{Comptes Rendus Physique}, vol.~6, no.~3, pp. 321--326, 2005.

\bibitem{francke2009}
J.~Francke and V.~Utsi, ``Advances in long-range gpr systems and their
  applications to mineral exploration, geotechnical and static correction
  problems,'' \emph{First Break}, vol.~27, no.~7, pp. 85--93, 2009.

\bibitem{colton1998}
D.~Colton and R.~Kress, \emph{Inverse Acoustic and Electromagnetic Scattering
  Theory}, ser. Applied Mathematical Sciences.\hskip 1em plus 0.5em minus
  0.4em\relax Springer, 1998.

\bibitem{pursiainen2015}
\BIBentryALTinterwordspacing
S.~Pursiainen and M.~Kaasalainen, ``Electromagnetic 3d subsurface imaging with
  source sparsity for a synthetic object,'' \emph{Inverse Problems}, vol.~31,
  no.~12, p. 125004, 2015. [Online]. Available:
  \url{http://stacks.iop.org/0266-5611/31/i=12/a=125004}
\BIBentrySTDinterwordspacing

\bibitem{miller2000}
D.~Miller, A.~Saenz-Otero, J.~Wertz, A.~Chen, G.~Berkowski, C.~Brodel,
  S.~Carlson, D.~Carpenter, S.~Chen, and S.~Cheng, ``{SPHERES}: A testbed for
  long duration satellite formation flying in micro-gravity conditions,''
  \emph{ADVANCES IN THE ASTRONAUTICAL SCIENCES}, vol. 105, pp. 167--180, 2000.

\bibitem{yeh2000}
H.-H. Yeh and A.~Sparks, ``Geometry and control of satellite formations,'' in
  \emph{American Control Conference, 2000. Proceedings of the 2000}, vol.~1,
  no.~6, Sep 2000, pp. 384--388 vol.1.

\bibitem{pongvthithum2005}
R.~Pongvthithum, S.~M. Veres, S.~B. Gabriel, and E.~Rogers, ``Universal
  adaptive control of satellite formation flying,'' \emph{International Journal
  of Control}, vol.~78, pp. 45--52, 2005.

\end{thebibliography}
\bibliographystyle{IEEEtran}

\begin{IEEEbiography}[{\includegraphics[width=1in,height=1.25in,clip,keepaspectratio]{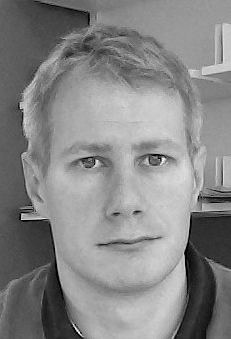}}]{Sampsa Pursiainen} received his MSc(Eng) 
and PhD(Eng) degrees (Mathematics)  in the Helsinki University of Technology  (Aalto University since 2010), Espoo, Finland, in 2003 and 2009. He focuses on various forward and inversion techniques of applied mathematics. In 2010--11, he stayed at the Department of Mathematics, University of Genova, Italy collaborating also with the Institute  for Biomagnetism and Biosignalanalysis (IBB), University of M\"{u}nster, Germany. In 2012--15, he worked at the  Department of Mathematics and  Systems Analysis, Tampere University of Technology Finland and also at the Department of Mathematics, Tampere University of Technology, Finland, where he currently holds an Assistant Professor position. 
\end{IEEEbiography}

\begin{IEEEbiography}[{\includegraphics[width=1in,height=1.25in,clip,keepaspectratio]{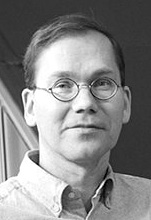}}]
{Prof.\ Mikko Kaasalainen} obtained his DPhil at Oxford in 1994, and he has been professor of applied mathematics at Tampere University of Technology since 2009. His main research fields are inverse problems and mathematical modelling, with interdisciplinary application projects ranging from biology to galactic dynamics. He is the vice director of the Centre of Excellence in Inverse Problems funded by the Academy of Finland. The website of his research group at TUT is at math.tut.fi/inversegroup
\end{IEEEbiography}

\end{document}